\documentclass[12pt]{article}
\usepackage{graphics}
\usepackage{makeidx}
\usepackage{epsf}
\usepackage{here}
\usepackage{feynmp}
\usepackage{mcite}
\usepackage{amsmath}
\usepackage{pstricks}
\usepackage{pst-node}

\def\sigmap{\sigma^{\prime}}
\def\mup{\mu^{\prime}}
\def\nup{\nu^{\prime}}
\def\rhop{\rho^{\, \prime}}

\def\cM{{\cal M}} 
\def\cO{{\cal O}}
\def\cK{{\cal K}} 
\def \ni {\noindent}
\def \be {\begin{equation}}
\def \e {\end{equation}}
\def \bea {\begin{eqnarray}}
\def \ea {\end{eqnarray}}
\def \eps {\epsilon}

\def \ga {\gamma}

\def \la {\lambda}
\def \no {\nonumber}
\def \G {{\rm g}}
\def \dd {{\rm d}}

\def \ps {p\hspace{-0.43em}/}
\def \mps {p\hspace{-0.45em}/}
\def \ns {n\hspace{-0.51em}/}
\def \mns {n\hspace{-0.53em}/}
\def \ks {k\hspace{-0.49em}/}
\def \es {\epsilon\hspace{-0.47em}/}

\def \Zw{\ensuremath{ {Z^{-\frac{1}{2}}_{\scriptscriptstyle W}\, } }}
\def \Zz{\ensuremath{ {Z^{-\frac{1}{2}}_{\scriptscriptstyle Z}\, } }}
\def \Zwn{\ensuremath{ \delta Z_{\scriptscriptstyle W,\,n} \,}}
\def \Zzn{\ensuremath{ \delta Z_{\scriptscriptstyle Z,\,n} \,}}

\newcommand{\To}[2]{\stackrel{#1}{\hbox to #2 pt{\rightarrowfill}}}

\def \pa {\partial}
\def \c {\hspace{-0.2em} \cdot}
\def\wp{\ifmmode W^+\else $W^+$\fi}
\def\wm{\ifmmode W^-\else $W^-$\fi}
\def\emm{\ifmmode e^-\else $e^-$\fi}
\def\ep{\ifmmode e^+\else $e^+$\fi}

\def\lsim{\;\raisebox{-.3em}{$\stackrel{\displaystyle <}{\sim}$}\;}

\def\sw{\ensuremath{ \sin \theta_{\rm w}}}
\def\swto{\ensuremath{ \sin^2 \theta_{\rm w}}} 
\def\swfor{\ensuremath{ \sin^4 \theta_{\rm w}}} 
\def\cw{\ensuremath{ \cos \theta_{\rm w}}} 
\def\cwto{\ensuremath{ \cos^2 \theta_{\rm w}}}

\def\ie{{\it i.e.~}}
\def\eg{{\it e.g.~}}

\def\mw{\ensuremath{ {M}_{\scriptscriptstyle W}} }
\def\mz{\ensuremath{ {M}_{\scriptscriptstyle Z}} }
\def\mh{\ensuremath{ {M}_{\scriptscriptstyle H}} }
\def\mn{\ensuremath{ {M}_{\scriptscriptstyle N}} }
\def\mwp{\ensuremath{ {M}_{{\scriptscriptstyle W},\,{\rm phys. }}}}

\def \Cw#1{\ensuremath { \delta Z^{\, (1)}_{f_{#1}}(W)}}
\def \Cz#1{\ensuremath { \delta Z^{\, (1)}_{f_{#1}}(Z)}}
\def \Cga#1{\ensuremath { \delta Z^{\, (1)}_{f_{#1}}(\ga) }}

\def\L{\ensuremath{ {\rm L}}}
\def\z#1{\ensuremath { \frac{\dd z_{#1}}{z_{#1}}  }}
\def\y#1{\ensuremath { \frac{\dd y_{#1}}{y_{#1}}  }}

\def \onshellW{\vphantom{\frac{A^2}{A^2}}\left|_{_{\scr{\,k^2=\mw^2}}}\right.}
\def \onshellZ{\vphantom{\frac{A^2}{A^2}}\left|_{_{\scr{\,k^2=\mz^2}}}\right.}
\def \onshellA{\vphantom{\frac{A^2}{A^2}}\left|_{_{\scr{\,k^2=0}}}\right.}
\def \onshellH{\vphantom{\frac{A^2}{A^2}}\left|_{_{\scr{\,k^2=\mh^2}}}\right.}

\def \onshellWp{\vphantom{\frac{A^2}{A^2}}\left|_{_{\scr{\,k^2=\mwp^2}}}
                \right.}

\def \h#1{ \hspace*{#1mm}}
\def \v#1{ \vspace*{#1mm}}

\def\dis#1{\ensuremath { {\displaystyle  #1}}}  
\def\scr#1{\ensuremath { { \scriptstyle #1}}}  
\def\sscr#1{\ensuremath { { \scriptscriptstyle #1}}}  

\def \myto#1#2{\ensuremath {\begin{array}[H]{c}
 {\scriptstyle {\rm #1}} \\[-2mm] {-\!\!\!-\!\!\!-\!\!\!-\!\!\!\longrightarrow}
 \\[-2mm] {\scriptstyle {\rm #2}}
\end{array} }}

\def \myk{\ensuremath {\begin{array}[H]{c} \\[-7mm] {-\!\!\!-\!\!\!
 \longrightarrow} \\[-2mm] k
\end{array} }}

\def \Myto#1{\ensuremath {\begin{array}[H]{c}
 {\scriptstyle { #1}} \\[-2mm] {-\!\!\!\longrightarrow} \end{array} }}

\def \e#1#2{\ensuremath { \eps_{\,#1}^{\,*}\,({#2})\,\, }}

\begin{document}

\baselineskip 7mm
\pagestyle{empty}

\vspace*{-2cm}
\renewcommand{\thefootnote}{\fnsymbol{footnote}}  
\begin{flushright}  
DESY 01--131\\  
hep-ph/0112030\\
December 2001 \\  
\end{flushright}  
\vspace{5mm} 
\begin{center}  
{\Large \bf Electroweak two-loop Sudakov logarithms \\[2mm] for on-shell
  fermions and bosons}\footnote{Work supported by the European Union under
  contract HPRN-CT-2000-00149 \\[-1mm]}  \\ 
\vspace{1cm} 
{  {\large
W. Beenakker}${}^{1}$\footnote{wimb@sci.kun.nl}  and
{\large A. Werthenbach}${}^2$\footnote{Anja.Werthenbach@desy.de} }\\  
\vspace{1cm}  
{\small 
{\sf 1) Theoretical Physics, Univ. of Nijmegen, P.O. Box 9010, NL-6500 GL
  Nijmegen, The Netherlands\\ }
{\sf 2) Deutsches Elektronen-Synchrotron, DESY Zeuthen, Platanenallee
  6, D-15738 Zeuthen, Germany  }  }
\end{center} 
   
\vspace{5mm} 

\begin{center}
{\bf \sc{Abstract}}
\end{center}
We calculate the virtual electroweak Sudakov (double) logarithms at one- and 
two-loop level for arbitrary on-shell/on-resonance particles in the Standard 
Model. The associated Sudakov form factors apply in a universal way to 
arbitrary non-mass-suppressed electroweak processes at high energies, although
this universality has to be interpreted with care. The actual calculation is 
performed in the temporal Coulomb gauge, where the relevant contributions from 
collinear-soft gauge-boson exchange are contained exclusively in the 
self-energies of the external on-shell/on-resonance particles. In view of the 
special status of the time-like components in this gauge, a careful analysis of
the asymptotic states of the theory is required. From this analysis we derive 
an all-order version of the Goldstone-boson Equivalence Theorem without the 
need for finite compensation factors. By exploiting conditions obtained from 
non-renormalization requirements, which are a consequence of our choice of 
gauge, we show that the Sudakov corrections can be extracted through a
combination of energy derivatives and projections by means of external sources.
We observe that the Standard Model behaves dynamically like an unbroken theory 
in the Sudakov limit, in spite of the fact that the explicit particle masses 
are needed at the kinematical (phase-space) level while calculating the Sudakov
form factors. 
\\

\setcounter{footnote}{0}  
\renewcommand{\thefootnote}{\arabic{footnote}}  
\vfill  
\clearpage

\tableofcontents

\newpage
\setcounter{page}{1}  
\pagestyle{plain} 
\section{Introduction}

At the next generation of colliders center-of-mass energies will be reached 
that largely exceed the electroweak scale. For instance, the energy at a future
linear $e^+e^-$ collider is expected to be in the TeV 
range~\cite{Accomando:1997wt}.
At these energies one enters the realm of large perturbative corrections.
Even the effects arising from weak corrections are expected to be of the order 
of $10\%$ or more~\cite{Kuroda:1991wn}--\cite{Denner:2000jv}, \ie just as
large as the well-known electromagnetic 
corrections. In order not to jeopardize any of the high-precision studies at 
these high-energy colliders, it is therefore indispensable to improve the 
theoretical understanding of the radiative corrections in the weak sector of 
the Standard Model (SM). In particular this will involve a careful analysis of 
effects beyond first order in the perturbative expansion in the 
(electromagnetic) coupling $\,\alpha=e^2/(4\pi)$.
\\[-5mm]

\ni 
The dominant source of radiative corrections at TeV-scale energies is given by 
logarithmically enhanced effects of the form $\,\alpha^n\log^m(M^2/s)\,$ for 
$\,m\le 2n$, involving particle masses $M$ well below the collider energy 
$\sqrt{s}$. A natural way of controlling the theoretical uncertainties would 
therefore consist in a comprehensive study of these large logarithms, taking 
into account all possible sources (i.e.\ ultraviolet, soft, and collinear). 
The potentially most important electroweak corrections are the so-called 
Sudakov logarithms $\,\propto \alpha^n\log^{2n}(M^2/s)$, arising from 
collinear-soft singularities~\cite{Sudakov:1956sw}. It should be noted,
however, that for pure fermionic final states (numerical) cancellations can 
take place between leading and subleading logarithms~\cite{Kuhn:2001hz}. 
For on-shell bosons in the final state, the Sudakov logarithms in general tend
to be dominant~\cite{Beenakker:1993tt,Denner:2000jv}.
\\[-5mm]

\ni 
Over the last few years various QCD-motivated methods have been applied to 
predict the electroweak Sudakov logarithms to all orders in perturbation 
theory~\cite{Ciafaloni:1999ub,Fadin:1999bq,Kuhn:1999nn}. The methods vary in 
the way that the QCD-motivated factorization and exponentiation properties are 
translated to the electroweak theory. This is caused by the fact that the 
electroweak theory is a spontaneously broken theory with two mass scales in 
the gauge-boson sector, whereas QCD is basically a single-scale theory. 
The main debate therefore focusses on the question ``to what extent does 
the SM behave like an unbroken theory at high energies?'' In fact, we already
know that the transition from QCD to electroweak theory does not come without 
surprises. In Ref.~\cite{Ciafaloni:2000rp} it was shown that the 
Bloch--Nordsieck cancellation between virtual and real collinear-soft 
gauge-boson radiation~\cite{Bloch:1937pw} is violated in the 
SM as soon as initial- or final-state particles carry 
an explicit weak charge (isospin) and summation over the partners within an
$SU(2)$ multiplet is not performed. At an electron--positron collider, for 
instance, the weak isospin of the initial-state particles is fixed by the 
accelerator and the Bloch--Nordsieck theorem is in general violated for 
left-handed initial states, even for fully inclusive cross-sections. 
The resulting electroweak effects can be very large, exceeding the QCD 
corrections for energies in the TeV range. With this in mind, explicit 
calculations of Sudakov corrections at two-loop level are needed to resolve any
ambiguity in the translation from QCD to SM. Up to know the explicit two-loop
calculations have been performed for pure fermionic processes, like  
$\,e^+e^- \to f\bar{f}\,$~\cite{Beenakker:2000kb} and fermion-pair production 
by an $\,SU(2)\times U(1)$-singlet source~\cite{Hori:2000tm,Melles:2000ed}.
\\[-5mm]

\ni 
In this paper we complete our previous analysis~\cite{Beenakker:2000kb} of 
virtual Sudakov logarithms for fermions by extending the explicit two-loop 
calculation to scalar particles as well as transverse and longitudinal gauge 
bosons. By means of this explicit calculation we try to establish to what 
extent the SM behaves like an unbroken theory at high energies. 
Like in Ref.~\cite{Beenakker:2000kb}, we perform the calculation in 
the (temporal) Coulomb gauge, exploiting the fact that the Sudakov form
factors are exclusively contained in the self-energies of the particles.
A detailed description of the Coulomb-gauge method for massive particles, 
including a discussion of the asymptotic states, is presented in 
Sects.~\ref{sec:born} and~\ref{sec:wave}. For longitudinal gauge bosons we
carefully study the effects of the broken nature of the SM on the asymptotic 
states. This will result in a special, particularly simple form of the
Goldstone-boson Equivalence Theorem where the longitudinal degrees of freedom 
can be substituted by the corresponding Goldstone-boson degrees of freedom 
without the need for finite compensation factors. Also in the transverse 
neutral gauge-boson sector, with its mixing between the $Z$ boson and the 
photon, a careful reanalysis of the asymptotic states beyond lowest order in 
perturbation theory is needed. By exploiting conditions obtained from 
non-renormalization requirements, which are a consequence of our choice of 
gauge, we will show that the Sudakov corrections can be extracted through a
combination of energy derivatives and projections by means of external sources.
In Sects.~\ref{sec:one-loop} and \ref{sec:two-loop} we explicitly perform the
one- and two-loop calculations of the Sudakov form factors. These explicit 
calculations, combined with the special properties of the Coulomb gauge, 
enable us to identify the similarities and differences between the SM and 
unbroken (single-scale) theories like QED and QCD.

\section{The Coulomb gauge}
\label{sec:born}

In order to facilitate the calculation of the one- and two-loop Sudakov 
logarithms, we work in the Coulomb gauge for both massless and massive gauge
bosons~\cite{Beenakker:2000kb}. In the Coulomb gauge the gauge-fixing 
Lagrangian for the gauge-boson mass eigenstates $\,V=W^{\pm},Z,\ga\,$ is given 
by 
\bea
\label{coulGF}
{\cal L}_{GF} &=& -\,\frac{\la}{2}\sum\limits_{V=W^{\pm},Z,\ga}
    \left[ \left( \pa_{\mu} -\frac{n \cdot \pa}{n^2}\,n_{\mu}\right) 
           V^{\dagger, \,\mu} \, \right]\,
    \left[ \left( \pa_{\nu} -\frac{n \cdot \pa}{n^2}\,n_{\nu}\right) 
           V^{\nu} \, \right]\,,
\ea
with the temporal gauge vector $\,n^{\mu}= (1,0,0,0)$. As a result of the gauge
choice, the massive gauge bosons ($W^{\pm}$ and $Z$) will mix at lowest-order
level with the corresponding would-be Goldstone bosons ($\phi^{\pm}$ and 
$\chi$), defined through the SM Higgs doublet 
\bea
  \Phi(x) = \left( \begin{array}{c}
                     \phi^+(x) \\
                     \frac{1}{\sqrt{2}}\,[v+H(x)+i\chi(x)] 
                   \end{array} \right)
\ea
with vacuum expectation value 
$\,\langle\Phi\rangle=\left( \begin{array}{c} 0 \\ v/\sqrt{2}
                             \end{array} \right)$
and hypercharge $Y_{\phi}=1$.
This has to be contrasted with the covariant $R_{\xi}$ gauges, which by 
construction do not exhibit such mixing at lowest order.
Selecting the bilinear interactions in the $\,W-\phi\,$ and $\,Z-\chi\,$
sectors, we obtain the following relevant pieces of Lagrangian in the Coulomb 
gauge: 
\bea
\label{L_Wphi}
{\cal L}_{W-\phi}^{\sscr{\rm bilinear}} &=& 
         (\pa^{\mu}W_{\mu}^+)(\pa^{\nu}W_{\nu}^-)
       - (\pa_{\mu}W_{\nu}^+)(\pa^{\mu}W^{-,\,\nu}) 
       + (\pa_{\mu} \phi^+)(\pa^{\mu}\phi^-) 
       +\mw^2\,W_{\mu}^+W^{-,\,\mu} \\[2mm] 
                                    &+& 
         i\mw \,[ (\pa^{\mu}W_{\mu}^+)\phi^- - (\pa^{\nu}W_{\nu}^-)\phi^+ ]
       - \lambda\left[ \left( \pa_{\mu} - \frac{n\c\pa}{n^2}\,n_{\mu} \right) 
                       W^{+,\,\mu} \right]  
                \left[ \left( \pa_{\nu} - \frac{n\c\pa}{n^2}\,n_{\nu} \right) 
                       W^{-,\,\nu} \right] \no\\[3mm]
\label{L_Zchi}
{\cal L}_{Z-\chi}^{\sscr{\rm bilinear}} &=&  
         \frac{1}{2}\,(\pa^{\mu}Z_{\mu})(\pa^{\nu}Z_{\nu}) 
       - \frac{1}{2}\,(\pa_{\mu}Z_{\nu})(\pa^{\mu}Z^{ \nu}) 
       + \frac{1}{2}\,(\pa_{\mu} \chi)(\pa^{\mu}\chi)
       + \frac{1}{2}\,\mz^2\,Z_{\mu}Z^{\mu} \\[2mm]
                                    &-& \mz\,(\pa^{\mu}Z_{\mu})\chi
       - \frac{\la}{2}\,\left[ \left( \pa_{\mu} - \frac{n\c\pa}{n^2}\,n_{\mu}
                       \right) Z^{\mu} \right]  
                \left[ \left( \pa_{\nu} - \frac{n\c\pa}{n^2}\,n_{\nu} \right) 
                       Z^{\nu} \right] \, \no.
\ea
Hence the lowest-order interaction matrix in the charged-boson sector can be 
written as
\bea
\label{coulint}
  \left( \begin{array}{cc}
           -i\,\left[ (k^2-\mw^2)\,\G^{\mu\nu} - k^{\mu}k^{\nu} 
                       + \la \left( k^{\mu}-k_0\,n^{\mu} \right)\, 
                             \left( k^{\nu}-k_0\,n^{\nu} \right) \right] 
           & \hspace*{5mm} \pm\,i\,\mw k^{\mu} \no\\[4mm]
           \pm\,i\,\mw k^{\nu} & i\,k^2
         \end{array} \right) 
\ea
\vspace*{-7mm}
\begin{figure}[H]
\bea  
  =\,\, \left(
   \begin{array}{cc}
      \hspace*{15mm}
     \begin{fmffile}{coulww}
       \begin{fmfgraph*}(60,20) \fmfpen{thin} \fmfleft{i1} \fmfright{o1}
         \fmf{boson}{i1,v1,o1} \fmfdot{v1} \fmf{phantom_arrow}{i1,v1}
         \fmf{phantom_arrow}{v1,o1}
         \fmflabel{$W^{\pm,\mu}, k$}{i1} \fmflabel{$W^{\pm,\nu}, k$}{o1}
       \end{fmfgraph*} 
      \end{fmffile} \hspace*{15mm} &  \hspace*{15mm}
      \begin{fmffile}{coulwphi}
       \begin{fmfgraph*}(60,20) \fmfpen{thin} \fmfleft{i1} \fmfright{o1}
         \fmf{boson,tension=1}{i1,v1} \fmf{phantom_arrow}{i1,v1} \fmf{scalar
              ,tension=2}{v1,o1} \fmfdot{v1} 
         \fmflabel{$W^{\pm,\mu}, k$}{i1} \fmflabel{$\phi^{\pm}, k$}{o1}
       \end{fmfgraph*}
      \end{fmffile}\\[1mm]
        \hspace*{15mm}
      \begin{fmffile}{coulphiw}
        \begin{fmfgraph*}(60,20) \fmfpen{thin} \fmfleft{i1} \fmfright{o1}
           \fmf{boson,tension=1}{v1,o1} \fmf{phantom_arrow}{v1,o1}
           \fmf{scalar,tension=2}{i1,v1} \fmfdot{v1} 
           \fmflabel{$W^{\pm,\nu}, k$}{o1} \fmflabel{$\phi^{\pm}, k$}{i1}
        \end{fmfgraph*}
       \end{fmffile}  \hspace*{15mm} &  \hspace*{15mm}
       \hspace*{15mm}
       \begin{fmffile}{coulphiphi} 
         \begin{fmfgraph*}(60,20) \fmfpen{thin} \fmfleft{i1} \fmfright{o1}
             \fmf{scalar}{i1,v1,o1} \fmfdot{v1} 
           \fmflabel{$\phi^{\pm}, k$}{i1} \fmflabel{$\phi^{\pm}, k$}{o1}
              \end{fmfgraph*} 
        \end{fmffile}
        \hspace*{15mm}
  \end{array}
 \right) \,,
\ea
\end{figure}

\vspace*{-8mm}
\ni
where the $\pm$ occurring in the first matrix of Eq.~(\ref{coulint}) 
correspond to $W^{\pm}$.
\\[-5mm]

\ni
The propagators in the Coulomb gauge are now obtained by inverting the 
interaction matrix and taking the limit $\lambda \to \infty$: 

\vspace*{-3mm}
\begin{minipage}[l]{3.5cm}
\begin{figure}[H]
\bea
 \hspace*{-5mm}
 \left(
   \h{2}
   \begin{array}{cc}
     \begin{fmffile}{cww}
       \begin{fmfgraph*}(60,20) \fmfpen{thin} \fmfleft{i1} \fmfright{o1}
         \fmf{boson}{i1,v1,o1} \fmfdot{v1} \fmf{phantom_arrow}{i1,v1}
         \fmf{phantom_arrow}{v1,o1}
         \fmfv{l=${\mu}$,l.a=80}{i1}
         \fmfv{l=${\nu}$,l.a=100}{o1}
       \end{fmfgraph*} 
      \end{fmffile} \hspace*{2mm}  
                   & \hspace*{2mm} 
      \begin{fmffile}{cwphi}
       \begin{fmfgraph*}(60,20) \fmfpen{thin} \fmfleft{i1} \fmfright{o1}
         \fmf{boson,tension=1}{i1,v1} \fmf{phantom_arrow}{i1,v1} \fmf{scalar
              ,tension=2}{v1,o1} \fmfdot{v1} 
         \fmfv{l=${\mu}$,l.a=80}{i1}
       \end{fmfgraph*}
      \end{fmffile}\\[1mm]
      \begin{fmffile}{cphiw}
        \begin{fmfgraph*}(60,20) \fmfpen{thin} \fmfleft{i1} \fmfright{o1}
           \fmf{boson,tension=1}{v1,o1} \fmf{phantom_arrow}{v1,o1}
           \fmf{scalar,tension=2}{i1,v1} \fmfdot{v1} 
         \fmfv{l=${\nu}$,l.a=100}{o1}
        \end{fmfgraph*}
       \end{fmffile}  \hspace*{2mm} 
                    & \hspace*{2mm}
       \begin{fmffile}{cphiphi} 
         \begin{fmfgraph*}(60,20) \fmfpen{thin} \fmfleft{i1} \fmfright{o1}
             \fmf{scalar}{i1,v1,o1} \fmfdot{v1} 
              \end{fmfgraph*} 
        \end{fmffile}
  \end{array}
 \right) \no 
\ea
\end{figure}
\end{minipage}
\begin{minipage}[r]{12.2cm}
\bea
\h{2}
\left(
  \begin{array}{cc}
    P^{\sscr{W}^{\pm}\sscr{W}^{\pm}}_{\nu \rho} 
    & P^{\sscr{W}^{\pm}\phi^{\pm}}_{\nu} 
    \\[1mm] 
    P^{\phi^{\pm}\sscr{W}^{\pm}}_{\rho}  
    & P^{\phi^{\pm}\phi^{\pm}}
  \end{array}
\right) = 
\left(
  \begin{array}{cc}
    -\,\delta^{\mu}_{\phantom{\mu} \rho} & 0  \\[1mm] 0  & -\,1
  \end{array}
\right) \, .
\ea
\end{minipage}

\vspace*{2mm}
\ni
This leads to the explicit form of the lowest-order propagators

\vspace*{-2mm}
\begin{figure}[H]
\begin{fmffile}{prop}
\begin{minipage}[l]{3cm}
\hspace*{4mm}
\begin{fmfgraph*}(80,40) \fmfpen{thin} \fmfleft{i1} \fmfright{o1}
 \fmf{boson,label=$W^{\pm}$,l.side=left}{i1,v1,o1} \fmfdot{i1,o1} 
 \fmf{phantom_arrow}{i1,v1}
 \fmf{phantom_arrow}{v1,o1} 
 \fmfv{l=${\mu}$,l.a=-130}{i1} \fmfv{l=${\nu}$,l.a=-50}{o1}
 \fmfv{l=$\myk$,l.a=-90}{v1} 
\end{fmfgraph*} \\[-0.5cm]

\hspace*{4mm}
\begin{fmfgraph*}(80,40) \fmfpen{thin} \fmfleft{i1} \fmfright{o1}
 \fmf{boson,tension=1,label=$W^{\pm}$,l.side=left}{i1,v1}
 \fmf{phantom_arrow}{i1,v1}
 \fmf{scalar,tension=2,label=$\phi^{\pm}$,l.side=left}{v1,o1}  \fmfdot{i1,o1} 
 \fmfv{l=$\myk$,l.a=-90}{v1} \fmfv{l=${\mu}$,l.a=-130}{i1}
\end{fmfgraph*} \\[-0.5cm]

\hspace*{4mm}
\begin{fmfgraph*}(80,40) \fmfpen{thin} \fmfleft{i1} \fmfright{o1}
 \fmf{boson,tension=1,label=$W^{\pm}$,l.side=left}{v1,o1}
 \fmf{phantom_arrow}{v1,o1}
 \fmf{scalar,tension=2,label=$\phi^{\pm}$,l.side=left}{i1,v1} \fmfdot{i1,o1} 
 \fmfv{l=$\myk$,l.a=-90}{v1}  \fmfv{l=${\nu}$,l.a=-50}{o1}
\end{fmfgraph*}\\[-0.5cm]

\hspace*{4mm}
\begin{fmfgraph*}(80,40) \fmfpen{thin} \fmfleft{i1} \fmfright{o1}
 \fmf{scalar,label=$\phi^{\pm}$,l.side=left}{i1,v1,o1}
 \fmfv{l=$\myk$,l.a=-90}{v1}  
\fmfdot{i1,o1}
\end{fmfgraph*} 
\end{minipage}
\begin{minipage}[r]{13.4cm}
\vspace*{-0.5cm}
\bea
\label{wphi0}
\hspace*{3mm}&:& 
  P^{\sscr{W}^{\pm}\sscr{W}^{\pm}}_{\mu\nu} = P_{\mu \nu}(k,\mw) = 
  \frac{-\,i}{k^2-\mw^2+i\,\eps}\; 
  \left( \G_{\mu \nu} + \frac{k_{\mu}k_{\nu}}{\vec{k}^2} 
         - k_0 \,\frac{k_{\mu}n_{\nu} + k_{\nu}n_{\mu}}{ \vec{k}^2} \right) 
  \no\\[0.3cm]
\hspace*{3mm}&:& 
  P^{\sscr{W}^{\pm}\phi^{\pm}}_{\mu} = \mp\,M_{\mu}(k,\mw) = 
  \frac{\mp\,i\,\mw}{k^2-\mw^2+i\,\eps} \; \frac{k_0}{\vec{k}^2}\,n_{\mu} 
  \no\\[0.3cm]
\hspace*{3mm}&:& 
  P^{\phi^{\pm}\sscr{W}^{\pm}}_{\nu} = \mp\,M_{\nu}(k,\mw) = 
  \frac{\mp\,i\,\mw}{k^2-\mw^2+i\,\eps} \; \frac{k_0}{\vec{k}^2}\,n_{\nu} 
  \no\\[0.3cm]
\hspace*{3mm}&:& 
  P^{\phi^{\pm}\phi^{\pm}} =  
  \frac{i}{k^2-\mw^2+i\,\eps} \; \left( 1 + \frac{\mw^2}{\vec{k}^2} \right)\,.
\ea
\end{minipage}
\end{fmffile} 
\end{figure}

\ni
In the neutral $\,Z-\chi\,$ sector the propagators are given by

\vspace*{-2mm}
\begin{figure}[H]
\begin{fmffile}{propZ}
\begin{minipage}[l]{3cm}
\hspace*{4mm}
\begin{fmfgraph*}(80,40) \fmfpen{thin} \fmfleft{i1} \fmfright{o1}
 \fmf{boson,label=$Z$,l.side=left}{i1,v1,o1} \fmfdot{i1,o1} 
 \fmf{phantom_arrow}{i1,v1}
 \fmf{phantom_arrow}{v1,o1} 
 \fmfv{l=${\mu}$,l.a=-130}{i1} \fmfv{l=${\nu}$,l.a=-50}{o1}
 \fmfv{l=$\myk$,l.a=-90}{v1} 
\end{fmfgraph*} \\[-0.5cm]

\hspace*{4mm}
\begin{fmfgraph*}(80,40) \fmfpen{thin} \fmfleft{i1} \fmfright{o1}
 \fmf{boson,tension=1,label=$Z$,l.side=left}{i1,v1}
 \fmf{phantom_arrow}{i1,v1}
 \fmf{scalar,tension=2,label=$\chi$,l.side=left}{v1,o1}  \fmfdot{i1,o1} 
 \fmfv{l=$\myk$,l.a=-90}{v1} \fmfv{l=${\mu}$,l.a=-130}{i1}
\end{fmfgraph*} \\[-0.5cm]

\hspace*{4mm}
\begin{fmfgraph*}(80,40) \fmfpen{thin} \fmfleft{i1} \fmfright{o1}
 \fmf{boson,tension=1,label=$Z$,l.side=left}{v1,o1}
 \fmf{phantom_arrow}{v1,o1}
 \fmf{scalar,tension=2,label=$\chi$,l.side=left}{i1,v1} \fmfdot{i1,o1} 
 \fmfv{l=$\myk$,l.a=-90}{v1}  \fmfv{l=${\nu}$,l.a=-50}{o1}
\end{fmfgraph*}\\[-0.5cm]

\hspace*{4mm}
\begin{fmfgraph*}(80,40) \fmfpen{thin} \fmfleft{i1} \fmfright{o1}
 \fmf{scalar,label=$\chi$,l.side=left}{i1,v1,o1}
 \fmfv{l=$\myk$,l.a=-90}{v1}  
\fmfdot{i1,o1}
\end{fmfgraph*} 
\end{minipage}
\begin{minipage}[r]{13.4cm}
\vspace*{-0.5cm}
\bea
\label{zchi0}
\hspace*{3mm}&:&
  P^{\sscr{ZZ}}_{\mu\nu} = P_{\mu \nu}(k,\mz) =  
  \frac{-\,i}{k^2-\mz^2+i\,\eps}\; 
  \left( \G_{\mu \nu} + \frac{k_{\mu}k_{\nu}}{\vec{k}^2} 
         - k_0 \,\frac{k_{\mu}n_{\nu}+k_{\nu}n_{\mu}}{ \vec{k}^2} \right) 
  \no\\[0.3cm]
\hspace*{3mm}&:&
  P^{\sscr{Z}\chi}_{\mu} = -\,i\,M_{\mu}(k,\mz) =  
  \frac{\mz}{k^2-\mz^2+i\,\eps} \; \frac{k_0}{\vec{k}^2}\,n_{\mu} 
  \no\\[0.3cm]
\hspace*{3mm}&:&
 P^{\chi\sscr{Z}}_{\nu} = i\,M_{\nu}(k,\mz) = 
  \frac{-\,\mz}{k^2-\mz^2+i\,\eps} \; \frac{k_0}{\vec{k}^2}\,n_{\nu} 
  \no\\[0.3cm]
\hspace*{3mm}&:&
  P^{\chi\chi} =  \frac{i}{k^2-\mz^2+i\,\eps} \; 
                  \left( 1 + \frac{\mz^2}{\vec{k}^2} \right) \,,
\end{eqnarray}
\end{minipage}
\end{fmffile} 
\end{figure}

\ni
and for the photon we obtain

\vspace*{-2mm}
\begin{figure}[H]
\begin{fmffile}{propga}
\begin{minipage}[l]{3cm}
\hspace*{4mm}
\begin{fmfgraph*}(80,40) \fmfpen{thin} \fmfleft{i1} \fmfright{o1}
 \fmf{boson,label=$\ga$,l.side=left}{i1,v1,o1} \fmfdot{i1,o1} 
 \fmf{phantom_arrow}{i1,v1}
 \fmf{phantom_arrow}{v1,o1} 
 \fmfv{l=${\mu}$,l.a=-130}{i1} \fmfv{l=${\nu}$,l.a=-50}{o1}
 \fmfv{l=$\myk$,l.a=-90}{v1} 
\end{fmfgraph*} 
\end{minipage}
\begin{minipage}[r]{13.4cm}
\vspace*{-0.5cm}
\begin{eqnarray}
\label{gaga0}
\hspace*{-2mm}&:&
  P^{\ga\ga}_{\mu\nu} = P_{\mu \nu}(k,0) =  
  \frac{-\,i}{k^2+i\,\eps}\; 
  \left( \G_{\mu \nu} + \frac{k_{\mu}k_{\nu}}{\vec{k}^2} 
         - k_0 \,\frac{k_{\mu}n_{\nu}+k_{\nu}n_{\mu}}{ \vec{k}^2} \right)\,.
\end{eqnarray}
\end{minipage}
\end{fmffile} 
\end{figure}

\vspace*{2mm}
\ni
The properties of the above-given propagators, like the relation between 
$\,P^{\sscr{Z}\chi}_{\mu}\,$ and $\,P^{\chi\sscr{Z}}_{\mu}\,$ etc., follow 
from the hermiticity of the bilinear Lagrangians (\ref{L_Wphi}) and 
(\ref{L_Zchi}).
\\[-5mm]

\ni
The power of choosing the Coulomb gauge lies in the fact that in the 
kinematical region of interest the gauge-boson propagators become effectively 
transverse:
\bea
\label{trans}
  P_{\mu\nu}(k,M) &=& -\,i\,\frac{\vec{k}^{\,2}\,\G_{\mu\nu}
        + k_{\mu}k_{\nu} - k_0\,( k_{\mu}n_{\nu}+n_{\mu}k_{\nu} )}
        {\vec{k}^{\,2}\,(k^2-M^2+i \epsilon)} 
        \ =\ \frac{-\,i}{k^2-M^2+i \epsilon}\,
        \Bigl[ Q_{\mu\nu}(k)-\frac{k^2}{\vec{k}^{\,2}}\,n_{\mu}n_{\nu}\Bigr].
        \quad\quad
\ea
The tensor
\bea
\label{polsum}
  Q_{\mu\nu}(k) = - \sum\limits_{\la=\pm}\eps_{\mu}(k,\la)\,
                                         \eps_{\nu}^{\star}(k,\la)
\ea
is the polarization sum for the transverse helicity states, characterized by
\begin{equation}
  \eps(k,\pm)\cdot n = 0 
  \quad , \quad
  \vec{\eps}\,(k,\pm)\cdot \vec{k} = 0 
  \quad , \quad
  \vec{\eps}\,(k,\pm)\cdot\vec{\eps}\,(k,\mp) = 0
  \quad {\rm and} \quad
  \vec{\eps}\,(k,\pm)\cdot\vec{\eps}\,(k,\pm) = 1 \,.
\end{equation}
Therefore the 
gauge bosons are effectively transverse if $\,k^2 \ll \vec{k}^{\,2}$, which
is the case for collinear gauge-boson emission at high energies 
($k^2 \propto M^2$ and $\vec{k}^{\,2} \approx k_0^2 \gg M^2$). As a result
of the effective transversality, the virtual Sudakov logarithms originating 
from vertex, box etc.\ corrections are suppressed\footnote{We will come back
    to that in sect.~\ref{sec:one-loop_fermions}, once we have established all
    the necessary ingredients} 
as long as the two defining conditions for Sudakov corrections are met, \ie
all kinematical invariants of the process under investigation have to be of 
the same order as the initial-state centre-of-mass (CM) energy squared and the 
lowest-order matrix element should not be suppressed by powers of 
$\,M/k_0\,$ to start with. \\[-5mm]

\ni
Hence, all virtual Sudakov logarithms are contained exclusively in the 
self-energies of the external on-shell particles (external wave-function 
factors)~\cite{Beenakker:2000kb,Frenkel:1976bj} or the self-energies of any 
intermediate particle that happens to be effectively on-shell.\footnote{Note 
    that similar simplifications can probably be obtained equally well by 
    working in an axial gauge, see for instance
    Ref.~\cite{Frenkel:1976ed} for massless particles}
The latter is, for instance, needed for the production of near-resonance 
unstable particles. In that case the leading contribution can be determined by 
employing the so-called pole scheme~\cite{Stuart:1991xk} in the leading-pole 
approximation, which restricts the calculation to the on-shell residue 
belonging to the unstable particle that is close to its mass-shell. For the 
explicit formulation of this approximation as well as its subtleties we refer 
to the literature~\cite{Beenakker:1998gr}. The elegance of the Coulomb-gauge 
method lies in the fact that, once all self-energies to all 
on-shell/on-resonance SM particles have been calculated, the prediction of the
Sudakov form factor for an {\it arbitrary} electroweak process becomes more or 
less straightforward. It should be noted, however, that for an electroweak 
process like $\,e^+e^- \to 4f\,$ it is in general not correct to assume 
universality and merely calculate the Sudakov form factors for the external 
particles (i.e.~the six fermions). Depending on the final state and the 
kinematical configuration, the process $\,e^+e^- \to 4f\,$ can have different 
near-resonance subprocesses~\cite{Grunewald:2000ju} 
(like $e^+e^-\!\to W^+W^-\! \to 4f\,$ or $\,e^+e^-\!\to ZZ \to 4f$). 
In that case the Sudakov correction factor is given
by the wave-function factors of the near-resonance intermediate particles 
rather than the four final-state fermions. The reason for this is that the
invariant mass of those intermediate particles is close to being on-shell and
therefore {\it not\,} of the same order as the CM energy squared. 
The subsequent decay of the intermediate particles into the final-state 
fermions does not involve a large invariant mass and will as such not give 
rise to Sudakov logarithms.
\\[-5mm]

\ni 
We finally note that the relevant 
self-energies for the calculation of the Sudakov logarithms involve the 
exchange of collinear-soft gauge bosons, including their potential mixing
with  the corresponding would-be Goldstone bosons. The collinear-soft exchange
of fermions and ghosts leads to suppressed contributions, since the propagators
of these particles do not have the required pole structure. The fermion 
propagators $\,\propto 1/[\ps-m+i\epsilon]\,$ are not affected by the gauge 
choice and therefore lack the additional $\,1/\vec{k}^{\,2}\,$ poles, whereas 
the ghost propagators $\,\propto 1/\vec{k}^{\,2}\,$ lack the 
$\,1/[k^2-M^2+i\epsilon]\,$ poles. As we will see later, both poles are 
required for obtaining the Sudakov logarithms.

\section{The Coulomb gauge: asymptotic states and external wave-function 
         factors in the Sudakov limit}
\label{sec:wave}

The calculation of the external wave-function factors for fermions is 
non-trivial~\cite{Aoki:1982ed}, but due to the absence of mixing between 
different fermions no major complications arise in the 
Coulomb gauge. For massive gauge bosons, however, the mixing with the 
corresponding component of the Higgs doublet introduces an additional 
complication~\cite{Beenakker:2000na}. 

\subsection{The charged-boson sector}
\label{sec:wave_charged}

Let us start off by considering the $W$ boson and the would-be Goldstone boson 
$\phi$. For a proper description of the on-shell $W$ bosons we need the 
asymptotic $W^{\rm as}$ field, which generates the asymptotic $W$-boson
states. It will have to be defined in terms of the interacting $W$ and $\phi$ 
fields:
\bea
\h{-8}
\label{as}
  W_{\mu}^{\pm,\,\rm as}(x) \!\!&=&\!\! 
          \Zw\,W_{\mu}^{\pm}(x)
          \,\pm\, i\,\delta Z_{1}\,\frac{\pa_{\mu}\phi^{\pm}(x)}{\mw}  
          + \Zwn\, n_{\mu}n\,\c W^{\pm}(x)  
          + \delta Z_{2}\,\frac{\pa_{\mu}\,\pa\,\c W^{\pm}(x)}{\mw^2}\,,
\ea
in such a way that the free-field propagators are retrieved for $W^{\rm as}$ 
in the on-shell limit. This fixes the wave-function factors $Z$ and 
$\delta Z$ in terms of the self-energies of the interacting fields.
The full expression in Eq.~(\ref{as}) is in fact only needed to guarantee that
the asymptotic vector field satisfies the physical polarization condition
\begin{align}
  \pa ^{\, \mu} \,  W_{\mu}^{\pm,\,\rm as}(x) =0
\end{align}
in the weak limit. For all practical purposes, \ie calculating $S$-matrix 
elements, the asymptotic state will be connected to a source term 
$\eps^{\mu}(k)$ and it is sufficient to consider 
\begin{align} 
\label{aseff}
  W_{\mu}^{\pm,\,\rm as}(x) \to \Zw W_{\mu}^{\pm}(x)  
                                + \Zwn n_{\mu}\, n\,\c W^{\pm}(x)\,,
\end{align}
since the two terms containing $\,\pa_{\mu}\,$ vanish owing to 
$\,\eps (k)\,\c k =0$. In the remainder of this section we will denote those 
irrelevant terms proportional to $\,k_{\mu}\,$ by $\,`\dots$'~. Note that for 
transverse $W$ bosons
($W_T$) the second term in Eq.~(\ref{aseff}) will also vanish, since 
$\,\eps_T(k)\c n = \eps^0_T(k) = 0$. For longitudinal $W$ bosons ($W_L$) the
full expression (\ref{aseff}) will be of relevance, since $\,\eps_L^{\mu}(k)\,$
lies in the plane spanned by $k^{\mu}$ and $n^{\mu}$:
\begin{equation}
  \eps_L^{\mu}(k) \ \equiv\ \eps^{\mu}(k,0) 
                  \ =\ \frac{k^0}{\mw\,|\vec{k}|}\,k^{\mu}
                       - \frac{\mw}{|\vec{k}|}\,n^{\mu} \,.
\end{equation}
Consequently, the wave-function factors for transverse and longitudinal $W$ 
bosons are different in the Coulomb gauge due to the special status of the 
time-like components.
\\[-5mm]

\ni
In order to actually determine the wave-function factors \,\Zw\, and \,\Zwn\,
we study the Fourier Transform (FT) of the asymptotic-field propagator 
\begin{align}
\label{FT}
  & {\rm FT}\, \langle\,0\,|{\rm T}\,( \,W_{\mu}^{+,\,\rm as}(x)\,
    W_{\nu}^{-,\,\rm as}(y)\, )\,|\,0\,\rangle\  = \no \\[2mm]
  & \hspace*{2mm} =\ {\rm FT}\, \langle\,0\,|{\rm T}\,\Big( 
    Z_{\sscr{W}}^{-1}\,W_{\mu}^{+}(x)\,W_{\nu}^{-}(y) 
    +  \Zwn\Zw  n_{\mu}\,n\c W^+(x) \,W_{\nu}^{-}(y) \no \\[2mm]
  & \hspace*{21mm} \ + \Zw\Zwn W_{\mu}^{+}(x)\,n_{\nu}\,n\c W^-(y) 
    + \Zwn^{\h{-4} 2}\h{2} n_{\mu}\,n\c W^+(x)\,n_{\nu}\,n\c W^-(y) 
    \Big)|\,0\,\rangle + \dots\,.
\end{align}
To further specify the above we need to gain knowledge about the dressed 
propagators for the interacting $W$ fields. Using the conventions introduced 
in the previous section, the interaction matrix can be written to all orders 
in perturbation theory as 
\v{2} 
\begin{align}
  \left(
  \begin{array}{cc}
    -\,i\,\Big[ \G^{\mu \nu} (k^2-\mw^2) - k^{\mu}k^{\nu} 
              + \la (k^{\mu}-k^0\,n^{\mu})(k^{\nu}-k^0\,n^{\nu})
              - \Sigma_{\sscr{WW}}^{\mu \nu} \Big] \,\,
    & \,\,\pm\,i\,k^{\mu}\mw + i\,\Sigma_{\sscr{W}^{\pm}\phi^{\pm}}^{\mu} 
    \\[5mm]
    \,\,\pm\,i\,k^{\nu}\mw + i\,\Sigma_{\phi^{\pm}\sscr{W}^{\pm}}^{\nu}
    & \,\, i\,\left[ k^2 + \Sigma_{\phi\phi} \right]
  \end{array}
 \right)~. \no \\[-3mm] \no
\end{align}
Here $\,\Sigma_{\sscr{WW}}^{\mu\nu} = 
\Sigma_{\sscr{W}^{\pm}\sscr{W}^{\pm}}^{\mu\nu}\,$ is the $W$-boson self-energy,
$\,\Sigma_{\sscr{W}^{\pm}\phi^{\pm}}^{\mu} = 
\Sigma_{\phi^{\pm}\sscr{W}^{\pm}}^{\mu}\,$ is the mixed $W$-boson/would-be 
Goldstone boson self-energy, and $\,\Sigma_{\phi\phi} = 
\Sigma_{\phi^{\pm}\phi^{\pm}}\,$ is the would-be Goldstone boson self-energy. 
For simplicity we will suppress the arguments (like $k$ and $n$) of these
self-energy functions. The Dyson-resummed (dressed) propagator matrix is 
obtained by inverting this interaction matrix and taking the limit 
$\la \to \infty$ (see previous section). In order to make the derivation of 
these dressed propagators as compact as possible we now use the transverse 
tensor $Q^{\mu \nu}$ as defined in Eq.~(\ref{polsum}) and introduce the 
space-like momentum
\bea
\label{qq}
  q^{\mu} \equiv k^{\mu}-k^0\,n^{\mu}\,.
\ea
These quantities have the following useful properties
\bea
\label{nQ}
  n_{\mu}\,Q^{\mu \nu} = q_{\mu}\,Q^{\mu \nu} = Q^{\mu \nu}\,n_{\nu}
  = Q^{\mu \nu}\,q_{\nu} = 0 \no \\[1mm]
  Q^{\mu \nu}\, Q_{\nu \rho} = Q^{\mu}_{\phantom{\mu}\rho}\,,
  \qquad n\c q = 0\,, \qquad q^2= -\,\vec{k}^2\,.
\ea 
Next we use Lorentz covariance and decompose the $W$-boson self-energy 
according to
\bea
  \Sigma_{\sscr{WW}}^{\mu \nu} = Q^{\mu \nu}\,\Sigma_{\sscr{WW},\,\G}
         + q^{\mu}q^{\nu}\,\Sigma_{\sscr{WW},\,q} 
         + (q^{\mu}n^{\nu} + q^{\nu} n^{\mu})\,\Sigma_{\sscr{WW},\,m} 
         + n^{\mu} n^{\nu}\,\Sigma_{\sscr{WW},\,n} \,,
\ea
bearing in mind that we have two independent four-vectors, $k$ and $n$, at our
disposal. Similarly the mixed $W$-boson/would-be Goldstone boson self-energy 
can be written as 
\bea
  \Sigma_{\sscr{W}^{\pm}\phi^{\pm}}^{\mu} = 
           \pm\,q^{\mu}\,\Sigma_{\sscr{W}^+\phi^+,\,q}
           \pm\,n^{\mu}\,\Sigma_{\sscr{W}^+\phi^+,\,n} 
\ea
by virtue of the hermiticity of the interaction Lagrangian.
Analogously the dressed-propagator matrix can be written in the generic form
\v{-3}
\begin{figure}[H]
\begin{align}
\vspace*{2mm}
\hspace*{5mm}
\left(
   \begin{array}{cc}
\begin{fmffile}{propdressedlong1}
\hspace*{8mm}
\begin{fmfgraph*}(80,40) \fmfpen{thin} \fmfleft{i1} \fmfright{o1}
 \fmf{boson,label=$W^{\pm}$,l.side=left}{i1,v1}
 \fmf{boson,label=$W^{\pm}$,l.side=left}{v1,o1} \fmfblob{10}{v1} 
 \fmf{phantom_arrow}{i1,v1}
 \fmf{phantom_arrow}{v1,o1}
 \fmfdot{i1,o1} \fmfv{l=${\mu}$,l.a=-130}{i1} \fmfv{l=${\nu}$,l.a=-50}{o1}
 \fmfv{l=$\myk$,l.a=-90}{v1} 
\end{fmfgraph*} 
\end{fmffile} 
 \hspace*{15mm} &  \hspace*{15mm}
\begin{fmffile}{propdressedlong2}
\begin{fmfgraph*}(80,40) \fmfpen{thin} \fmfleft{i1} \fmfright{o1}
 \fmf{boson,tension=1,label=$W^{\pm}$,l.side=left}{i1,v1}
 \fmf{phantom_arrow}{i1,v1} 
 \fmf{scalar,tension=2,label=$\phi^{\pm}$,l.side=left}{v1,o1} \fmfblob{10}{v1} 
 \fmfdot{i1,o1} \fmfv{l=${\mu}$,l.a=-130}{i1} 
 \fmfv{l=$\myk$,l.a=-90}{v1} 
\end{fmfgraph*} 
\end{fmffile} 
 \\[1mm]
\hspace*{8mm}
\begin{fmffile}{propdressedlong3}
\begin{fmfgraph*}(80,40) \fmfpen{thin} \fmfleft{i1} \fmfright{o1}
 \fmf{boson,tension=1,label=$W^{\pm}$,l.side=left}{v1,o1}
 \fmf{phantom_arrow}{v1,o1} 
 \fmf{scalar,tension=2,label=$\phi^{\pm}$,l.side=left}{i1,v1} \fmfblob{10}{v1} 
 \fmfdot{i1,o1} 
\fmfv{l=${\nu}$,l.a=-50}{o1}
 \fmfv{l=$\myk$,l.a=-90}{v1} 
\end{fmfgraph*}
\end{fmffile} 
 \hspace*{15mm} &  \hspace*{15mm}
\begin{fmffile}{propdressedlong4}
\begin{fmfgraph*}(80,40) \fmfpen{thin} \fmfleft{i1} \fmfright{o1}
 \fmf{scalar,label=$\phi^{\pm}$,l.side=left}{i1,v1} 
 \fmf{scalar,label=$\phi^{\pm}$,l.side=left}{v1,o1} 
\fmfblob{10}{v1} \fmfdot{i1,o1} 
 \fmfv{l=$\myk$,l.a=-90}{v1} 
\end{fmfgraph*}
\end{fmffile} 
  \end{array}
 \right)   \h{3} = \no 
\vspace*{-1cm}
\end{align}
\end{figure}

\vspace*{-1.4cm}
\begin{align}
\left(
  \begin{array}[H]{cc}
    -\,i\,\left[ A_{\sscr{WW}}\,Q_{\mu \nu} 
               + B_{\sscr{WW}}\,q_{\mu}\,q_{\nu} 
               + C_{\sscr{WW}}( q_{\mu}\,n_{\nu} + n_{\mu}\,q_{\nu} ) 
               + D_{\sscr{WW}}\,n_{\mu}\,n_{\nu} \right]\,\,
     & \,\,i\,\left[ E_{\sscr{W}^{\pm}\phi^{\pm}}\,q_{\mu} 
                   + F_{\sscr{W}^{\pm}\phi^{\pm}}\,n_{\mu} \right] 
     \\[4mm] 
     \,\,i\,\left[ E_{\phi^{\pm}\sscr{W}^{\pm}}\,q_{\nu} 
                 + F_{\phi^{\pm}\sscr{W}^{\pm}}\,n_{\nu} \right]\,\, 
     & \,\,i\,G_{\phi\phi}   
  \end{array}
\right) \,. \no \\[-3mm] \no 
\end{align}
Making use of the Ward identities in the Coulomb gauge, which state that 
Green's functions with a single gauge-boson line contracted with the 
corresponding space-like vector $\,\sqrt{\la}\,q^{\mu}\,$ should vanish for
$\lambda \to \infty$, we immediately obtain $B_{\sscr{WW}} = C_{\sscr{WW}} 
= E_{\sscr{W}^{\pm}\phi^{\pm}} = E_{\phi^{\pm}\sscr{W}^{\pm}} = 0$. 
For the other coefficients we have to solve separate equations for the 
transverse sector,
\bea
\label{AWW}
  A_{\sscr{WW}} = \left[ k^2-\mw^2-\Sigma_{\sscr{WW},\,\G} \right]^{-1}\,,
\ea
as well as for the longitudinal/scalar sector, 
\v{3}
\bea  
\label{DWW} 
  \left(
  \begin{array}{cc} 
       \vec{k}^2 + \mw^2 + \Sigma_{\sscr{WW},\,n}  
       & \pm \left[ k_0\mw + \Sigma_{\sscr{W}^+\phi^+,\,n} \right]
       \\[2mm]
       \pm \left[ k_0\mw + \Sigma_{\sscr{W}^+\phi^+,\,n} \right]
       & k^2 + \Sigma_{\phi\phi} 
  \end{array}
  \right)\,
  \left( 
  \begin{array}{cc} 
       -\,D_{\sscr{WW}} & F_{\sscr{W}^{\pm}\phi^{\pm}} \\[2mm]
       F_{\phi^{\pm}\sscr{W}^{\pm}} & G_{\phi\phi}
  \end{array}
  \right)
  =
  \left( 
  \begin{array}{cc} 
       1 & 0 \\ 0 & 1
  \end{array}
  \right). \\ \no
\ea
 
\ni
Up to now the discussion has been completely general. At this point the
calculation can be simplified by exploiting the special properties of the 
self-energies in the Sudakov limit ($k^2,\mw^2 \ll k_0^2$). In this limit 
Eq.~(\ref{DWW}) can be approximated by
\v{3}
\bea  
\label{DWWsud} 
  \left(
  \begin{array}{cc} 
       k_0^2 
       & \pm \left[ k_0\mw + \Sigma_{\sscr{W}^+\phi^+,\,n} \right]
       \\[2mm]
       \pm \left[ k_0\mw + \Sigma_{\sscr{W}^+\phi^+,\,n} \right]
       & k^2 + \Sigma_{\phi\phi} 
  \end{array}
  \right)\,
  \left( 
  \begin{array}{cc} 
       -\,D_{\sscr{WW}} & F_{\sscr{W}^{\pm}\phi^{\pm}} \\[2mm]
       F_{\phi^{\pm}\sscr{W}^{\pm}} & G_{\phi\phi}
  \end{array}
  \right)
  =
  \left( 
  \begin{array}{cc} 
       1 & 0 \\ 0 & 1
  \end{array}
  \right) \\[-3mm] \no
\ea
as a result of the fact that the self-energy $\,\Sigma_{\sscr{WW},n} =
\cO(k^2,\mw^2)\,$ is suppressed with respect to $k_0^2$. 
The other self-energies are of the same order as the corresponding lowest-order
terms, \ie $\,\Sigma_{\sscr{W}^+\phi^+,\,n} = \cO(k_0\mw)$, 
$\,\Sigma_{\phi\phi} = \cO(k^2,\mw^2)\,$ and neither self-energy contains 
inverse powers of $k^2$ or $\mw^2$ (see Appendix~\ref{app:selfenergy}), as 
required by hermiticity and analyticity.
The resulting solutions for the dressed propagator functions read
\bea
\label{DFGsol}
  G_{\phi\phi} &=& \frac{k_0^2}{k_0^2\,\left[ k^2+\Sigma_{\phi\phi} \right]
           - \left[ k_0\mw + \Sigma_{\sscr{W}^+\phi^+,\,n} \right]^2}
           \no \\[2mm]
  D_{\sscr{WW}} &=& -\,\frac{G_{\phi\phi}}{k_0^2}\,
           \left[k^2+\Sigma_{\phi\phi}\right]
           \no \\[2mm]
  F_{\sscr{W}^{\pm}\phi^{\pm}} &=& \mp\,\frac{G_{\phi\phi}}{k_0^2}\,
           \left[k_0\mw + \Sigma_{\sscr{W}^+\phi^+,\,n}\right] 
  \ =\  F_{\phi^{\pm}\sscr{W}^{\pm}}\,.
\ea 
Note the explicit hierarchy in the Sudakov limit, $|D_{\sscr{WW}}| \ll 
|F_{\sscr{W}^{\pm}\phi^{\pm}}| \ll |G_{\phi\phi}|$. This will play an important
role later in the derivation of the Goldstone-boson Equivalence Theorem. 
\\[-5mm]

\ni
With the help of the dressed propagator functions and Eq.~(\ref{nQ}), the
asymptotic-field propagator in Eq.~(\ref{FT}) can be expressed as
\bea
\label{FTa}
  {\rm FT}\, \langle\,0\,|{\rm T}\,( \,W_{\mu}^{+,\,\rm as}(x)\,
    W_{\nu}^{-,\,\rm as}(y)\, )\,|\,0\,\rangle 
  = -\,i\,Z_{\sscr{W}}^{-1}\,A_{\sscr{WW}}\,Q_{\mu \nu} 
    - i\,\left[ \Zw + \Zwn \right]^2 D_{\sscr{WW}}\,n_{\mu}\,n_{\nu}  
    + \dots\,.
\ea
The wave-function factors are subsequently obtained from the free-field 
constraint
\begin{align}
  & i\,(k^2-\mwp^2)\,{\rm FT}\,\langle\,0\,|{\rm T}\,(\,W_{\mu}^{+,\,\rm as}(x)
  \,W_{\nu}^{-,\,\rm as}(y)\, )\,|\,0\,\rangle\onshellWp \no \\[1mm] 
  & \h{20} \equiv\ 
    - \sum_{\la=\pm,0}\,\eps_{\mu}(k,\la)\,\eps_{\nu}^{\,\,*}(k,\la)\ 
    =\ \left[ Q_{\mu\nu}(k) - \frac{k^2}{\vec{k}^{\,2}}\,n_{\mu}n_{\nu}
       + \dots \right]\onshellWp\,,
\end{align}
where $\mwp$ is the physical pole mass of the $W$ boson. The $W$-boson mass 
does not receive any corrections in the Sudakov limit, \ie $\mwp=\mw$, since 
a non-zero Sudakov correction to the mass would imply that either the pole mass
or the mass counterterm of the
$W$ boson would become energy dependent. This special property of the Coulomb 
gauge can be understood in another way by realizing that in covariant 
$R_{\xi}$ gauges no Sudakov logarithms occur in the self-energies. 
Consequently, the masses of the particles will not be shifted by the Sudakov 
corrections. This ``non-renormalization'' condition has far-reaching 
consequences for the determination of the wave-function factors. 
Applied to the above-given dressed propagators it leads to two identities:
\bea
\label{nonrenMW}
  \Sigma_{\sscr{WW},\,\G} \propto (k^2-\mw^2)\ 
  &,&\ 
  \Sigma_{\phi\phi} - \frac{2\mw}{k_0}\,\Sigma_{\sscr{W}^+\phi^+,\,n}
  - \frac{1}{k_0^2}\,\Sigma^{\,2}_{\sscr{W}^+\phi^+,\,n} \propto (k^2-\mw^2)\,,
\ea
which hold to all orders. In Appendix~\ref{app:selfenergy} we have verified 
explicitly that these identities hold at the one-loop level. 
Making use of the ``non-renormalization'' conditions we finally obtain for the
wave-function factors:
\begin{align}
\label{ZW}
  & Z_{\sscr{W}}^{-1}\ 
  =\ \frac{k^2 - \mw^2 - \Sigma_{\sscr{WW},\,\G}}{k^2-\mw^2}\onshellW\
  =\ 1 - \frac{\pa\,\Sigma_{\sscr{WW},\,\G}}{\pa\,k^2}\onshellW \no \\[2mm]
  & \h{7} \ =\ 1 - \frac{i}{2\,k_0}\,\eps_{T,\,\mu}(k) 
    \biggl\{ \frac{\pa}{\pa\,k_0}\Bigl[ i\,\Sigma_{\sscr{WW}}^{\mu\nu} \Bigr] 
    \biggr\}\, \eps_{T,\,\nu}^{*}(k)\onshellW \no \\[3mm] 
  & Z_{\phi}^{-1}\ 
  =\ \frac{k_0^2(k^2-\mw^2) + k_0^2\,\Sigma_{\phi\phi} 
          - 2\mw k_0\,\Sigma_{\sscr{W}^+\phi^+,\,n}
          - \Sigma^{\,2}_{\sscr{W}^+\phi^+,\,n}}
         {k_0^2(k^2-\mw^2)}\onshellW
  =\ 1 + \frac{\pa\,\Sigma_{\phi\phi}}{\pa\,k^2}\onshellW \no \\[2mm] 
  & \h{7} \ =\ 1 - \frac{i}{2k_0}\biggl\{ \frac{\pa}{\pa\,k_0}
    \Bigl[ i\,\Sigma_{\phi\phi} \Bigr]\biggr\}\onshellW \no \\[3mm] 
  & \Zw+\Zwn\ 
  =\ Z_{\phi}^{-\frac{1}{2}}\,
     \frac{k_0\mw}{k_0\mw+\Sigma_{\sscr{W}^+\phi^+,\,n}}\onshellW\,.
\end{align}
Here $\,Z_{\phi}^{-\frac{1}{2}}\,$ is the wave-function factor that enters the
definition of an asymptotic state for the would-be Goldstone bosons: 
\bea 
  \phi^{\pm,\,\rm as}(x) = Z_{\phi}^{-\frac{1}{2}}\,\phi^{\pm}(x)
\ea
with
\begin{align}
  -\,i\,(k^2-\mwp^2)\,{\rm FT}\, \langle\,0\,|{\rm T}\,( 
  \,\phi^{+,\,\rm as}(x)\,\phi^{-,\,\rm as}(y)\, )\,|\,0\,\rangle\onshellWp
  \equiv\ 1 \,. \\ \no
\end{align}
In the first expression of Eq.~(\ref{ZW}) we have used the fact that the 
contraction with the transverse polarization vectors (sources)
$\eps^{\mu}_T(k)$ and $\eps^{* \,\nu}_T(k)$ projects on $-\G_{\mu\nu}$,
since $\,\eps_T(k)\c k = \eps_T(k)\c n = 0\,$ and 
$\,\eps_T(k)\c\eps_T^{*}(k) = -1$, whereas the derivative
$\frac{1}{2\,k_0}\,\frac{\pa}{\pa k_0}\,$ projects on the on-shell 
wave-function factor by virtue of the ``non-renormalization'' condition. 
The drastic simplification of the second expression is due to the fact that 
the leading $k^2$ dependence is contained exclusively in $\,\Sigma_{\phi\phi}$,
since $\,\Sigma_{\sscr{W}^+\phi^+,\,n} = \cO(k_0\mw)$. In fact, if we would add
the unit sources for scalar particles, this second expression would bear a 
close similarity to the first one. As we will see later in this section
and in Sect.~\ref{sec:one-loop}, such a projection by means of sources and
energy derivatives occurs in all bosonic and fermionic sectors.
\\[-5mm]

\ni
Now we have all the ingredients to consider the $S$-matrix elements in the 
charged-boson sector. First some conventions. In the following we will denote 
the amputated Green's functions generically by open circles. Only one external
line will be given explicitly, namely the one that belongs to the asymptotic 
state under investigation. These asymptotic states are represented by double 
lines and the corresponding dressed propagators by hatched circles. 
We start with the asymptotic $W$-boson fields, which give rise to two 
diagrams. The first diagram involves the pure $W$-boson propagator:
\vspace*{-7mm}
\begin{figure}[H]
\hspace*{10mm} 
 \begin{minipage}[l]{7cm}
     \begin{fmffile}{smatrixgw1}
     \begin{fmfgraph*}(80,80) \fmfpen{thin} \fmfleft{i1} \fmfright{o1}
     \fmf{phantom}{i1,v1,v2,o1} 
     \fmf{plain,left,tension=1.5}{i1,v1} \fmf{plain,right,tension=1.5}{i1,v1}
     \fmf{boson,tension=4}{v1,v2} \fmfblob{.15w}{v2}
     \fmf{double,tension=4}{v2,o1}
     \fmfdot{o1}
     \fmfv{label={$ \nu$},l.a=-90,l.d=7}{o1}
     \fmfv{label={$ W^{\pm ,, \mathrm as}$},l.a=15,l.d=45}{v1}
     \fmfv{label={$ W^{\pm}$},l.a=113,l.d=9}{v2}
     \end{fmfgraph*}
     \end{fmffile}
       \vspace*{-2.37cm}
       \begin{align}
        \hspace*{3.4cm} i\,(k^2-\mw^2)\,\,\eps^{\nu}(k)\onshellW \no
       \hspace*{2mm} = 
       \end{align}
\end{minipage}
\vspace*{1mm}

\begin{minipage}[r]{15cm}
     \hspace*{20mm}
    \begin{fmffile}{smatrixgw3}
     \begin{fmfgraph*}(80,80) \fmfpen{thin} \fmfleft{i1} \fmfright{o1}
     \fmf{phantom}{i1,v1,v2,o1} 
     \fmf{plain,left,tension=1.5}{i1,v1} \fmf{plain,right,tension=1.5}{i1,v1}
     \fmf{boson,tension=4}{v1,v2} \fmfblob{.15w}{v2}
     \fmf{boson,tension=4}{v2,o1}
     \fmfdot{o1}
     \fmfv{label={$ \nu$},l.a=-90,l.d=7}{o1}
     \fmfv{label={$ W^{\pm}$},l.a=15,l.d=45}{v1}
     \fmfv{label={$ W^{\pm}$},l.a=113,l.d=9}{v2}
     \end{fmfgraph*}
     \end{fmffile} 
      \vspace*{-2.37cm}
    \begin{align}
     \hspace*{5.8cm} i\,(k^2-\mw^2)\, \left[ \Zw \, \eps^{\nu}(k) + \Zwn \,
     n^{\nu} \, \eps_{0}(k) \right] \onshellW  \,\, .\no
    \end{align}
\end{minipage}
\end{figure} 

\vspace*{1mm}
\ni
Here we have left the polarization state unspecified, bearing in mind that
for transverse $W$ bosons $\eps_T(k)\,\c n=0\,$ and for longitudinal $W$ bosons
$\,\eps_L(k)\,\c n \approx k_0/\mw$ in the high-energy limit. Upon amputation 
of the external legs we find in the Sudakov limit
\vspace*{-4mm}

\begin{figure}[H]
\hspace*{-2mm}
\begin{minipage}[l]{10cm}
     \begin{fmffile}{sgw2}
      \begin{fmfgraph*}(65,65) \fmfpen{thin} \fmfleft{i1} \fmfright{o1}
      \fmf{phantom}{i1,v1,v2,o1}
      \fmf{plain,left,tension=1.5}{i1,v1} \fmf{plain,right,tension=1.5}{i1,v1}
      \fmf{phantom,tension=4}{v1,v2} \fmf{phantom,tension=4}{v2,o1}
      \fmf{boson}{v1,v2}
      \fmfv{label={$\textstyle W^{\pm}$},l.a=102,l.d=6}{v2}
      \fmfv{label={$\textstyle \mu$},l.a=-160,l.d=35}{o1}
      \end{fmfgraph*} 
      \end{fmffile}
     \vspace*{-2.1cm}
    {\small 
      \begin{align}
        \hspace*{1.8cm}  
        \left\{\! 
        \frac{(k^2\!-\!\mw^2)\,Q^{\mu}_{\phantom{\mu}\nu}}
             {k^2\!-\!\mw^2\!-\!\Sigma_{\sscr{WW},\,\G}} 
      - \frac{(k^2\!-\!\mw^2)\,\left[k^2\!+\!\Sigma_{\phi\phi}\right]\,
              n^{\mu}n_{\nu}}
             {k_0^2\left[ k^2\!+\!\Sigma_{\phi\phi} \right]
              \!-\!\left[ k_0\mw\!+\!\Sigma_{\sscr{W}^+\phi^+,\,n} \right]^2}
        \!\right\}\!\left[\!\Zw\eps^{\nu }(k) 
        \!+\!\Zwn n^{\nu}\eps_{0}(k) \right]\!\!\! \onshellW
        \no 
     \end{align}
}
 \end{minipage}

\vspace*{8mm}
 \hspace*{1.5cm} =
\vspace*{-13mm}

 \begin{minipage}[r]{10cm}
     \begin{fmffile}{sgw2}
       \hspace*{2.7cm} 
      \begin{fmfgraph*}(65,65) \fmfpen{thin} \fmfleft{i1} \fmfright{o1}
      \fmf{phantom}{i1,v1,v2,o1}
      \fmf{plain,left,tension=1.5}{i1,v1} \fmf{plain,right,tension=1.5}{i1,v1}
      \fmf{phantom,tension=4}{v1,v2} \fmf{phantom,tension=4}{v2,o1}
      \fmf{boson}{v1,v2}
      \fmfv{label={$\textstyle W^{\pm}$},l.a=102,l.d=6}{v2}
      \fmfv{label={$\textstyle \mu$},l.a=-160,l.d=35}{o1}
      \end{fmfgraph*} 
      \end{fmffile}
     \vspace*{-2.07cm}
     \begin{align}
      \hspace*{4.7cm} 
      \left[ 
       Z_{\sscr{W}}^{\frac{1}{2}}\,\eps^{\nu}(k)\,Q^{\mu}_{\phantom{\mu}\nu}  
       + Z_{\phi}^{\frac{1}{2}}\,n^{\mu}
       \left( -\,\frac{\mw^2}{k_0^2} \right)
       \frac{k_0\mw+\Sigma_{\sscr{W}^+\phi^+,\,n}}{k_0\mw}\,\eps_{0}(k)  
      \right]\! \onshellW \,. \no 
     \end{align}
\end{minipage}
\end{figure}

\vspace*{1mm}
\ni
{}From this general result we deduce that for transverse $W$ bosons, with  
$\,\eps^{\nu}_T(k)\,Q^{\mu}_{\phantom{\mu}\nu} = \eps_T^{\mu}(k)\,$ and 
$\,\eps_T(k)\,\c n =0$, the contribution of Sudakov corrections simply amounts 
to multiplying each external transverse $W$-boson line of the matrix element 
by the factor $Z_{\sscr{W}}^{\frac{1}{2}}$. For longitudinal $W$ bosons, with 
$\,\eps^{\nu}_L(k)\,Q^{\mu}_{\phantom{\mu}\nu} = 0\,$ and 
$\,\eps_L(k)\,\c n \approx k_0/ \mw$, one obtains a mass-suppressed 
contribution $\propto \mw/k_0$.

\ni
The second diagram involves the mixed $W$-boson/would-be Goldstone boson 
propagator:
\vspace*{-3mm}
 \vspace*{-5mm}
 \begin{figure}[H]
 \hspace*{-1mm} 
  \begin{minipage}[l]{7cm}
      \begin{fmffile}{gphi1}
      \begin{fmfgraph*}(80,80) \fmfpen{thin} \fmfleft{i1} \fmfright{o1}
      \fmf{phantom}{i1,v1,v2,o1} 
      \fmf{plain,left,tension=1.5}{i1,v1} \fmf{plain,right,tension=1.5}{i1,v1}
     \fmf{dashes,tension=4}{v1,v2} \fmfblob{.15w}{v2}
      \fmf{double,tension=4}{v2,o1}
      \fmfdot{o1}
      \fmfv{label={$ \nu$},l.a=-90,l.d=7}{o1}
      \fmfv{label={$ W^{\pm \mathrm,, as}$},l.a=15,l.d=45}{v1}
      \fmfv{label={$ \phi^{\pm}$},l.a=113,l.d=9}{v2}
      \end{fmfgraph*}
      \end{fmffile}
        \vspace*{-2.37cm}
        \begin{align}
         \hspace*{3.4cm} i\,(k^2-\mw^2)\,\,\eps^{\nu}(k) \onshellW \no
        \end{align}
 \end{minipage} 

\vspace*{12mm}
 \hspace*{1.5cm} =
\vspace*{-13mm}

 \begin{minipage}[r]{7cm}
     \begin{fmffile}{smatrixgphi3}
       \vspace*{-2mm}
       \hspace*{23mm}
      \begin{fmfgraph*}(80,80) \fmfpen{thin} \fmfleft{i1} \fmfright{o1}
      \fmf{phantom}{i1,v1,v2,o1} 
      \fmf{plain,left,tension=1.5}{i1,v1} \fmf{plain,right,tension=1.5}{i1,v1}
       \fmf{dashes,tension=4}{v1,v2} \fmfblob{.15w}{v2}
       \fmf{boson,tension=4}{v2,o1}
      \fmfdot{o1}
      \fmfv{label={$ \nu$},l.a=-90,l.d=7}{o1}
      \fmfv{label={$ W^{\pm}$},l.a=15,l.d=45}{v1}
       \fmfv{label={$ \phi^{\pm}$},l.a=113,l.d=9}{v2}
      \end{fmfgraph*}
      \end{fmffile} 
       \vspace*{-2.37cm}
    \begin{align}
        \hspace*{58mm} i\,(k^2-\mw^2)\left[ \Zw\,\eps^{\nu}(k) 
        + \Zwn\,n^{\nu}\,\eps_0(k) \right] \onshellW \,. \no
     \end{align}
 \end{minipage}
 \end{figure} 

\vspace*{1mm}
\ni
Upon amputation of the external legs we find in the Sudakov limit a vanishing 
contribution for transversely polarized $W$ bosons and
\vspace*{-6mm}
\begin{figure}[H]
\hspace*{6mm}
\begin{minipage}[l]{10cm}
     \begin{fmffile}{sgphi2}
      \begin{fmfgraph*}(65,65) \fmfpen{thin} \fmfleft{i1} \fmfright{o1}
      \fmf{phantom}{i1,v1,v2,o1}
      \fmf{plain,left,tension=1.5}{i1,v1} \fmf{plain,right,tension=1.5}{i1,v1}
      \fmf{phantom,tension=4}{v1,v2} \fmf{phantom,tension=4}{v2,o1}
      \fmf{dashes}{v1,v2}
      \fmfv{label={$\textstyle \phi^{\pm}$},l.a=102,l.d=6}{v2}
      \end{fmfgraph*} 
      \end{fmffile}
     \vspace*{-2.2cm}
    \begin{align}
      \hspace*{1.8cm}
      \left\{\pm\, 
      \frac{(k^2-\mw^2)\,\left[ k_0\mw+\Sigma_{\sscr{W}^+\phi^+,\,n}\right]}
           {k_0^2\,\left[ k^2+\Sigma_{\phi\phi} \right]
            - \left[ k_0\mw + \Sigma_{\sscr{W}^+\phi^+,\,n} \right]^2} \right\}
      \,n_{\nu} \left[ \Zw\,\eps^{\nu}_{L}(k) + \Zwn\,n^{\nu}\,\frac{k_0}{\mw}
      \right] \onshellW \no 
    \end{align}
    \vspace*{2.1cm}
 \end{minipage}
\end{figure}

\vspace*{-17mm}
 \hspace*{1.5cm} =
\vspace*{-18mm}

\begin{figure}[H]
\hspace*{28mm}
\begin{minipage}[l]{10cm}
     \begin{fmffile}{sgphi2}
      \begin{fmfgraph*}(65,65) \fmfpen{thin} \fmfleft{i1} \fmfright{o1}
      \fmf{phantom}{i1,v1,v2,o1}
      \fmf{plain,left,tension=1.5}{i1,v1} \fmf{plain,right,tension=1.5}{i1,v1}
      \fmf{phantom,tension=4}{v1,v2} \fmf{phantom,tension=4}{v2,o1}
      \fmf{dashes}{v1,v2}
      \fmfv{label={$\textstyle \phi^{\pm}$},l.a=102,l.d=6}{v2}
      \end{fmfgraph*} 
      \end{fmffile}
     \vspace*{-2.1cm}
    \begin{align}
      \hspace*{-3.1cm} \left[ \pm\,Z_{\phi}^{\frac{1}{2}} \right]\onshellW \no 
    \end{align}
     \vspace*{2.1cm}
 \end{minipage}
\end{figure}
\vspace*{-2.8cm}
\ni
for longitudinally polarized $W$ bosons. In other words, we find for 
longitudinal $W$ bosons that the dominant contribution to any physical process
originates from the amputated Green's function where the amputated leg is a 
would-be Goldstone boson $\phi$, provided of course that the matrix element is 
not mass-suppressed to start with (which is anyhow one of the basic defining
conditions for the Sudakov corrections). The other contribution, where the 
amputated leg is a $W$ boson contracted with the temporal gauge vector, is 
mass-suppressed. This is in fact the Goldstone-boson Equivalence 
Theorem~\cite{Kallosh:1973ap},
which is hence obtained quite naturally in the Coulomb-gauge approach as a 
result of the explicit mixing between gauge bosons and would-be Goldstone 
bosons\footnote{In covariant $R_{\xi}$ gauges there is by construction no 
                mixing between gauge bosons and would-be Goldstone bosons
                at lowest order}. 
\\[-5mm]

\ni 
Owing to the same mixing between gauge bosons and would-be Goldstone bosons, 
we have to address one more question before being able to wrap up the 
discussion of the Equivalence Theorem. Namely, we have to show that the
$S$-matrix element with the asymptotic state $\phi^{\pm,\,\rm as}$ will not
exhibit a leading contribution for the amputated Green's function where the
amputated leg is a $W$ boson. The dominant contribution again has to be the one
where the amputated leg is a would-be Goldstone boson.  
For the  mixed $S$-matrix element we find
\vspace*{-1mm}
\vspace*{-5mm}
\begin{figure}[H]
\hspace*{1mm} 
 \begin{minipage}[l]{7cm}
     \begin{fmffile}{smatrixetgw1}
     \begin{fmfgraph*}(80,80) \fmfpen{thin} \fmfleft{i1} \fmfright{o1}
     \fmf{phantom}{i1,v1,v2,o1} 
     \fmf{plain,left,tension=1.5}{i1,v1} \fmf{plain,right,tension=1.5}{i1,v1}
     \fmf{boson,tension=4}{v1,v2} \fmfblob{.15w}{v2}
     \fmf{dbl_dashes,tension=4}{v2,o1}
     \fmfdot{o1}
     \fmfv{label={$ \phi^{\pm \mathrm,, as}$},l.a=15,l.d=45}{v1}
     \fmfv{label={$ W^{\pm}$},l.a=113,l.d=9}{v2}
     \end{fmfgraph*}
     \end{fmffile}
       \vspace*{-2.37cm}
       \begin{align}
         \hspace*{3.3cm} (-i)\,(k^2-\mw^2) \onshellW\,\, \no
       \hspace*{-1mm} = 
       \end{align}
\end{minipage}
\begin{minipage}[r]{7cm}
    \begin{fmffile}{smatrixetgw3}

      \hspace*{3mm}
     \begin{fmfgraph*}(80,80) \fmfpen{thin} \fmfleft{i1} \fmfright{o1}
     \fmf{phantom}{i1,v1,v2,o1} 
     \fmf{plain,left,tension=1.5}{i1,v1} \fmf{plain,right,tension=1.5}{i1,v1}
     \fmf{boson,tension=4}{v1,v2} \fmfblob{.15w}{v2}
     \fmf{dashes,tension=4}{v2,o1}
     \fmfdot{o1}
     \fmfv{label={$ \phi^{\pm}$},l.a=15,l.d=45}{v1}
     \fmfv{label={$ W^{\pm}$},l.a=113,l.d=9}{v2}
     \end{fmfgraph*}
     \end{fmffile} 
      \vspace*{-2.37cm}
    \begin{align}
       \hspace*{37mm} (-i)\,(k^2-\mw^2)\,\,Z_{\phi}^{- \frac{1}{2}} \onshellW
        \no
    \end{align}
\end{minipage}
\end{figure}

\vspace*{2mm}
\ni
and amputating the legs leads in the Sudakov limit to 
\vspace*{-6mm}
\begin{figure}[H]
\hspace*{6mm}
\begin{minipage}[l]{10cm}
     \begin{fmffile}{gw2}
      \begin{fmfgraph*}(65,65) \fmfpen{thin} \fmfleft{i1} \fmfright{o1}
      \fmf{phantom}{i1,v1,v2,o1}
      \fmf{plain,left,tension=1.5}{i1,v1} \fmf{plain,right,tension=1.5}{i1,v1}
      \fmf{phantom,tension=4}{v1,v2} \fmf{phantom,tension=4}{v2,o1}
      \fmf{boson}{v1,v2}
      \fmfv{label={$\textstyle W^{\pm}$},l.a=102,l.d=6}{v2}
      \fmfv{label={$\textstyle \mu$},l.a=-160,l.d=35}{o1}
      \end{fmfgraph*} 
      \end{fmffile}
     \vspace*{-2.17cm}
    \begin{align}
      \hspace*{2cm}
      \left\{\mp\,
      \frac{(k^2-\mw^2)\,\left[ k_0\mw+\Sigma_{\sscr{W}^+\phi^+,\,n}\right]}
           {k_0^2\,\left[ k^2+\Sigma_{\phi\phi} \right]
            - \left[ k_0\mw + \Sigma_{\sscr{W}^+\phi^+,\,n} \right]^2}\right\}
      \,n^{\mu}\,Z_{\phi}^{-\frac{1}{2}} \onshellW \no 
    \end{align}
 \end{minipage}
\end{figure}

\vspace*{10mm}
 \hspace*{1.5cm} =
\vspace*{-17.4mm}

\begin{figure}[H]
\hspace*{30mm}
\begin{minipage}[l]{10cm}
     \begin{fmffile}{gw2}
      \begin{fmfgraph*}(65,65) \fmfpen{thin} \fmfleft{i1} \fmfright{o1}
      \fmf{phantom}{i1,v1,v2,o1}
      \fmf{plain,left,tension=1.5}{i1,v1} \fmf{plain,right,tension=1.5}{i1,v1}
      \fmf{phantom,tension=4}{v1,v2} \fmf{phantom,tension=4}{v2,o1}
      \fmf{boson}{v1,v2}
      \fmfv{label={$\textstyle W^{\pm}$},l.a=102,l.d=6}{v2}
      \fmfv{label={$\textstyle \mu$},l.a=-160,l.d=35}{o1}
      \end{fmfgraph*} 
      \end{fmffile}
     \vspace*{-2.1cm}
    \begin{align}
      \hspace*{2cm} Z_{\phi}^{\frac{1}{2}}\,n^{\mu}
      \left( \mp\,\frac{\mw}{k_0} \right)
      \frac{k_0\mw+\Sigma_{\sscr{W}^+\phi^+,\,n}}{k_0\mw} \onshellW \,. \no 
    \end{align}
     \vspace*{2.1cm}
 \end{minipage}
\end{figure}

\vspace*{-2.2cm}
\ni 
This indeed leads to a mass-suppressed contribution.
Similarly we obtain for the diagram involving the pure scalar propagator
\vspace*{-2mm}
 \vspace*{-5mm}
 \begin{figure}[H]
 \hspace*{5mm} 
  \begin{minipage}[l]{7cm}
      \begin{fmffile}{etgphi1}
      \begin{fmfgraph*}(80,80) \fmfpen{thin} \fmfleft{i1} \fmfright{o1}
      \fmf{phantom}{i1,v1,v2,o1} 
      \fmf{plain,left,tension=1.5}{i1,v1} \fmf{plain,right,tension=1.5}{i1,v1}
     \fmf{dashes,tension=4}{v1,v2} \fmfblob{.15w}{v2}
      \fmf{dbl_dashes,tension=4}{v2,o1}
      \fmfdot{o1}
       \fmfv{label={$ \phi^{\pm ,,\mathrm as}$},l.a=15,l.d=45}{v1}
      \fmfv{label={$ \phi^{\pm}$},l.a=113,l.d=9}{v2}
      \end{fmfgraph*}
      \end{fmffile}
        \vspace*{-2.37cm}
        \begin{align}
          \hspace*{3.3cm} (-i)\,(k^2-\mw^2) \onshellW \no
        \hspace*{1mm} = 
        \end{align}
 \end{minipage}
 \begin{minipage}[r]{7cm}
     \begin{fmffile}{etgphi3}

       \vspace*{1.2mm}
       \hspace*{2.5mm}
      \begin{fmfgraph*}(80,80) \fmfpen{thin} \fmfleft{i1} \fmfright{o1}
      \fmf{phantom}{i1,v1,v2,o1} 
      \fmf{plain,left,tension=1.5}{i1,v1} \fmf{plain,right,tension=1.5}{i1,v1}
      \fmf{dashes,tension=4}{v1,v2} \fmfblob{.15w}{v2}
      \fmf{dashes,tension=4}{v2,o1}
      \fmfdot{o1}
       \fmfv{label={$ \phi^{\pm}$},l.a=15,l.d=45}{v1}
       \fmfv{label={$ \phi^{\pm}$},l.a=113,l.d=9}{v2}
      \end{fmfgraph*}
      \end{fmffile} 
       \vspace*{-2.37cm}
     \begin{align}
       \hspace*{36mm} (-i) \,(k^2-\mw^2)\,Z_{\phi}^{-\frac{1}{2}} \onshellW \no
     \end{align}
 \end{minipage}
 \end{figure} 

\vspace*{3mm}
\ni
and amputation yields in the Sudakov limit
\vspace*{-6mm}
\begin{figure}[H]
\hspace*{6mm}
\begin{minipage}[l]{10cm}
     \begin{fmffile}{gphi2}
       \v{2}
      \begin{fmfgraph*}(65,65) \fmfpen{thin} \fmfleft{i1} \fmfright{o1}
      \fmf{phantom}{i1,v1,v2,o1}
      \fmf{plain,left,tension=1.5}{i1,v1} \fmf{plain,right,tension=1.5}{i1,v1}
      \fmf{phantom,tension=4}{v1,v2} \fmf{phantom,tension=4}{v2,o1}
      \fmf{dashes}{v1,v2}
      \fmfv{label={$\textstyle \phi^{\pm}$},l.a=102,l.d=6}{v2}
      \end{fmfgraph*} 
      \end{fmffile}
     \vspace*{-2.27cm}
    \begin{align}
      \hspace*{1.8cm}
      \left\{  
      \frac{k_0^2\,(k^2-\mw^2)}{k_0^2\,\left[ k^2+\Sigma_{\phi\phi} \right]
            - \left[ k_0\mw + \Sigma_{\sscr{W}^+\phi^+,\,n} \right]^2} \right\}
      \,Z_{\phi}^{-\frac{1}{2}} \onshellW \h{2} = \no
       \vspace*{-50mm}
    \end{align}
     \begin{fmffile}{gphi2}

     \vspace*{-22.7mm}
      \h{124}
      \begin{fmfgraph*}(65,65) \fmfpen{thin} \fmfleft{i1} \fmfright{o1}
      \fmf{phantom}{i1,v1,v2,o1}
      \fmf{plain,left,tension=1.5}{i1,v1} \fmf{plain,right,tension=1.5}{i1,v1}
      \fmf{phantom,tension=4}{v1,v2} \fmf{phantom,tension=4}{v2,o1}
      \fmf{dashes}{v1,v2}
      \fmfv{label={$\textstyle \phi^{\pm}$},l.a=102,l.d=6}{v2}
      \end{fmfgraph*} 
      \end{fmffile}

          \vspace*{-31.7mm}
    \begin{align}
      \hspace*{14.4cm} Z_{\phi}^{\frac{1}{2}}\onshellW \,\,. \no 
    \end{align}
     \vspace*{2.1cm}
 \end{minipage}
\end{figure}

\vspace*{-2.4cm}
\ni
This is not only the leading contribution, but {\it identical\,} to the leading
contribution that we obtained from the asymptotic 
$W_{\sscr{L},\,\mu}^{\pm,\, \rm as}$ field. Applying the high-energy Sudakov 
limit to the Coulomb gauge, we do not only find the Equivalence Theorem
$W_{\sscr{L},\,\mu}^{\pm,\,\rm as} \to \pm\,C\,\phi^{\pm,\,\rm as}$ to hold for
massive particles, but we find a very special case of the Equivalence 
Theorem, \ie $C=1$ to all orders in perturbation theory. This implies the 
{\it identity} of the two particles rather than mere 
{\it proportionality}.\footnote{In fact, the $C=1$ property is a general 
      high-energy feature of the Coulomb gauge, even for non-double-logarithmic
      corrections. This has to be contrasted with covariant $R_{\xi}$ gauges
      where the Goldstone-boson fields occur in the gauge-fixing Lagrangian.
      This in general necessitates the introduction of a finite 
      (renormalization-scheme- and $\xi$-dependent) factor 
      $\,C_{\scr{mod}}\neq 1\,$ in order to link the unphysical would-be 
      Goldstone boson with the physical asymptotic Goldstone-boson state 
      beyond lowest order~\cite{Yao:1988aj}--\cite{Espriu:1995ep}. It requires 
      a very special renormalization scheme to get $C=1$ in that 
      case~\cite{He:1994yd,He:1995br}}
\\[-5mm]

\ni
We close this discussion in the charged-boson sector with the following 
observations and conclusions. For transversely polarized external $W$ 
bosons the mixing with the $\phi$ field vanishes and the Sudakov correction 
factor amounts to multiplying each external transverse $W$-boson line of the 
matrix element by the factor $Z_{\sscr{W}}^{\frac{1}{2}}$. For longitudinally 
polarized external $W^{\pm}$ bosons the correction factor is not equal to
$Z_{\sscr{W}}^{\frac{1}{2}}$, instead the dominant Sudakov correction factor 
amounts to multiplying each external longitudinal gauge-boson line of the 
matrix element by the factor $\pm\,Z_{\phi}^{\frac{1}{2}}$ (provided that the 
lowest-order matrix element is not mass-suppressed to start with). 
This statement is a special case of the Equivalence Theorem in the sense that 
the longitudinal $W$ bosons can be substituted by their would-be Goldstone 
boson counterparts, $\phi$, in the high-energy limit. Hence, we have 
effectively returned to the situation before spontaneous symmetry breaking
where the Goldstone bosons represent the physical degrees of freedom. In this 
respect we could say that the SM behaves 
(dynamically) like an unbroken theory in the Sudakov limit, in spite of the 
fact that we cannot neglect the $\,W/\phi\,$ mass at the kinematical 
(phase-space) level while calculating the Sudakov correction factors.

\subsection{The neutral-boson sector}
\label{sec:wave_neutral}

In the neutral-boson sector of the SM we have to deal with
four particles. The physical Higgs boson can be treated in a trivial way,
since it does not mix with any of the other neutral particles. 
In the Sudakov limit the corresponding external wave-function factor is simply
given by
\bea
  Z_{\sscr{H}}^{-1} \ =\ 1 + \frac{\pa\,\Sigma_{\sscr{HH}}}{\pa\,k^2}\onshellH
  =\ 1 - \frac{i}{2k_0}\biggl\{ \frac{\pa}{\pa\,k_0}
    \Bigl[ i\,\Sigma_{\sscr{HH}} \Bigr]\biggr\}\onshellH \,, 
\ea
with $\,\Sigma_{\sscr{HH}}\,$ the Higgs-boson self-energy. The Sudakov 
correction factor then amounts to multiplying each external Higgs-boson line
of the matrix element by the factor $Z_{\sscr{H}}^{\frac{1}{2}}$.
The remaining three particles are the photon ($\ga$), the $Z$ boson, and the
corresponding would-be Goldstone boson $\chi$. At lowest order the situation is
equivalent to the charged-boson sector, since in that case only the $Z$ boson 
and $\chi$ mix. 
However, beyond lowest order all three particles mix, which adds an extra
level of complication. Before presenting the corresponding asymptotic states,
we first address the propagator functions in the Sudakov limit. They can be
derived using the methods developed for the charged-boson sector.
Again only the $A,\,F$ and $G$ propagator functions survive, yielding in the
Sudakov limit:
\bea
  A_{\ga\ga} &=& \frac{k^2-\mz^2-\Sigma_{\sscr{ZZ},\,\G}}
                      {\left[k^2-\mz^2-\Sigma_{\sscr{ZZ},\,\G}\right]
                       \left[k^2-\Sigma_{\ga\ga,\,\G}\right]
                       - \Sigma_{\ga\sscr{Z},\,\G}^{\,2}}
  \no \\[2mm]
  A_{\ga\sscr{Z}} &=& \frac{\Sigma_{\ga\sscr{Z},\,\G}}
                      {\left[k^2-\mz^2-\Sigma_{\sscr{ZZ},\,\G}\right]
                       \left[k^2-\Sigma_{\ga\ga,\,\G}\right]
                       - \Sigma_{\ga\sscr{Z},\,\G}^{\,2}}
  \ =\ A_{\sscr{Z}\ga}
  \no \\[2mm]
  A_{\sscr{ZZ}} &=& \frac{k^2-\Sigma_{\ga\ga,\,\G}}
                      {\left[k^2-\mz^2-\Sigma_{\sscr{ZZ},\,\G}\right]
                       \left[k^2-\Sigma_{\ga\ga,\,\G}\right]
                       - \Sigma_{\ga\sscr{Z},\,\G}^{\,2}}
\ea
in the transverse sector, and
\bea
\label{DFGsol_neutral}
  G_{\chi\chi} &=& \frac{k_0^2}{k_0^2\,\left[k^2+\Sigma_{\chi\chi}\right]
          + \left[i\,k_0\,\mz+\Sigma_{\sscr{Z}\chi,\,n}\right]^2
          + \Sigma_{\ga\chi,\,n}^{\,2}}
          \no \\[2mm]
  D_{\ga\ga} &=& -\,\frac{G_{\chi\chi}}{k_0^4}\,
          \left\{ k_0^2\,\left[ k^2+\Sigma_{\chi\chi} \right]
                  + \left[i\,k_0\,\mz+\Sigma_{\sscr{Z}\chi,\,n}\right]^2
          \right\}
          \no \\[2mm]
  D_{\ga\sscr{Z}} &=& \frac{G_{\chi\chi}}{k_0^4}\,
          \Sigma_{\ga\chi,\,n}\,
          \left[i\,k_0\,\mz+\Sigma_{\sscr{Z}\chi,\,n}\right]
  \ =\ D_{\sscr{Z}\ga}
          \no \\[2mm]
  D_{\sscr{ZZ}} &=& -\,\frac{G_{\chi\chi}}{k_0^4}\,
          \left\{ k_0^2\,\left[ k^2+\Sigma_{\chi\chi} \right]
                  + \Sigma_{\ga\chi,\,n}^{\,2} 
          \right\}
          \no \\[2mm]
  F_{\ga\chi} &=& -\,\frac{G_{\chi\chi}}{k_0^2}\,\Sigma_{\ga\chi,\,n}
  \ =\ -\,F_{\chi\ga}
          \no \\[2mm]
  F_{\sscr{Z}\chi} &=& -\,\frac{G_{\chi\chi}}{k_0^2}\,
          \left[i\,k_0\,\mz+\Sigma_{\sscr{Z}\chi,\,n}\right]
  \ =\ -\,F_{\chi\sscr{Z}} 
\ea 
in the longitudinal/scalar sector. The functions occurring in these 
expressions have been obtained by decomposing the various self-energies
in the same way as prescribed for the charged-boson sector.
These self-energies have the following properties, by virtue of the hermiticity
of the interaction Lagrangian: 
$\Sigma_{\ga\sscr{Z}}^{\mu\nu} = \Sigma_{\sscr{Z}\ga}^{\mu\nu}$, 
$\Sigma_{\ga\chi}^{\mu} = -\,\Sigma_{\chi\ga}^{\mu}\,$ and 
$\Sigma_{\sscr{Z}\chi}^{\mu} = -\,\Sigma_{\chi\sscr{Z}}^{\mu}$. 
\\[-5mm]

\ni
The ``non-renormalization'' conditions for the photon and $Z$-boson masses
give rise to the identities
\bea
\label{nonrenN}
  \left[k^2-\mz^2-\Sigma_{\sscr{ZZ},\,\G}\right]
  \left[k^2-\Sigma_{\ga\ga,\,\G}\right] - \Sigma_{\ga\sscr{Z},\,\G}^{\,2} 
  &\propto& k^2\,(k^2-\mz^2)\no \\[3mm]
  \Sigma_{\chi\chi} + \frac{2\,i\,\mz}{k_0}\,\Sigma_{\sscr{Z}\chi,\,n}
  + \frac{1}{k_0^2}\,\left[ \Sigma_{\sscr{Z}\chi,\,n}^{\,2} 
                          + \Sigma_{\ga\chi,\,n}^{\,2} \right]
  &\propto& (k^2-\mz^2) \,.
\ea 
These identities have been verified explicitly in 
Appendix~\ref{app:selfenergy} at the one-loop level. 
Note that the photon-mass condition only applies to the transverse sector,
since the on-shell photon only has transverse degrees of freedom. At this point
we note that there will be, in fact, one more ``non-renormalization'' 
condition, related to the electromagnetic charge. We will come back to this 
later.
\\[-5mm]

\ni
According to the observations made in the charged-boson case, the 
wave-function factors in the transverse and longitudinal/scalar sectors are
best treated separately. We start with the longitudinal sector. Bearing in 
mind that the asymptotic on-shell photon field $A_{\mu}^{\rm as}$ is transverse
and decouples from the asymptotic $Z_{\mu}^{\rm as}$ and $\chi^{\rm as}$ 
fields, the asymptotic states can be defined as\footnote{Eq.~(\ref{aseffN}) is
     valid in double-logarithmic approximation. In more general situations one 
     should replace $Z_{\sscr{L}}$ by $\,Z_{\sscr{L}} + A_{\sscr{L}}\,
     \Sigma_{\ga\sscr{Z},\,n}\,/[k_0^2+\Sigma_{\sscr{ZZ},\,n}]\,$ on the 
     right-hand side of Eq.~(\ref{aseffN})} 
\begin{align} 
\label{aseffN}
  A_{\sscr{L},\,\mu}^{\rm as}(x) &\ =\ 0 \no \\[3mm]
  Z_{\sscr{L},\,\mu}^{\rm as}(x) &\ \to\ \Zz Z_{\sscr{L},\,\mu}(x)  
                                      + \Zzn n_{\mu}\,n\,\c Z_{\sscr{L}}(x)
                                  \no \\[3mm]
  \chi^{\rm as}(x) &\ =\ Z_{\chi}^{-\frac{1}{2}} \chi(x) \,.  
\end{align}
The additional derivative terms needed for a proper asymptotic $Z$-boson state
have again been left out, since these terms will not contribute to the physical
$S$-matrix elements. By applying the ``non-renormalization'' conditions, the 
wave-function factors can be determined from the free-field constraints in the
usual way:
\begin{align}
\label{ZZL}
  & Z_{\chi}^{-1}\ =\ 1 + \frac{\pa\,\Sigma_{\chi\chi}}{\pa\,k^2}\onshellZ\
    =\ 1 - \frac{i}{2k_0}\biggl\{ \frac{\pa}{\pa\,k_0}
       \Bigl[ i\,\Sigma_{\chi\chi} \Bigr]\biggr\}\onshellZ \no \\[2mm] 
  & \Zz+\Zzn\ 
    =\ Z_{\chi}^{-\frac{1}{2}}\,
       \frac{k_0\mz}{k_0\mz-i\,\Sigma_{\sscr{Z}\chi,\,n}}\onshellZ \,.
\end{align}
The relation between the $S$-matrix elements for outgoing longitudinal $Z$ 
bosons and outgoing would-be Goldstone bosons $\chi$ reads in the Sudakov limit
(with $N=\ga,\,Z$)
\vspace*{-8mm}
\begin{figure}[H]
 \begin{minipage}[l]{7cm}
     \begin{fmffile}{smatrixgn1}
     \begin{fmfgraph*}(80,80) \fmfpen{thin} \fmfleft{i1} \fmfright{o1}
     \fmf{phantom}{i1,v1,v2,o1} 
     \fmf{plain,left,tension=1.5}{i1,v1} \fmf{plain,right,tension=1.5}{i1,v1}
     \fmf{boson,tension=4}{v1,v2} \fmfblob{.15w}{v2}
     \fmf{double,tension=4}{v2,o1}
     \fmfdot{o1}
     \fmfv{label={$ \nu$},l.a=-90,l.d=7}{o1}
     \fmfv{label={$ Z^{\mathrm as}$},l.a=15,l.d=45}{v1}
     \fmfv{label={$ N$},l.a=145,l.d=12}{v2}
     \end{fmfgraph*}
     \end{fmffile}
       \vspace*{-2.37cm}
       \begin{align}
        \hspace*{3.3cm} i\,(k^2-\mz^2)\,\,\eps_{\sscr{L}}^{\nu}(k)\onshellZ \no
       \hspace*{2mm} + 
       \end{align}
  \end{minipage}
  \hspace*{4mm}
  \begin{minipage}[l]{7cm}
      \vspace*{0.5mm}   
      \begin{fmffile}{gchi1}
      \begin{fmfgraph*}(80,80) \fmfpen{thin} \fmfleft{i1} \fmfright{o1}
      \fmf{phantom}{i1,v1,v2,o1} 
      \fmf{plain,left,tension=1.5}{i1,v1} \fmf{plain,right,tension=1.5}{i1,v1}
     \fmf{dashes,tension=4}{v1,v2} \fmfblob{.15w}{v2}
      \fmf{double,tension=4}{v2,o1}
      \fmfdot{o1}
      \fmfv{label={$ \nu$},l.a=-90,l.d=7}{o1}
      \fmfv{label={$ Z^{\mathrm as}$},l.a=15,l.d=45}{v1}
      \fmfv{label={$ \chi$},l.a=145,l.d=12}{v2}
      \end{fmfgraph*}
      \end{fmffile}
        \vspace*{-2.37cm}
        \begin{align}
         \hspace*{3.2cm} i\,(k^2-\mz^2)\,\,\eps_{\sscr{L}}^{\nu}(k) \onshellZ 
         \no
        \end{align}
 \end{minipage} 
\end{figure}

\vspace*{7mm}
\h{-5}=
\vspace*{-18mm}

\begin{figure}[H]
\hspace*{9mm}
\begin{minipage}[l]{10cm}
     \begin{fmffile}{gchi2}
      \begin{fmfgraph*}(65,65) \fmfpen{thin} \fmfleft{i1} \fmfright{o1}
      \fmf{phantom}{i1,v1,v2,o1}
      \fmf{plain,left,tension=1.5}{i1,v1} \fmf{plain,right,tension=1.5}{i1,v1}
      \fmf{phantom,tension=4}{v1,v2} \fmf{phantom,tension=4}{v2,o1}
      \fmf{dashes}{v1,v2}
      \fmfv{label={$\textstyle \chi$},l.a=113,l.d=9}{v2}
      \end{fmfgraph*} 
      \end{fmffile}
     \vspace*{-2.1cm}
    \begin{align}
      \hspace*{-3.1cm} \left[ -\,i\,Z_{\chi}^{\frac{1}{2}} \right]\onshellZ
      \h{2}+ \no 
    \end{align}
\end{minipage}
\h{-47}
\vspace*{1mm}
\begin{minipage}[l]{10cm}
     \begin{fmffile}{gn2}
      \begin{fmfgraph*}(65,65) \fmfpen{thin} \fmfleft{i1} \fmfright{o1}
      \fmf{phantom}{i1,v1,v2,o1}
      \fmf{plain,left,tension=1.5}{i1,v1} \fmf{plain,right,tension=1.5}{i1,v1}
      \fmf{phantom,tension=4}{v1,v2} \fmf{phantom,tension=4}{v2,o1}
      \fmf{boson}{v1,v2}
      \fmfv{label={$\textstyle \gamma$},l.a=102,l.d=6}{v2}
      \fmfv{label={$\textstyle \mu$},l.a=-160,l.d=35}{o1}
      \end{fmfgraph*} 
      \end{fmffile}
     \vspace*{-2.1cm}
    \begin{align}
      \hspace*{-0.7cm} Z_{\chi}^{\frac{1}{2}}\,n^{\mu}\,
      \left( \frac{\mz}{k_0} \right)\,\frac{i\,\Sigma_{\ga\chi,\,n}}{k_0\mz}
      \onshellZ \no 
    \end{align}
 \end{minipage}
\end{figure}

\vspace*{3mm}
\h{52} $-$
\vspace*{-18mm}

\begin{figure}[H]
\hspace*{66mm}
\begin{minipage}[l]{10cm}
     \begin{fmffile}{gn3}
      \begin{fmfgraph*}(65,65) \fmfpen{thin} \fmfleft{i1} \fmfright{o1}
      \fmf{phantom}{i1,v1,v2,o1}
      \fmf{plain,left,tension=1.5}{i1,v1} \fmf{plain,right,tension=1.5}{i1,v1}
      \fmf{phantom,tension=4}{v1,v2} \fmf{phantom,tension=4}{v2,o1}
      \fmf{boson}{v1,v2}
      \fmfv{label={$\textstyle Z$},l.a=102,l.d=6}{v2}
      \fmfv{label={$\textstyle \mu$},l.a=-160,l.d=35}{o1}
      \end{fmfgraph*} 
      \end{fmffile}
     \vspace*{-2.1cm}
    \begin{align}
      \hspace*{1cm} Z_{\chi}^{\frac{1}{2}}\,n^{\mu}\,
      \left( \frac{\mz}{k_0} \right)\,
      \frac{k_0\mz-i\,\Sigma_{\sscr{Z}\chi,\,n}}{k_0\mz}
      \onshellZ \no 
    \end{align}
 \end{minipage}
\end{figure}

\vspace*{6mm}
\h{-5}=
\vspace*{-20mm}

\begin{figure}[H]
\hspace*{10mm} 
 \begin{minipage}[l]{7cm}
     \begin{fmffile}{smatrixetgn5}
     \begin{fmfgraph*}(80,80) \fmfpen{thin} \fmfleft{i1} \fmfright{o1}
     \fmf{phantom}{i1,v1,v2,o1} 
     \fmf{plain,left,tension=1.5}{i1,v1} \fmf{plain,right,tension=1.5}{i1,v1}
     \fmf{boson,tension=4}{v1,v2} \fmfblob{.15w}{v2}
     \fmf{dbl_dashes,tension=4}{v2,o1}
     \fmfdot{o1}
     \fmfv{label={$\chi^{\mathrm as}$},l.a=15,l.d=45}{v1}
     \fmfv{label={$ N$},l.a=145,l.d=12}{v2}
     \end{fmfgraph*}    
     \end{fmffile}
       \vspace*{-2.37cm} 
       \begin{align}
         \hspace*{3.2cm} (-i)^2\,(k^2-\mz^2) \onshellZ\,\, \no
       \hspace*{2mm} + 
       \end{align}
\end{minipage}
\h{3.5}
\begin{minipage}[l]{7cm}
      \begin{fmffile}{etgchi1}
      \begin{fmfgraph*}(80,80) \fmfpen{thin} \fmfleft{i1} \fmfright{o1}
      \fmf{phantom}{i1,v1,v2,o1} 
      \fmf{plain,left,tension=1.5}{i1,v1} \fmf{plain,right,tension=1.5}{i1,v1}
     \fmf{dashes,tension=4}{v1,v2} \fmfblob{.15w}{v2}
      \fmf{dbl_dashes,tension=4}{v2,o1}
      \fmfdot{o1}
       \fmfv{label={$ \chi^{\mathrm as}$},l.a=15,l.d=45}{v1}
      \fmfv{label={$ \chi$},l.a=145,l.d=12}{v2}
      \end{fmfgraph*}
      \end{fmffile}
        \vspace*{-2.37cm}
        \begin{align}
          \hspace*{3.2cm} (-i)^2\,(k^2-\mz^2) \onshellZ .\no
        \end{align}
\end{minipage}
\vspace*{1mm}
\end{figure}
\ni 
Hence, for outgoing longitudinal $Z$ bosons the dominant Sudakov correction
factor amounts to multiplying each outgoing longitudinal $Z$-boson line of the
matrix element by the factor $-\,i\,Z_{\chi}^{\frac{1}{2}}$ (provided that the 
matrix element is not mass-suppressed to start with). For incoming longitudinal
$Z$ bosons this Sudakov factor becomes $\,i\,Z_{\chi}^{\frac{1}{2}}$. So, just 
like in the charged-boson sector a special case of the Equivalence Theorem is
obtained, $Z_{\sscr{L},\,\mu}^{\rm as} \to -\!/\!\!+\,i\,\chi^{\rm as}\,$ for
outgoing/incoming particles.
\\[-5mm]

\ni
In the transverse sector the situation is quite different, since now the
{\it gauge bosons\/} mix explicitly. The corresponding terms in the asymptotic
states do not involve derivatives and therefore play an explicit role in the
$S$-matrix elements. The easiest procedure to deal with this $\,Z\,$--$\,\ga\,$
mixing is to first diagonalize the propagator matrix in the transverse sector
according to 
\bea 
  \left( \begin{array}{c} Z_{\sscr{T},\,\mu}^{^{\sscr{diag}}} \\
                          A_{\sscr{T},\,\mu}^{^{\sscr{diag}}}
         \end{array} \right)
  =
  \left( \begin{array}{cc} \cos\theta(k^2)    & \sin\theta(k^2) \\
                           -\,\sin\theta(k^2) & \cos\theta(k^2) 
         \end{array} \right)
  \,
  \left( \begin{array}{c} Z_{\sscr{T},\,\mu} \\
                          A_{\sscr{T},\,\mu}
         \end{array} \right) \,,
\ea
with
\bea
  \theta(k^2) = \frac{1}{2}\,\arctan\left( 
                \frac{2\,\Sigma_{\ga\sscr{Z},\,\G}}
                     {\mz^2+\Sigma_{\sscr{ZZ},\,\G}-\Sigma_{\ga\ga,\,g}} 
                \right)\,.
\ea
Subsequently the asymptotic states are defined in terms of these diagonal 
interaction states through the relation
\bea 
  \left( \begin{array}{c} Z_{\sscr{T},\,\mu}^{\rm as} \\
                          A_{\sscr{T},\,\mu}^{\rm as}
         \end{array} \right)
  =
  \left( \begin{array}{cc} 
           C_{\sscr{ZZ}}^{-\,\frac{1}{2}}\,\cos\theta(\mz^2) &  0 \\
           0 & C_{\ga\ga}^{-\,\frac{1}{2}}\,\cos\theta(0)
         \end{array} \right)
  \,
  \left( \begin{array}{c} Z_{\sscr{T},\,\mu}^{^{\sscr{diag}}} \\
                          A_{\sscr{T},\,\mu}^{^{\sscr{diag}}}
         \end{array} \right) \,.
\ea                 
Using the ``non-renormalization'' conditions in Eq.~(\ref{nonrenN}) and the
transverse part $\propto Q_{\mu\nu}$ of the free-field constraint, one obtains
\begin{subequations}
\begin{align}
\label{Cgamma}
  \tan\theta(0)\     &=\ \frac{\Sigma_{\ga\sscr{Z},\,\G}}
                              {\mz^2+\Sigma_{\sscr{ZZ},\,\G}}\onshellA 
  \quad , \quad
  C_{\ga\ga}^{-1}
  \ =\ 1 - \frac{\pa}{\pa\,k^2}\left.\!\Biggl[ \Sigma_{\ga\ga,\,\G}
       + \frac{\Sigma^{\,2}_{\ga\sscr{Z},\,\G}}
              {k^2-\mz^2-\Sigma_{\sscr{ZZ},\,\G}} \Biggr]\right|_{k^2=0}
  \\[1mm]
\label{CZ}
  \tan\theta(\mz^2)\ &=\ \frac{\Sigma_{\ga\sscr{Z},\,\G}}
                              {\mz^2-\Sigma_{\ga\ga,\,\G}}\onshellZ
  \quad\!\!\!\! , \quad 
  C_{\sscr{ZZ}}^{-1}
  \ =\ 1 - \frac{\pa}{\pa\,k^2}\left.\!\Biggl[ \Sigma_{\sscr{ZZ},\,\G}
       + \frac{\Sigma^{\,2}_{\ga\sscr{Z},\,\G}}
              {k^2-\Sigma_{\ga\ga,\,\G}} \Biggr]\right|_{k^2=\mz^2} \,.
\end{align}
\end{subequations}             
This leads to the following $S$-matrix elements for transverse neutral gauge 
bosons in the Sudakov limit (with $N=\ga,\,Z$):
\vspace*{-7mm}
\begin{figure}[H]
 \begin{minipage}[l]{7cm}
     \begin{fmffile}{smatrixgn2}
     \begin{fmfgraph*}(80,80) \fmfpen{thin} \fmfleft{i1} \fmfright{o1}
     \fmf{phantom}{i1,v1,v2,o1} 
     \fmf{plain,left,tension=1.5}{i1,v1} \fmf{plain,right,tension=1.5}{i1,v1}
     \fmf{boson,tension=4}{v1,v2} \fmfblob{.15w}{v2}
     \fmf{double,tension=4}{v2,o1}
     \fmfdot{o1}
     \fmfv{label={$ \nu$},l.a=-90,l.d=7}{o1}
     \fmfv{label={$ \gamma^{ \mathrm as}$},l.a=15,l.d=45}{v1}
     \fmfv{label={$ N$},l.a=145,l.d=12}{v2}
     \end{fmfgraph*}
     \end{fmffile}
       \vspace*{-2.5cm}
       \begin{align}
        \hspace*{3.3cm} i\,k^2\,\eps_{\sscr{T}}^{\nu}(k)\onshellA \no
       \hspace*{2mm} 
       =\ C_{\ga\ga}^{\frac{1}{2}}\,\eps_{\sscr{T}}^{\mu}(k)\h{2} \Biggl[
       \end{align}
  \end{minipage}
\begin{minipage}[l]{4cm}
     \begin{fmffile}{gn2}
      \begin{fmfgraph*}(65,65) \fmfpen{thin} \fmfleft{i1} \fmfright{o1}
      \fmf{phantom}{i1,v1,v2,o1}
      \fmf{plain,left,tension=1.5}{i1,v1} \fmf{plain,right,tension=1.5}{i1,v1}
      \fmf{phantom,tension=4}{v1,v2} \fmf{phantom,tension=4}{v2,o1}
      \fmf{boson}{v1,v2}
      \fmfv{label={$\textstyle \gamma$},l.a=102,l.d=6}{v2}
      \fmfv{label={$\textstyle \mu$},l.a=-160,l.d=35}{o1}
      \end{fmfgraph*} 
      \end{fmffile}
     \vspace*{-1.97cm}
    \begin{align}
      \h{15} -  \h{1} \tan\theta(0) \no
    \end{align}
 \end{minipage} 
\v{4.5}\h{-3}
\begin{minipage}[l]{14cm}
     \vspace*{4mm}
     \begin{fmffile}{gn3}
      \begin{fmfgraph*}(65,65) \fmfpen{thin} \fmfleft{i1} \fmfright{o1}
      \fmf{phantom}{i1,v1,v2,o1}
      \fmf{plain,left,tension=1.5}{i1,v1} \fmf{plain,right,tension=1.5}{i1,v1}
      \fmf{phantom,tension=4}{v1,v2} \fmf{phantom,tension=4}{v2,o1}
      \fmf{boson}{v1,v2}
      \fmfv{label={$\textstyle Z$},l.a=102,l.d=6}{v2}
      \fmfv{label={$\textstyle \mu$},l.a=-160,l.d=35}{o1}
      \end{fmfgraph*} 
      \end{fmffile}
     \vspace*{-2.3cm}
    \begin{align}
      \h{-52.5} \left. \Biggr]\right|_{k^2=0} \no
    \end{align}
 \end{minipage}
\vspace*{-7mm}
\end{figure}
\ni for the photon and
\vspace*{-7mm}
\begin{figure}[H]
 \begin{minipage}[l]{7cm}
     \begin{fmffile}{smatrixgn1}
     \begin{fmfgraph*}(80,80) \fmfpen{thin} \fmfleft{i1} \fmfright{o1}
     \fmf{phantom}{i1,v1,v2,o1} 
     \fmf{plain,left,tension=1.5}{i1,v1} \fmf{plain,right,tension=1.5}{i1,v1}
     \fmf{boson,tension=4}{v1,v2} \fmfblob{.15w}{v2}
     \fmf{double,tension=4}{v2,o1}
     \fmfdot{o1}
     \fmfv{label={$ \nu$},l.a=-90,l.d=7}{o1}
     \fmfv{label={$ Z^{ \mathrm as}$},l.a=15,l.d=45}{v1}
     \fmfv{label={$ N$},l.a=145,l.d=12}{v2}
     \end{fmfgraph*}
     \end{fmffile}
       \vspace*{-2.5cm}
       \begin{align}
        \hspace*{3.2cm} i(k^2\!-\!\mz^2)\,\eps_{\sscr{T}}^{\nu}(k)\!\onshellZ 
        \no
       =\ C_{\sscr{ZZ}}^{\frac{1}{2}}\,\eps_{\sscr{T}}^{\mu}(k) \Biggl[
       \end{align}
  \end{minipage}
\begin{minipage}[l]{4cm}
     \begin{fmffile}{gn3}
      \begin{fmfgraph*}(65,65) \fmfpen{thin} \fmfleft{i1} \fmfright{o1}
      \fmf{phantom}{i1,v1,v2,o1}
      \fmf{plain,left,tension=1.5}{i1,v1} \fmf{plain,right,tension=1.5}{i1,v1}
      \fmf{phantom,tension=4}{v1,v2} \fmf{phantom,tension=4}{v2,o1}
      \fmf{boson}{v1,v2}
      \fmfv{label={$\textstyle Z$},l.a=102,l.d=6}{v2}
      \fmfv{label={$\textstyle \mu$},l.a=-160,l.d=35}{o1}
      \end{fmfgraph*} 
      \end{fmffile}
     \vspace*{-1.97cm}
    \begin{align}
      \h{18} + \h{1} \tan\theta(\mz^2) \h{0.7}\no
    \end{align}
 \end{minipage} 
\v{4.5}
\begin{minipage}[l]{14cm}
     \vspace*{4mm}
     \begin{fmffile}{gn2}
      \begin{fmfgraph*}(65,65) \fmfpen{thin} \fmfleft{i1} \fmfright{o1}
      \fmf{phantom}{i1,v1,v2,o1}
      \fmf{plain,left,tension=1.5}{i1,v1} \fmf{plain,right,tension=1.5}{i1,v1}
      \fmf{phantom,tension=4}{v1,v2} \fmf{phantom,tension=4}{v2,o1}
      \fmf{boson}{v1,v2}
      \fmfv{label={$\textstyle \gamma$},l.a=102,l.d=6}{v2}
      \fmfv{label={$\textstyle \mu$},l.a=-160,l.d=35}{o1}
      \end{fmfgraph*} 
      \end{fmffile}
     \vspace*{-2.3cm}
    \begin{align}
      \h{-53} \left. \Biggr]\right|_{k^2=\mz^2} \no
    \end{align}
 \end{minipage}
\vspace*{-7mm}
\end{figure}
\ni for the $Z$ boson. So $C_{\ga\ga}^{\frac{1}{2}}$ and 
$C_{\sscr{ZZ}}^{\frac{1}{2}}$ act as overall normalization factors, whereas
$\tan\theta(0)$ and $\tan\theta(\mz^2)$ account for the fact that the
asymptotic neutral gauge-boson states have been obtained from a mixture of 
(interacting) photonic and $Z$-boson components. 
With the help of 
Eq.~(\ref{nonrenN}) we can bring the expressions for $\,C_{\ga\ga}^{-1}\,$
and $\,C_{\sscr{ZZ}}^{-1}\,$ in the familiar form of a projection by means of
sources:
\begin{eqnarray}
  C_{\ga\ga}^{-1} &=& 1 + \tan^2\theta(0) - \frac{i}{2\,k_0}\,\eps_{T,\,\mu}(k)
    \biggl\{ i\,\frac{\pa}{\pa\,k_0}\Bigl[ \Sigma_{\ga\ga}^{\mu\nu} 
    - 2\,\tan\theta(0)\,\Sigma_{\ga\sscr{Z}}^{\mu\nu}
    + \tan^2\theta(0)\,\Sigma_{\sscr{ZZ}}^{\mu\nu} \Bigr] 
    \biggr\}\, \eps_{T,\,\nu}^{*}(k)\onshellA \no \\[1mm]
  C_{\sscr{ZZ}}^{-1} &=& 1 - \frac{i}{2\,k_0}\,\eps_{T,\,\mu}(k) 
    \biggl\{ i\,\frac{\pa}{\pa\,k_0}\Bigl[ \Sigma_{\sscr{ZZ}}^{\mu\nu} 
    + 2\,\tan\theta(\mz^2)\,\Sigma_{\ga\sscr{Z}}^{\mu\nu}
    + \tan^2\theta(\mz^2)\,\Sigma_{\ga\ga}^{\mu\nu} \Bigr] 
    \biggr\}\, \eps_{T,\,\nu}^{*}(k)\onshellZ \no \\[3mm]
                     & & +\ \tan^2\theta(\mz^2) \,.
\end{eqnarray}
\\[-5mm]

\ni
As promised we come back to the ``non-renormalization'' condition for the 
electromagnetic charge, which follows automatically from the requirement that
the electromagnetic charge should not become energy-dependent. Combining the 
$S$-matrix element for the photon and this ``non-renormalization''
condition for the electromagnetic charge\footnote{This condition applies to the
          electromagnetic, non-isospin part of the coupling to the amputated
          Green's functions, e.g.~for couplings to fermions this consists of 
          the complete $ff\ga$ coupling and the 
          $\,-\,Q_f\,\gamma_{\mu}\,\sw/\cw\,$ part of the $ffZ$ coupling}, 
we obtain 
\bea
\label{nonrene}
  C_{\ga\ga}^{-\,\frac{1}{2}} = 1-\tan\theta(0)\,\frac{\sw}{\cw} \,,
\ea
where $\theta_{\rm w}$ is the weak mixing angle.
This condition will be crucial for limiting the calculation of the Sudakov
correction factors to the calculation of derivatives of self-energies. 
An explicit check of Eq.~(\ref{nonrene}) at the one-loop level can be found in 
Appendix~\ref{app:selfenergy}.

\section{Electroweak one-loop Sudakov logarithms}
\label{sec:one-loop}

To establish the formalism that will be used in the following
sections we are presenting here the one-loop calculation of the Sudakov
logarithms in the Coulomb gauge~\cite{Beenakker:2000kb}. For arbitrary 
on-shell/on-resonance SM particles our calculations are in agreement with the 
well known one-loop contributions to the external wave-function factors 
$Z=1+\delta Z$. These one-loop contributions will be denoted by 
$\,\delta Z^{(1)}$.

\subsection{The fermionic self-energy at one-loop level}
\label{sec:one-loop_fermions}

\ni As mentioned above, in order to determine the Sudakov logarithms in 
$s$-channel processes like $\,e^+e^-\!\to f\bar{f}\,$ ($f\!\neq\!e,\nu_e$), 
one has to calculate the external self-energies (i.e.\ the wave-function 
factors) of all four fermions involved in the process. Consider to this end 
the fermionic one-loop self-energy $\,\Sigma_f^{\,(1)}(p,n,M_1)$, originating 
from the emission of a gauge boson $V_1$ with loop-momentum $k_1$ and mass 
$M_1$ from an effectively
massless\footnote{Whenever possible the fermion mass will be neglected. 
                  The massive case (\eg top-quarks) can be treated in a similar
                  way in view of the ``non-renormalization'' condition for the 
                  fermion mass (see the discussion in Sect.~\ref{sec:wave}).} 
fermion $f$ with momentum $p$:\\
\begin{minipage}[c]{15.5cm}
\vspace{-1cm}
\begin{figure}[H]
\hspace{7cm}
  \begin{fmffile}{self}
  \begin{fmfgraph*}(140,140) \fmfpen{thin} \fmfleft{i1} \fmfright{o1}
  \fmf{fermion,tension=0.7,label=$f(p)$,l.side=right,width=0.6}{i1,v1} 
  \fmf{fermion,tension=0.5,label=$f_1(p\!-\!k_1)$,l.side=right,width=0.6}
      {v1,v3} 
  \fmf{fermion,tension=0.7,label=$f(p)$,l.side=right,width=0.6}{v3,o1} 
  \fmfdot{v1} \fmfdot{v3} 
  \fmf{boson,left,tension=-1,label=$V_1(k_1)$,l.side=left,width=0.6}{v1,v3}
  \fmf{phantom,left}{v1,v3}
  \fmffreeze
  \fmf{phantom_arrow}{v1,v3}
  \end{fmfgraph*}
  \end{fmffile}
\end{figure}
{}\vspace*{-4.3cm}
\bea
\hspace{-5cm}  -\,i\,\, \Sigma_f^{\,(1)}(p,n,M_1)\,\,\,\,\, =\,   \no
\ea
{}\vspace*{0.2cm}
\end{minipage}\\
\ni Again $n$ is the unit vector in the time direction, which enters by virtue 
of using the Coulomb gauge. In the high-energy limit the fermion mass in the 
numerator of the fermion propagator can be neglected with respect to $\ps$
and similarly the contribution involving a mixed gauge-boson -- Goldstone-boson
propagator can be discarded. The self-energy 
$\Sigma_f^{\,(1)}$ then contains an odd number of $\ga$-matrices, leading to 
the following natural decomposition in terms of the two possible structures 
$\ps$ and $\ns$:
\bea
  \Sigma_f^{\,(1)}(p,n,M_1) &\approx& 
      \Bigl[ \,\mps\,\Sigma_p^{\,(1)}(n\,\c p,p^2,M_1)
           + \mns\:\frac{p^2}{n\,\c p}\,\Sigma_n^{\,(1)}(n\,\c p,p^2,M_1)
      \Bigr] e^2\,\Gamma_{\!\!ff_1\!V_1}^{\,^{\scriptstyle 2}}~,
\ea
with the proportionality factor of the second term being dictated by the 
``non-renormalization'' condition for the fermion mass.
The coupling factor $\Gamma_{\!\!ff_1\!V_1}$ is defined according to
\bea
   \Gamma_{\!\!ff_1\!V_1} &=& V_{\!\!ff_1\!V_1} - \ga_5\,A_{\!ff_1\!V_1}~,
\ea
where $V_{\!\!ff_1\!V_1}$ and $A_{\!ff_1\!V_1}$ are the vector and axial-vector
couplings of the fermion $f$ to the exchanged gauge boson $V_1$. In our 
convention these coupling factors read
\begin{equation}
 \Gamma_{\!\!ff\,\ga} \ =\ -\,Q_f, \qquad
 \Gamma_{\!\!ffZ}\ =\ \frac{(1-\ga_5)\,I_f^3 - 2\,Q_f\,\sin^2\theta_{\rm w}}
                         {2\cos\theta_{\rm w}\,\sin\theta_{\rm w}}, \qquad
 \Gamma_{\!\!ff'\,\!W}\ =\ \frac{(1-\ga_5)}{2\sqrt{2}\,\sin\theta_{\rm w}}~.
\end{equation} 
Here $I_f^3$ is the quantum number corresponding to the third component of the
weak isospin, $e\,Q_f$ is the electromagnetic charge, and $\theta_{\rm w}$ is 
the weak mixing angle. We have denoted the isospin partner of $f$ by 
$f^{\prime}$.
\\[-5mm]

\ni The contribution to the external wave-function factor now amounts to
multiplying the self-energy by $\,i/\mps\,$ on the side where it is 
attached to the rest of the scattering diagram and by the appropriate fermion 
source on the other side. Finally the square root should be taken of the
external wave-function factor, i.e.\ the one-loop contribution should be 
multiplied by the usual factor $1/2$. For an initial-state fermion, 
for example, one obtains%
\footnote{For an outgoing fermion one obtains $\frac{1}{2}\,\bar{u}_f(p)\,
          \delta \,\widetilde{Z}_f^{\,(1)}$, where $\delta \,
          \widetilde{Z}_f^{\,(1)}$ can be derived from 
          ${\delta\,Z}_f^{\,(1)}$ by reversing the sign in front of $\ga_5$.}
\bea
  \frac{1}{2}\,\,\frac{i}{\mps}\,\Bigl[ -i\,\Sigma_f^{\,(1)}(p,n,M_1) \Bigr]\,
  u_f(p) 
  &\approx& \frac{e^2}{2}\,\Gamma_{\!\!ff_1\!V_1}^{\,^{\scriptstyle 2}}\, 
      \Bigl[ \Sigma_p^{\,(1)}(n\,\c p,m_f^2,M_1)
             + 2\,\Sigma_n^{\,(1)}(n\,\c p,m_f^2,M_1)
      \Bigr]\,u_f(p) \no\\[1mm]
                &\equiv& \frac{1}{2}\,\delta\, Z_f^{\,\,(1)}(V_1)\,u_f(p)~,
\ea
where $m_f$ is the mass of the external fermion and $\sqrt{s}=2\,p_0$ is the 
center-of-mass energy of the process $e^+e^- \to f \bar{f}$. 
This contribution to the external wave-function factor
$\,Z_f=1+\delta\,Z_f\,$ can be extracted from the full fermionic self-energy 
by applying a projection by means of sources (see Sect.~\ref{sec:wave})
\begin{align}
\label{f1loopa}
& \delta\, Z_f^{\,\,(1)}(V_1)\ 
  =\ \frac{i}{2\,p_0}\,\bar{u}_f(p)\,
  \biggl\{ \frac{\pa}{\pa p_0}\,\Bigl[ -i\,\Sigma_f^{\,(1)}(p,n,M_1) \Bigr] 
  \biggr\}\,u_f(p) \no\\[2mm]
& \h{5} =\ \frac{i}{2\,p_0}\,\bar{u}_f(p)\,\biggl\{ \frac{\pa}{\pa \, p_0}\int
  \frac{\dd^4 k_1}{(2\,\pi)^4}\,(ie\,\gamma_{\nu}\,\Gamma_{\!\!ff_1\!V_1})\,
  \frac{i}{(\ps-\ks_1)-m_{f_1}+i\eps}\,(ie\,\gamma_{\mu}\,
  \Gamma_{\!\!ff_1\!V_1})\,P^{\mu\nu}(k_1,M_1)\biggr\}\,u_f(p)\, \no\\[2mm] 
& \h{5} \approx\ -\,e^2\,\Gamma_{\!\!ff_1\!V_1}^{\,^{\scriptstyle 2}}\,\int 
  \frac{\dd^4 k_1}{(2\,\pi)^4}\,
  \frac{4\,p_{\mu}\,p_{\nu}}{[(p-k_1)^2-m_{f}^2+i\epsilon]\,^2}\,
  P^{\mu\nu}(k_1,M_1)~,
\end{align}
where we have made use of the Dirac equation for the spinor $u_f(p)$, its 
normalization condition $\bar{u}_f(p)\, \gamma^0 u_f(p) = 2\, p_0$, as well as 
\bea
\frac{\pa}{\pa \, p_{\mu}} \, \frac{1}{\ps}\ =\
          -\,\frac{1}{\ps} \, \gamma^{\mu} \, \frac{1}{\ps}\,.
\ea
Note also that the loop-momentum $k_1$ has been neglected in the numerator of 
the fermion propagator, since only collinear-soft gauge-boson momenta will 
give rise to the Sudakov logarithms. Therefore it comes as no big surprise that
we observe an eikonal factor in the integrand of the last 
integral in Eq.~(\ref{f1loopa}). The mass of the fermion inside the loop, 
$m_{f_1}$, is at best of the order of the $Z$-boson mass (for the top-quark). 
At the leading-logarithmic level it therefore only enters as an independent 
mass scale if the exchanged gauge boson is a photon [i.e.\ $m_{f_1}=m_f$, as
implemented in the last step of Eq.~(\ref{f1loopa})], where the fermion mass 
is needed for the regularization of the collinear singularities. In the last 
step of Eq.~(\ref{f1loopa}) we have also exploited the fact that
$\delta Z^{\,(1)}_f$ will be multiplied on the right by $u_f(p)$, so
writing $ \Gamma_{\!\!ff_1\!V_1}^{\,^{\scriptstyle 2}}$ or its projection 
on left/right-handed chiral couplings $(V_{ff_1V_1} \pm A_{ff_1V_1})^2 $ is 
effectively equivalent.
\\[-5mm]

\ni
Making use of the explicit form of the gauge-boson propagator in the Coulomb 
gauge, given in Eq.~(\ref{trans}), the numerator of the last integral in 
Eq.~(\ref{f1loopa}) can be simplified as follows
\begin{align}
 i\,\Big[ k_1^2-M_1^2+i\eps\, \Big]\,4\,p_{\mu}\,p_{\nu}\,P^{\mu\nu}(k_1,M_1)\ 
 &\approx\ \frac{4}{\vec{k_1}^2}\,\Big[ (p\c k_1)^2  
          - 2\,k_{1\,0}\,p_0\,(p\c k_1) \Big] \no \\
 &\h{-20} =\ \frac{1}{\vec{k_1}^2}\,\Big[ (p-k_1)^2-p^2 -k_1^2 \Big]^2 
   + \frac{4\,k_{1\,0}\,p_0}{\vec{k_1}^2}\,\Big[ (p-k_1)^2-p^2-k_{1}^2 \Big]\,.
\end{align}
As we will see below, in order to obtain two logarithms both the fermion and 
the gauge-boson propagator are needed. Now $\,p^2=m_f^2\,$ can be neglected
and the terms $\,\propto k_{1}^2\,$ and $\,(p-k_1)^4\,$ will kill one of the 
types of denominators. Thus we are left with
\bea
\label{4ppP}
  i\,\Big[ k_1^2-M_1^2+i\eps\, \Big]\,4\,p_{\mu}\,p_{\nu}\,P^{\mu\nu}(k_1,M_1) 
  &\approx& \frac{4\,k_{1\,0}\,p_0}{\vec{k_1}^2}\,
            \Big[ (p-k_1)^2-m_f^2 \Big] \,.
\ea
Therefore 
\bea
\label{f1loop}
  \delta\, Z_f^{\,\,(1)}(V_1) \!\!   &\approx& 
  -\,e^2\,\Gamma_{\!\!ff_1\!V_1}^{\,^{\scriptstyle 2}}\,\int 
  \frac{\dd^4 k_1}{(2\,\pi)^4}\,\frac{4\,k_{1\,0}\,p_0}{\vec{k_1}^2}\, 
  \frac{-\,i}{[(p-k_1)^2-m_{f}^2+i\epsilon]\,[ k_1^2-M_1^2+i\eps\,]} \,.
\ea 

\ni Having two canonical momenta at our disposal, i.e.\ $p$ and
$n$, we define the following Sudakov parametrisation of the gauge-boson 
loop-momentum $k_1$:
\bea
  k_1= v_1\,q + u_1\,\bar{q} + k_{1_\bot}~,
\ea
with 
\bea
\label{fourmom}
  p^{\mu} \equiv (E,\beta_f E,0,0)~, 
  & & \qquad\qquad \beta_f = \sqrt{1-m_f^2/E^2}~, 
      \qquad\qquad s=4\, E^2, \no \\[1mm] 
  q^\mu = (E,E,0,0)~,                
  & & \qquad\qquad \bar{q}^\mu = (E,- E,0,0)~,
      \qquad\qquad k_{1_\bot}^{\mu} = (0,0,\vec{k}_{1_\bot})~.
\ea
In terms of this parametrisation, the integration measure $\,\dd^4 k_1\,$,
the invariants $\,(p\, \c k_1)\,$ and $\,k_1^2\,$, and the gauge-boson energy 
$k_1^0$ read   
\bea
  \dd^4 k_1    &=& \pi\,\frac{s}{2}\,\dd v_1\,\dd u_1\, 
                   \dd \vec{k}_{1_\bot}^{\,2}~, \no\\[1mm]
  (p\, \c k_1) &=& \frac{s}{4}\, [\,v_1\,(1-\beta_f) + u_1\,(1+\beta_f)\,]\
                   \approx \ 
                   \frac{s}{2}\,\Bigl( u_1+\frac{m_f^2}{s}\,v_1 \Bigr)~,
                   \no\\[2mm]
  k_1^2        &=& s\,v_1\,u_1 - \vec{k}_{1_\bot}^{\,2}
                   \qquad {\rm and} \qquad 
                   k_1^0 \ =\ \frac{\sqrt{s}}{2}\,(v_1+u_1)~.
\ea
The term containing the fermion mass $m_f$ is needed for the exchange of 
photons only, regulating the collinear singularity at $u_1=0$. For the exchange
of a massive gauge boson the mass $M_1$ will be the dominant collinear as well
as infrared regulator.
\\[-5mm]  

\ni The $v_1$-integration is restricted to the interval $\,0\le v_1 \le 1$, as 
a result of the requirement of having poles in both hemispheres of the complex 
$u_1$-plane. The residue is then taken in the lower hemisphere in the pole of
the gauge-boson propagator: $s\,v_1\,u_1^{\rm res} = \vec{k}_{1_\bot}^{\,2} 
+ M_1^2 \equiv s\,v_1\,y_1$. Finally, $\vec{k}_{1_\bot}^{\,2}$ is substituted 
by $y_1$, with the condition $\,\vec{k}_{1_\bot}^{\,2} \ge 0\,$ translating 
into $\,v_1\,y_1 \ge M_1^2/s$. The one-loop Sudakov contribution to 
$\delta Z_f$ now reads 
\bea
\label{kernel1}
  \delta\, Z^{\,(1)}_f (V_1)&\approx& -\,\frac{\alpha}{\pi}\,
     \Gamma_{\!\!ff_1\!V_1}^{\,^{\scriptstyle 2}}\,\int_0^{\infty} \dd y_1 \,
     \int_0^1 \dd v_1 \ \frac{\Theta(v_1 y_1-\frac{M_1^2}{s})}
                              {(y_1+\frac{m_f^2}{s}v_1)\,(v_1+y_1)} 
     \no \\[1mm]
                                &\approx& -\,\frac{\alpha}{\pi}\,
     \Gamma_{\!\!ff_1\!V_1}^{\,^{\scriptstyle 2}}\,\int_0^1 \frac{\dd y_1}{y_1}
     \int_{y_1}^1 \frac{\dd z_1}{z_1}\ \cK^{\,(1)}(s,m_f^2,M_1,y_1,z_1)~,
\ea
with the integration kernel $\cK^{\,(1)}$ given by 
\bea
\label{kernel}
  \cK^{\,(1)}(s,m_f^2,M_1,y_1,z_1) &=& 
           \Theta\Bigl( y_1 z_1-\frac{M_1^2}{s} \Bigr)\,
           \Theta\Bigl (y_1-\frac{m_f^2}{s}\,z_1\Bigr)~.
\ea
Here we introduced the energy variable $\,z_1 = v_1 + y_1\,$ and made use of 
the fact that only collinear-soft gauge-boson momenta with 
$\,y_1 \ll z_1 \ll 1\,$ are responsible for the quadratic large-logarithmic 
effects. This can be read off directly from the first expression of
Eq.~(\ref{kernel1}), since for $\,v_1 \lsim {\cal O}(y_1)\,$ the integrand of
the $v_1$-integral does not exhibit a logarithmic $\,1/v_1\,$ type of 
evolution. Furthermore, one can use as rule of thumb that, in order to 
determine whether a certain term is negligible or not, the relevant 
kinematical region for quadratic large-logarithmic effects is given by
``lower integration bound'' $\ll$ integration variable $\ll$ ``upper 
integration bound'' (\eg $\,M_1/\sqrt{s} \ll z_1 \ll 1$, $y_1 \ll z_1$, or
$\,y_1 \gg M_1^2/s$).
As a result, the 
gauge boson inside the loop is effectively on-shell and transversely 
polarized (see Eq.~(\ref{trans}) with $k^2 \ll \vec{k}^2$ in the collinear 
regime). The same result can be obtained by means of the dispersion method. 
The dispersion method proceeds via the computation of the absorptive part by 
applying the Cutkosky cutting rule, which effectively puts both the internal 
gauge boson and fermion on-shell, whereas the external fermion becomes 
off-shell. Subsequently the real part is obtained by using dispersion-integral
(Cauchy-integral) techniques, turning the internal fermion off-shell and 
allowing the external fermion to be on-shell.
\\[-5mm]

\ni The exchanged gauge boson can either be a massless photon ($\ga$) or one 
of the massive weak bosons ($W^{\pm}$ or $Z$). The associated mass gap gives 
rise to distinctive differences in the two types of contributions. Bearing in 
mind that the SM is not parity conserving and making use of (\ref{App1LMA}) 
and (\ref{App1LlaA}) we present the one-loop Sudakov correction factors for 
right- and left-handed  fermions/antifermions separately:
\begin{subequations}  
\label{delta1}
\bea
\label{photon1+}
  \delta Z^{\,(1)}_{f_R}\,(\gamma) &=& \, \biggr[
        \left(\frac{Y^R_f}{2}\right)^{\!2}\, \biggr]\,{\rm L_{\ga}}(\la,m_f)
        \ =\ Q_f^2\,{\rm L_{\ga}}(\la,m_f)~, 
        \\
\label{photon1-}
  \delta Z^{\,(1)}_{f_L}\,(\gamma) &=& \,\biggl[ \left(I_f^{\,3}\right)^2 +
        I^{\,3}_f Y^L_f + \left(\frac{Y^L_f}{2}\right)^{\!2}\, \biggr]\, 
        {\rm L_{\ga}}(\la,m_f)\ =\ Q_f^2\,{\rm L_{\ga}}(\la,m_f)~,\\[1mm]
\label{W1+}
  \delta Z^{\,(1)}_{f_R}\,(W) &=& \,0 ~, \\[2mm]
\label{W1-}
  \delta Z^{\,(1)}_{f_L}\,(W) &=& \,\frac{1}{2\,\swto}\,{\rm L}(M,M)~,\\
\label{Z1+}
  \delta Z^{\,(1)}_{f_R}\,(Z) &=& \,\frac{\swto}{\cwto}\,
        \left( \frac{Y^R_f}{2} \right)^2\,{\rm L}(M,M) 
        \ =\ \biggl[ \left( \frac{Y_f^R}{2\,\cw} \right)^2 \!
                  - Q_f^2 \biggr]\L(M,M)~,\\
\label{Z1-}
  \delta Z^{\,(1)}_{f_L}\,(Z) &=& \,\biggl[ \frac{\cwto}{\swto}\,
        \bigl(I_f^{\,3}\bigr)^2 - I^{\,3}_f Y^L_f + \frac{\swto}{\cwto}
        \left(\frac{Y^L_f}{2}\right)^{\!2}\, \biggr]\,{\rm L}(M,M) \no\\
                              &=& \biggl[ \frac{1}{4\,\swto} 
        + \left( \frac{Y_f^L}{2\,\cw} \right)^2 \! - Q_f^2 \biggr]\,\L(M,M)~,
\ea 
\end{subequations} 
with
\bea
\label{L}
 {\rm L}(M_1,M_2)              &=& -\,\frac{\alpha}{4\,\pi}\,
          \log\left( \frac{M_1^2}{s} \right) 
          \log\left( \frac{M_2^2}{s} \right)~, \\[1mm]
\label{Lgamma}
 {\rm L_{\ga}}(\la,M_1) &=& -\,\frac{\alpha}{4\,\pi}\, 
          \left[ \log^{\, 2}\!\left( \frac{\la^2}{s} \right) 
               - \log^{\, 2}\!\left( \frac{\la^2}{M_1^2} \right) 
          \right]~,
\ea
and $\,\delta Z^{\,(1)}_{f_R} = \delta Z^{\,(1)}_{\bar{f}_L}\,$ as well as
$\,\delta Z^{\,(1)}_{f_L} = \delta Z^{\,(1)}_{\bar{f}_R}\,$ for all three gauge
bosons. Note that these correction factors are the same for incoming as well as
outgoing particles. In Eq.~(\ref{delta1}) $Y_f^{R,L}$ denotes the right- and 
left-handed hypercharge of the external fermion, which is connected to the 
third component of the weak isospin $I^3_f$ and the electromagnetic charge 
$e\,Q_f$ through the Gell-Mann -- Nishijima relation 
$\,Q_f = I^{\,3}_f + Y_f^{R,L}/2$. The parameter $\la$ is the fictitious 
(infinitesimally small) mass of the photon needed for regularizing the 
infrared singularity at $z_1=0$. For the sake of calculating the leading 
Sudakov logarithms, the masses of the $W$ and $Z$ bosons can be represented by 
one generic mass scale $M$. 
\\[-5mm]

\ni In the process $\,e^+e^- \to f \bar{f}\,$ the one-loop correction 
factors presented in Eq.~(\ref{delta1}) contribute in the following way to the 
polarized matrix element, bearing in mind that at high energies the helicity 
eigenstates are equivalent to the chiral eigenstates:
\bea
 \cM_{e^+_R e^-_L \to f_L \bar{f}_R}^{\rm 1-loop,\,\,sudakov} &=&
      \frac{1}{2}\,\Bigl[ \delta Z^{\,(1)}_{e^+_R} + \delta Z^{\,(1)}_{e^-_L}
                        + \delta Z^{\,(1)}_{f_L} + \delta Z^{\,(1)}_{\bar{f}_R}
                   \Bigr]\,\cM_{e^+_R e^-_L \to f_L \bar{f}_R}^{\rm born}~,
\ea
and similar expressions for the other possible helicity combinations.
\\[-5mm]

\ni
As promised, we come back to two aspects of Sudakov logarithms in the Coulomb 
gauge that were anticipated in Sect.~\ref{sec:born}. First of all there was the
question whether one could expect contributions to the Sudakov correction 
factor from self-energies with fermions or ghosts in the loop. We saw in 
this section that the $1/\vec{k}^2$ part of the gauge-boson propagator in the 
Coulomb gauge is crucial for obtaining double logarithmic contributions. 
Obviously the fermion propagator does not possess such part. The ghost 
propagator does contain the required $1/\vec{k}^2$ pole, but lacks the pole 
structure $\,1/[k^2-M^2+i\epsilon]\,$ and hence no contribution to the Sudakov 
correction factor can be obtained. The second issue was the suppression of 
Sudakov logarithms originating from vertex corrections. To this end we
consider the following vertex correction, where we assume for simplicity that
the exchanged gauge boson as well as the incoming gauge boson are both photons
and that the fermion is massless\\
\begin{minipage}[l]{4cm}
\vspace*{-12mm}
\begin{figure}[H]
\begin{fmffile}{vcorr}
\hspace*{5mm}
\begin{fmfgraph*}(85,65) \fmfpen{thin} \fmfleft{i1} \fmfright{o1,o2}
\fmf{boson,width=0.6,label=$\stackrel{p}{\rightarrow}$,l.side=left}{i1,v1}
\fmf{plain,width=0.6}{o2,v3,v1,v2,o1}
\fmffreeze
\fmfdot{v1,v2,v3}
\fmf{photon,right,tension=0,width=0.6,label=$\downarrow k_1$}{v2,v3}
\fmfv{label=$p_1 {\scriptstyle \nearrow}$,l.a=160,l.d=3}{o2} 
\fmfv{label=$p_2 {\scriptstyle \searrow}$,l.a=-160,l.d=3}{o1}
\fmfv{label=${\scriptstyle \nu}$,l.a=-180,l.d=6}{v2}
\fmfv{label=${\scriptstyle \mu}$,l.a=180,l.d=6}{v3} 
\end{fmfgraph*}
\end{fmffile}
\end{figure}
\end{minipage}
\begin{minipage}[r]{12cm}
\vspace*{3mm}
\begin{align}
& \hspace*{-9mm} 
  \propto \,\int\frac{\dd^4 k_1}{(2\,\pi)^4}\,\bar{u}(p_2)\,
  \ga_{\nu}\,\frac{1}{\ps_2-\ks_1}\,\es(p)\,\frac{1}{\ps_1+\ks_1}\, 
  \ga_{\mu}\,P^{\mu \nu}(k_1,0)\,u(p_1) \no\\[4mm]
& \hspace*{-9mm} 
  \approx \ \bar{u}(p_2)\,\es(p)\,u(p_1)\int\frac{\dd^4 k_1}{(2\,\pi)^4}\,
  \frac{4\,p_{1\,\mu}\,p_{2\,\nu}\,P^{\mu\nu}(k_1,0)}{(p_1+k_1)^2\,(p_2-k_1)^2}
  \,.
\end{align}
\end{minipage}

\vspace*{7mm}
\ni
With the Sudakov parametrisation $k_1 = x\,p_1 + y\,p_2 + k_{1_\bot}$ and
say $\,p_{1}^{\,\mu} = E (1,1,0,0)\,$ and $\,p_{2}^{\,\mu} = E (1,-1,0,0)\,$
we obtain
\begin{align}
\label{p1p2}
 i\,[k_1^2+i\eps]\,p_{1\,\mu}\,p_{2\,\nu}\,P^{\mu\nu}(k_1,0)
 & = 4\,E^2\,\frac{x\,y}{(x+y)^2} \,,
\end{align}
where we have made use of the on-shell condition 
$\vec{k}_{1_\bot}^{\,2} \approx 4\,E^2\,x\,y$. The remaining term will 
{\it not\,} lead to Sudakov logarithms since the numerator will kill both 
poles originating from the fermion propagators. Hence we conclude that the 
piece of the gauge-boson propagator that would usually lead to Sudakov 
logarithms, \ie $k_0\,(k^{\mu}n^{\nu}+n^{\mu}k^{\nu})/\vec{k}^2$, 
is effectively rendered inactive for vertex corrections by the $\G_{\mu\nu}$ 
part of the same 
gauge-boson propagator.\footnote{Recall that in the case of the self-energy 
     $p_{\mu}p_{\nu}\G^{\mu\nu}=0$ and that therefore the terms relevant for 
     Sudakov logarithms survive}
The same argument holds for box or higher-point corrections.

\subsection{The bosonic self-energies at one-loop level}
\label{sec:boson1loop}

As we have seen in Sect.~\ref{sec:wave}, the transverse and longitudinal gauge
bosons have to be treated separately. To all orders in perturbation theory
the Sudakov correction factors for longitudinal $W$ and $Z$ bosons are given by
$Z_{\phi}^{\frac{1}{2}}$  and $Z_{\chi}^{\frac{1}{2}}$, respectively (provided 
that the matrix element is not mass-suppressed to start with). These 
wave-function factors are obtained from the scalar $\phi$ and $\chi$ 
self-energies through the relations
\bea
  Z_{\phi}^{-1} = 1 + \frac{1}{2k_0}\,\frac{\pa\,\Sigma_{\phi\phi}}{\pa\,k_0}
                  \onshellW
  \quad \mbox{and} \quad
  Z_{\chi}^{-1} = 1 + \frac{1}{2k_0}\,\frac{\pa\,\Sigma_{\chi\chi}}{\pa\,k_0}
                  \onshellZ \,. \no
\ea
The corresponding one-loop corrections can be calculated in a trivial way with
the help of the (derivative) method described in 
Sect.~\ref{sec:one-loop_fermions}.\footnote{The full (\ie non-derivative) 
     scalar self-energies can be found in Appendix~\ref{app:selfenergy}}
\\[-5mm]

\ni
Next we sketch the calculation of the transverse gauge-boson self-energies. 
We start with the charged sector ($W^{\pm}$ bosons) and then move on to the 
neutral sector ($Z$ bosons and photons). According to the discussion in
Sect.~\ref{sec:wave}, the Sudakov correction factor for transverse $W$ bosons 
amounts to multiplying the corresponding external line of the
matrix element by $Z_{\sscr{W}}^{\frac{1}{2}}$. Recalling that 
\begin{align}
Z_{\sscr{W}}^{-1} & = 1 - \frac{i}{2\,k_0}\,\eps_{T,\,\mu}(k) 
    \biggl\{ \frac{\pa}{\pa\,k_0}\Bigl[ i\,\Sigma_{\sscr{WW}}^{\mu\nu} \Bigr] 
    \biggr\}\, \eps_{T,\,\nu}^{*}(k)\onshellW \,,\no
\end{align}
the one-loop contribution to the external wave-function factor 
$\,Z_{\sscr{W}} \equiv 1+\delta Z_{\sscr{W}_T}\,$ can be extracted from the 
full one-loop $W$-boson self-energy by means of the projection
\bea
\label{b1loop}
  \delta Z_{\sscr{W}_T}^{\,(1)} = \frac{i}{2\,k_0}\,\eps_{T,\,\mu}(k) 
       \biggl\{ \frac{\pa}{\pa\,k_0} 
       \Bigl[ i\,\Sigma_{\sscr{WW}}^{\mu\nu,\,(1)} \Bigr] 
       \biggr\}\, \eps_{T,\,\nu}^{*}(k)\onshellW \,.
\ea
Note again that the transverse polarization vectors $\,\eps^{\mu}_T(k)\,$ and 
$\,\eps^{* \,\nu}_T(k)\,$ project on $-\G_{\mu\nu}$.
The vertex structures present in $\,\Sigma_{\sscr{WW},\,\G}^{\,(1)}\,$ 
will give rise to the usual eikonal factors, since we can neglect the
loop-momentum with respect to the $W$-boson momentum $k$. The rest of the 
calculation proceeds in the same way as worked out in 
Sect.~\ref{sec:one-loop_fermions} (for more details we refer to
Appendix~\ref{app:selfenergy}), resulting in 
\begin{subequations}
\begin{align}
\delta Z^{\,(1)}_{\sscr{W}_T}(\ga) &\ =\ Q_{\sscr{W}}^2\,\L_{\ga}(\lambda,M)
     \\[2mm]
\delta Z^{\,(1)}_{\sscr{W}_T}(Z) &\ =\ \frac{\cwto}{\swto}\,\L(M,M)
     \ =\ \biggl[ \frac{1}{\swto} - Q_{\sscr{W}}^2 \biggr] \L(M,M) \\
\delta Z^{\,(1)}_{\sscr{W}_T}(W) &\ =\ \frac{1}{\swto}\,\L(M,M) 
\end{align}
\end{subequations}
with $\,\L_{\ga}(\lambda,M)\,$ and $\,\L(M,M)\,$ as defined in
Eqs.~(\ref{Lgamma}) and (\ref{L}), respectively. 
\\[-5mm]

\ni 
We have applied these one-loop Sudakov corrections to the reactions
$\,e^+e^- \to W_T^+W_T^-,\,W_L^+W_L^-\,$ and found perfect agreement with the
high-energy approximation in Ref.~\cite{Beenakker:1993tt}. This indeed 
confirms the fundamental differences between transverse and longitudinal 
degrees of freedom.
\\[-5mm]

\ni
In the neutral gauge-boson sector we have to follow a step-wise procedure in 
order to express everything in terms of derivatives. First of all we exploit 
the fact that for $\,N_{1,2}=\ga,Z\,$ the self-energies
$\,\Sigma_{\sscr{N}_1\sscr{N}_2,\,g} = \cO(k^2,\mz^2)\,$ do not contain 
inverse powers of $\,k^2\,$ or $\,\mz^2$, as required by analyticity. 
The higher-order terms in $k^2$ will therefore be suppressed in the Sudakov 
limit, leading to the decomposition
\begin{equation}
\label{taylor}
  \Sigma_{\sscr{N}_1\sscr{N}_2,\,g} 
  = 
  k^2\,\Sigma_{\sscr{N}_1\sscr{N}_2,\,g}^{\prime}
  + \mz^2\,C_{\sscr{N}_1\sscr{N}_2,\,g} \,,
\end{equation}
with both $\,\Sigma_{\sscr{N}_1\sscr{N}_2,\,g}^{\prime}\,$ and 
$\,C_{\sscr{N}_1\sscr{N}_2,\,g}\,$ being independent of $k^2$. 
Next we expand Eqs.~(\ref{Cgamma}) and (\ref{CZ}) to one-loop:
\begin{align}
\label{Cgamma1l}
  C_{\ga\ga}^{-1} & \myto{1-loop}{} 1 - \frac{1}{2\,k_0} \,
  \frac{\pa\,\Sigma_{\ga\ga,\,g}^{\,(1)}}{\pa\,k_0} \onshellA 
  \ =\ 1 - \Sigma_{\ga\ga,\,g}^{\prime\,(1)} 
  \ \equiv\ 1 - \delta C_{\ga\ga}^{\,(1)} \\[2mm]
\label{CZ1l}
  C_{\sscr{ZZ}}^{-1} & \myto{1-loop}{} 1-\frac{1}{2\,k_0} \, 
  \frac{\pa\,\Sigma_{\sscr{ZZ},\,g}^{\,(1)}}{\pa\,k_0}\onshellZ
  \ =\ 1 - \Sigma_{\sscr{ZZ},\,g}^{\prime\,(1)} 
  \ \equiv\ 1 - \delta C_{\sscr{ZZ}}^{\,(1)} \,. 
\end{align}
Both self-energies can be calculated by means of the derivative
method explained in Sect.~\ref{sec:one-loop_fermions}, resulting in
\begin{equation}
  \Sigma_{\sscr{ZZ},\,g}^{\prime\,(1)} 
  \ =\ -\,\frac{\cw}{\sw}\,\Sigma_{\ga\sscr{Z},\,g}^{\prime\,(1)}
  \ =\ \frac{\cwto}{\swto}\,\Sigma_{\ga\ga,\,g}^{\prime\,(1)}
  \ =\ \cwto\,\biggl[ \frac{2}{\swto}\,\L(M,M) \biggr] 
  \ \equiv\ \cwto\,\Sigma_{33,\,g}^{\prime\,(1)} \,.
\end{equation}
Only collinear-soft gauge-boson exchange contributes to the Sudakov correction
(see Sect.~\ref{sec:one-loop_fermions}), hence only the non-abelian $W^3$
components of the external neutral gauge bosons participate.
To calculate the Sudakov correction factors occurring in the $S$-matrix 
elements for neutral gauge bosons, we need one more ingredient according to
Sect.~\ref{sec:wave}: the tangent of the running $\ga\,$--$\,Z$ mixing angle, 
$\tan\theta(k^2)$, at $\,k^2=0\,$ and $\,k^2=\mz^2$. From the 
``non-renormalization'' condition (\ref{nonrene}) for the electromagnetic
charge and Eq.~(\ref{Cgamma}) we derive
\begin{equation}
  \tan\theta(0)_{\,\, \rm 1-loop}
  \ =\ C_{\ga\sscr{Z},\,g}^{\,(1)} 
  \ =\ \frac{\cw}{2\sw}\,\Sigma_{\ga\ga,\,g}^{\prime\,(1)}
  \ =\ \frac{\cw}{\sw}\,\L(M,M) \,.
\end{equation}
By means of Eqs.~(\ref{taylor}) and (\ref{CZ}), $\,\tan\theta(\mz^2)\,$ can be 
written at one-loop as
\begin{equation}
  \tan\theta(\mz^2)_{\,\, \rm 1-loop}
  \ =\ \Sigma_{\ga\sscr{Z},\,g}^{\prime\,(1)} + C_{\ga\sscr{Z},\,g}^{\,(1)} 
  \ =\ {} - \frac{\cw}{\sw}\,\L(M,M) \,.
\end{equation}
Thus, due to the non-renormalization condition Eq.~(\ref{nonrene}) we do not 
have to explicitly calculate the full $\ga\,$--$\,Z$ self energy. Instead it 
is sufficient to know the derivatives 
$\,\Sigma_{\sscr{N}_1\sscr{N}_2,\,g}^{\prime}\,$ up to the relevant 
order.\footnote{For completeness we give in Appendix~\ref{app:selfenergy} the 
                full (non-derivative) one-loop $\ga\,$--$\,Z$ self-energy, 
                which is found to be in agreement with the results presented 
                above}
\\[-5mm]

\ni
Now we have all the necessary ingredients for calculating the Sudakov 
correction factors that enter the $S$-matrix elements for transverse neutral 
gauge bosons (see Sect.~\ref{sec:wave}). To this end we replace the $Z$-boson
and photon fields in the amputated Green's functions by the unbroken gauge 
fields $\,B\,$ [belonging to $U(1)_Y$]\, and $\,W^3\,$ 
[belonging to $SU(2)_L$]:
\begin{align}
  A_{\mu} &= \cw\,B_{\mu} - \sw\,W_{\mu}^3 \no \\[2mm]
  Z_{\mu} &= \sw\,B_{\mu} + \cw\,W_{\mu}^3 \,.
\end{align}
In this way we obtain different multiplicative Sudakov correction factors 
$\,Z_{\sscr{N}_T,\,B}^{\frac{1}{2}}\,$ and 
$\,Z_{\sscr{N}_T,\,W^3}^{\frac{1}{2}}\,$
for the $\,B\,$ and $\,W^3\,$ components of an (asymptotic) transverse neutral 
gauge boson $N_T$. Writing as usual $Z=1+\delta Z$, the corresponding one-loop 
corrections read
\begin{align}
  \delta Z_{\ga,\,B}^{\,(1)}(W)          &\ =\ \delta C_{\ga\ga}^{\,(1)}
       - 2\,\frac{\sw}{\cw}\,\tan\theta(0)_{\,\, \rm 1-loop}
       \ =\ 0 \no \\[1mm]
  \delta Z_{\sscr{Z}_T,\,B}^{\,(1)}(W)   &\ =\ \delta C_{\sscr{ZZ}}^{\,(1)}
       + 2\,\frac{\cw}{\sw}\,\tan\theta(\mz^2)_{\,\, \rm 1-loop}
       \ =\ 0 \no \\[1mm]
  \delta Z_{\ga,\,W^3}^{\,(1)}(W)        &\ =\ \delta C_{\ga\ga}^{\,(1)}
       + 2\,\frac{\cw}{\sw}\,\tan\theta(0)_{\,\, \rm 1-loop}
       \ =\ \frac{2}{\swto}\,\L(M,M) \no \\[1mm]
  \delta Z_{\sscr{Z}_T,\,W^3}^{\,(1)}(W) &\ =\ \delta C_{\sscr{ZZ}}^{\,(1)}
       - 2\,\frac{\sw}{\cw}\,\tan\theta(\mz^2)_{\,\, \rm 1-loop}
       \ =\ \frac{2}{\swto}\,\L(M,M) \,.
\end{align}
This can be represented generically by
\begin{align}
  \delta Z_{\sscr{N}_T,\,B}^{\,(1)}(W) &\ =\ 0 \no \\[1mm]
  \delta Z_{\sscr{N}_T,\,W^3}^{\,(1)}(W) &\ =\ \frac{2}{\swto}\,\L(M,M) \,,
\end{align}
{\it irrespective} of the particular on-shell limit. In this respect 
the SM behaves (dynamically) like an unbroken theory in the 
Sudakov limit, with the unbroken gauge fields $B$ and $W^3$ being the relevant
physical degrees of freedom.

\subsection{General one-loop Sudakov logarithms}

Gathering the knowledge from the previous subsections we can now make general 
statements. Upon summation over the allowed gauge-boson exchanges, one obtains 
the following expression for the full one-loop Sudakov correction to the 
external wave-function factor for an arbitrary on-shell/on-resonance particle 
with mass $m$, charge $Q$ and hypercharge $Y$:
\begin{align}
\label{deltaZ1}
\hspace*{5mm} \delta Z^{\,(1)} = \left[ \frac{C_2(R)}{\sin^2\theta_{\rm w}} 
     + \left( \frac{Y}{2\cos\theta_{\rm w}} \right)^{\!2}\, 
     \right]\! {\rm L}(M,M) 
    + Q^2\,\bigg[ {\rm L_{\ga}}(\la,m)
                        - {\rm L}(M,M) \bigg] \,\, .
\end{align}
Here $\,C_2(R)\,$ is the $SU(2)$ Casimir operator of the particle.  
So, $\,C_2(R) = C^{\sscr{SU(2)}}_F = 3/4\,$ for particles in the fundamental 
representation: the left-handed fermions ($f_L/\bar{f}_R$), the physical Higgs 
boson ($H$) and the longitudinal gauge bosons ($W^{\pm}_L$ and $Z_L$, being 
equivalent to the Goldstone bosons $\phi^{\pm}$ and $\chi$). For the particles
in the adjoint representation of $SU(2)$, \ie the transverse $W$ bosons 
($W^{\pm}_T$) and the $W^3$ components of both the photon (at $\,k^2=0$) and 
the transverse $Z$ boson (at $\,k^2=\mz^2$), one obtains 
$\,C_2(R) = C^{\sscr{SU(2)}}_A = 2$. For the $SU(2)$ singlets, \ie the 
right-handed fermions ($f_R/\bar{f}_L$) and the $B$ components of both the 
photon (at $\,k^2=0$) and the transverse $Z$ boson (at $\,k^2=\mz^2$), 
the $SU(2)$ Casimir operator
vanishes, $C_2(R) = 0$. Note that the terms proportional to $Q^2$ in 
Eq.~(\ref{deltaZ1}) are the result of the mass gap between the photon and the 
weak bosons.

\section{Electroweak two-loop Sudakov logarithms}
\label{sec:two-loop}

Having established the method to calculate the Sudakov logarithms in the 
Coulomb gauge, we now perform the explicit two-loop calculation.
Again the calculation is very similar for the various types of external 
particles. We use the fermion case as the major example to 
illustrate all the subtleties and then briefly give the results for the bosons.

\subsection{The fermionic self-energy at two-loop level}
\label{sec:two-loop_fermions}

\ni At two-loop accuracy one has to take the following five generic 
sets of diagrams into account: \vspace*{2mm}\\
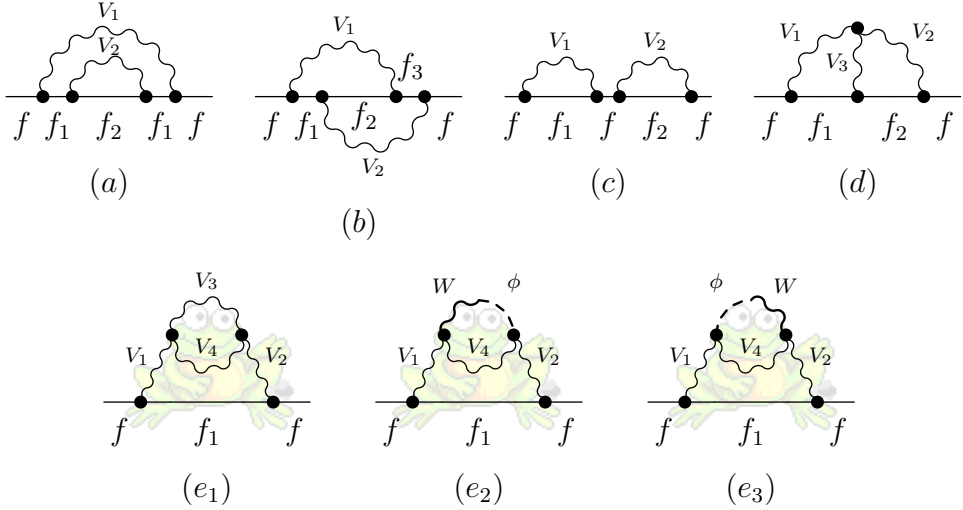
\begin{figure}[H]
\vspace*{3.2cm}

\hspace*{3cm}{\epsfysize=1.65cm{\epsffile{ frog4.epsi}}}
\vspace*{-2cm}
\hspace*{1.3cm}{\epsfysize=1.65cm{\epsffile{ frog4.epsi}}}
\vspace*{-1.65cm}
\hspace*{1.4cm}{\epsfysize=1.65cm{\epsffile{ frog4.epsi}}}
\vspace*{-1.7cm}

\begin{fmffile}{2l}
\hspace*{1.2cm}
\begin{fmfgraph*}(78,52) \fmfpen{thin} \fmfleft{i1} \fmfright{o1}
\fmf{phantom,tension=1}{i1,v1,v2,v5,v3,v4,o1}
\fmf{plain,tension=1,label=$f$,l.side=right,width=0.6}{i1,v1} 
\fmf{plain,tension=1.5,label=$f_1$,l.side=right,width=0.6}{v1,v2}
\fmf{plain,tension=0.5,label=$f_2$,l.side=right,width=0.6}{v2,v3}
\fmf{plain,tension=1.5,label=$f_1$,l.side=right,width=0.6}{v3,v4}
\fmf{plain,tension=1,label=$f$,l.side=right,width=0.6}{v4,o1}
\fmffreeze 
\fmf{boson,left,tension=-1,label=${\scriptstyle
V_1}$,l.side=left,l.dist=4,
     width=0.6}{v1,v4}
\fmf{boson,left,tension=-2,label=${\scriptstyle
V_2}$,l.side=left,l.dist=2,
     width=0.6}{v2,v3}
\fmfdot{v1} \fmfdot{v2} \fmfdot{v3} \fmfdot{v4} 
\fmfv{label=$(a)$,l.dist=27,l.a=-90}{v5}
\end{fmfgraph*}
\hspace{0.3cm}
\begin{fmfgraph*}(78,52) \fmfpen{thin} \fmfleft{i1} \fmfright{o1}
\fmf{phantom,tension=1}{i1,v1,v2,v5,v3,v4,o1}
\fmf{plain,tension=1,label=$f$,l.side=right,width=0.6}{i1,v1} 
\fmf{plain,tension=1.5,label=$f_1$,l.side=right,width=0.6}{v1,v2}
\fmf{plain,tension=0.5,width=0.6}{v2,v3}
\fmfv{label=$f_2$,l.dist=19,l.a=-157.55}{v4}
\fmf{plain,tension=0.1,width=0.6}{v3,v4}
\fmf{plain,tension=1.5,label=$f_3$,l.side=left,width=0.6}{v3,v4}
\fmf{plain,tension=1,label=$f$,l.side=right,width=0.6}{v4,o1}
\fmffreeze 
\fmf{boson,left,tension=-1.5,label=${\scriptstyle
V_1}$,l.side=left,l.dist=4,
     width=0.6}{v1,v3}
\fmf{boson,right,tension=-1.5,label=${\scriptstyle
V_2}$,l.side=right,l.dist=4,
     width=0.6}{v2,v4}
\fmfdot{v1} \fmfdot{v2} \fmfdot{v3} \fmfdot{v4} 
\fmfv{label=$(b)$,l.dist=42,l.a=-90}{v5}
\end{fmfgraph*}
\hspace{0.3cm}
\begin{fmfgraph*}(78,52) \fmfpen{thin} \fmfleft{i1} \fmfright{o1}
\fmf{phantom,tension=1}{i1,v1,v2,v5,v3,v4,o1}
\fmf{plain,tension=3,label=$f$,l.side=right,width=0.6}{i1,v1} 
\fmf{plain,tension=0.1,label=$f_1$,l.side=right,width=0.6}{v1,v2}
\fmf{plain,tension=3,label=$f$,l.side=right,width=0.6}{v2,v3}
\fmf{plain,tension=0.1,label=$f_2$,l.side=right,width=0.6}{v3,v4}
\fmf{plain,tension=3,label=$f$,l.side=right,width=0.6}{v4,o1}
\fmffreeze 
\fmf{boson,left,tension=1,label=${\scriptstyle V_1}$,l.side=left,l.dist=4, 
     width=0.6}{v1,v2}
\fmf{boson,left,tension=1,label=${\scriptstyle V_2}$,l.side=left,l.dist=4,
     width=0.6}{v3,v4}
\fmfdot{v1} \fmfdot{v2} \fmfdot{v3} \fmfdot{v4} 
\fmfv{label=$(c)$,l.dist=27,l.a=-90}{v5}
\end{fmfgraph*}
\hspace{0.3cm}
\begin{fmfgraph*}(78,52) \fmfpen{thin} \fmfleft{i1} \fmfright{o1}
\fmf{phantom,tension=1}{i1,v1,v2,v5,v3,v4,o1}
\fmf{plain,tension=1,label=$f$,l.side=right,width=0.6}{i1,v1} 
\fmf{plain,tension=1.5,width=0.6}{v1,v2}
\fmf{plain,tension=0.5,width=0.6}{v2,v3}
\fmf{plain,tension=1.5,width=0.6}{v3,v4}
\fmf{plain,tension=1,label=$f$,l.side=right,width=0.6}{v4,o1}
\fmffreeze
\fmftop{v6}
\fmf{boson,left,tension=-1,width=0.6}{v1,v4}
\fmf{boson,tension=2,label=${\scriptstyle
V_3}$,l.side=left,l.dist=3,width=0.6}
    {v5,v6}
\fmfv{label=${\scriptstyle V_1}$,l.a=90,l.dist=21}{v1}
\fmfv{label=${\scriptstyle V_2}$,l.a=90,l.dist=21}{v4}
\fmfv{label=${f_1}$,l.a=-90,l.dist=6}{v2}
\fmfv{label=${f_2}$,l.a=-90,l.dist=6}{v3}
\fmfv{label=$(d)$,l.dist=27,l.a=-90}{v5}
\fmfdot{v1} \fmfdot{v5} \fmfdot{v4} \fmfdot{v6} 
\end{fmfgraph*}
\vspace*{2.2cm}

\hspace*{2.5cm}
\begin{fmfgraph*}(78,52) \fmfpen{thin} \fmfleft{i1} \fmfright{o1}
\fmf{phantom,tension=1}{i1,v1,v2,v5,v3,v4,o1}
\fmf{plain,tension=1,label=$f$,l.side=right,width=0.6}{i1,v1} 
\fmf{plain,tension=1.5,width=0.6}{v1,v2}
\fmf{plain,tension=0.5,label=$f_1$,l.side=right,width=0.6}{v2,v3}
\fmf{plain,tension=1.5,width=0.6}{v3,v4}
\fmf{plain,tension=1,label=$f$,l.side=right,width=0.6}{v4,o1}
\fmffreeze
\fmftop{v6,v7,v8,v9}
\fmf{boson,tension=-1,label=${\scriptstyle V_1}$,l.side=left,l.dist=3,
     width=0.6}{v1,v7}
\fmf{boson,tension=-1,label=${\scriptstyle V_2}$,l.side=left,l.dist=3,
     width=0.6}{v8,v4}
\fmf{boson,left,tension=2,label=${\scriptstyle V_3}$,l.side=left,l.dist=4,
     width=0.6}{v7,v8}
\fmf{boson,right,tension=2,label=${\scriptstyle
V_4}$,l.side=left,l.dist=5,
     width=0.6}{v7,v8}
\fmfdot{v1} \fmfdot{v7} \fmfdot{v4} \fmfdot{v8} 
\fmfv{label=$(e_1)$,l.dist=27,l.a=-90}{v5}
\end{fmfgraph*}
\hspace{0.6cm}
\begin{fmfgraph*}(78,52) \fmfpen{thin} \fmfleft{i1} \fmfright{o1}
\fmf{phantom,tension=1}{i1,v1,v2,v5,v3,v4,o1}
\fmf{plain,tension=1,label=$f$,l.side=right,width=0.6}{i1,v1} 
\fmf{plain,tension=1.5,width=0.6}{v1,v2}
\fmf{plain,tension=0.5,label=$f_1$,l.side=right,width=0.6}{v2,v3}
\fmf{plain,tension=1.5,width=0.6}{v3,v4}
\fmf{plain,tension=1,label=$f$,l.side=right,width=0.6}{v4,o1}
\fmffreeze
\fmftop{v6,v7,v8,v9}
\fmf{boson,tension=-1,label=${\scriptstyle V_1}$,l.side=left,l.dist=3,
     width=0.6}{v1,v7}
\fmf{boson,tension=-1,label=${\scriptstyle V_2}$,l.side=left,l.dist=3,
     width=0.6}{v8,v4}
\fmf{phantom,left,tension=2,width=0.6,tag=1}{v7,v8}
\fmfposition
\fmfipath{p[]}
\fmfiset{p1}{vpath1(__v7,__v8)}
\fmfi{boson}{subpath (0,length(p1)/2) of p1}
\fmfi{dashes}{subpath (length(p1)/2,length(p1)) of p1}
\fmf{boson,right,tension=2,label=${\scriptstyle
V_4}$,l.side=left,l.dist=5,
     width=0.6}{v7,v8}
\fmfv{label=${\scriptstyle W}$,l.a=90,l.dist=16}{v7}
\fmfv{label=${\scriptstyle \phi}$,l.a=90,l.dist=16}{v8}
\fmfdot{v1} \fmfdot{v7} \fmfdot{v4} \fmfdot{v8} 
\fmfv{label=$(e_2)$,l.dist=27,l.a=-90}{v5}
\end{fmfgraph*}
\hspace*{0.6cm}
\begin{fmfgraph*}(78,52) \fmfpen{thin} \fmfleft{i1} \fmfright{o1}
\fmf{phantom,tension=1}{i1,v1,v2,v5,v3,v4,o1}
\fmf{plain,tension=1,label=$f$,l.side=right,width=0.6}{i1,v1} 
\fmf{plain,tension=1.5,width=0.6}{v1,v2}
\fmf{plain,tension=0.5,label=$f_1$,l.side=right,width=0.6}{v2,v3}
\fmf{plain,tension=1.5,width=0.6}{v3,v4}
\fmf{plain,tension=1,label=$f$,l.side=right,width=0.6}{v4,o1}
\fmffreeze
\fmftop{v6,v7,v8,v9}
\fmf{boson,tension=-1,label=${\scriptstyle V_1}$,l.side=left,l.dist=3,
     width=0.6}{v1,v7}
\fmf{boson,tension=-1,label=${\scriptstyle V_2}$,l.side=left,l.dist=3,
     width=0.6}{v8,v4}
\fmf{phantom,left,tension=2,width=0.6,tag=1}{v7,v8}
\fmfposition
\fmfipath{p[]}
\fmfiset{p1}{vpath1(__v7,__v8)}
\fmfi{dashes}{subpath (0,length(p1)/2) of p1}
\fmfi{boson}{subpath (length(p1)/2,length(p1)) of p1}
\fmf{boson,right,tension=2,label=${\scriptstyle
V_4}$,l.side=left,l.dist=5,
     width=0.6}{v7,v8}
\fmfv{label=${\scriptstyle W}$,l.a=90,l.dist=16}{v8}
\fmfv{label=${\scriptstyle \phi}$,l.a=90,l.dist=16}{v7}
\fmfdot{v1} \fmfdot{v7} \fmfdot{v4} \fmfdot{v8} 
\fmfv{label=$(e_3)$,l.dist=27,l.a=-90}{v5}
\end{fmfgraph*}
\end{fmffile}
\vspace*{2mm} 
\end{figure}
\ni The fermions $f_{i}$ are fixed by the exchanged gauge bosons $V_{i}$.
Various cancellations are going to take place between all these diagrams. 
In unbroken theories like QED and QCD merely the so-called `rainbow'
diagrams of set (a) survive.
The same holds if all gauge bosons of the theory would have
a similar mass. The unique feature of the SM is that it is only partially
broken, with the electromagnetic gauge group $\,U(1)_{\rm em} \neq U(1)_Y\,$ 
remaining unbroken. As such three of the four gauge bosons will acquire a mass,
whereas the photon remains massless and will interact with the charged massive
gauge bosons ($W^{\pm}$). As a consequence, merely calculating the `rainbow'
diagrams will {\it not} lead to the correct result.
To illustrate the above we study each of the generic five topologies
separately, indicating the corresponding two-loop contributions to the 
fermionic external wave-function factor by 
$\,\delta Z_f^{(2)}(a)\,$--$\,\delta Z_f^{(2)}(e)$, respectively
\\[-5mm]

\ni
First the `rainbow' diagrams of set (a). Let the outer loop-momentum in the 
`rainbow' be denoted by $k_1$ and the inner loop-momentum by $k_2$. 
For simplicity we use the generic mass
$m_f$ for every fermion and do not distinguish between different fermion
species. At one-loop level we learned that the fermion mass is only needed as
a cut-off parameter to regularize the collinear singularity if the soft
exchanged gauge boson is a photon. If this photon is attached to $f^{\prime}$
rather then $f$, then it has to be preceded by the emission of a $W$ boson and,
as we will see in the following, the heavy gauge-boson mass scale $M$ will 
replace $m_{f^{\prime}}$ as dominant collinear cut-off. So, for all
practical purposes we can forget about $m_{f^{\prime}}$. 
The `rainbow' contribution 
$\,\delta Z_f^{(2)}(a)\,$ can then be written as 
\bea
\delta\, Z_f^{\,(2)}(a) & \approx & -\,(ie)^4
   \int\frac{\dd ^4 k_1}{(2\,\pi)^4}\,\int\frac{\dd ^4 k_2}{(2\,\pi)^4}\,
   \frac{\Gamma_{\!\!ff_1\!V_1}^{\,^{\scriptstyle 2}}\,
         \Gamma_{\!\!f_1f_2\!V_2}^{\,^{\scriptstyle 2}}\,\,
         4\,p_{\mu}\,p_{\nu}\,P^{\mu\nu}(k_1,M_1)\,\,
         4\,p_{\mu^{\prime}}\,p_{\nu^{\prime}}\,
         P^{\mu^{\prime}\nu^{\prime}}(k_2,M_2)}
        {[(p-k_1)^2-m_{f}^2+i\eps\,]^2\,[(p-k_1-k_2)^2-m_{f}^2+i\eps\,]}  
   \no \\[2mm]
&& \hspace*{8mm} {\scriptstyle \times} \hspace*{2mm}
   \left( \frac{2}{[(p-k_1)^2-m_{f}^2+i\eps\,]} 
        + \frac{1}{[(p-(k_1+k_2))^2-m_{f}^2+i\eps\,]} \right) \,.
\ea
\ni
For the gauge-boson momentum $k_2$ we choose a Sudakov parametrisation 
equivalent to the one used for $k_1$, \ie
\vspace*{-4mm}
\bea
k_2 = v_2\, q + u_2 \, \bar{q} + k_{2_\bot} \,,
\ea
with $q$ and $\bar{q}$ defined in Eq.~(\ref{fourmom}). The calculation 
simplifies if we perform the $u_2$ integration first, taking the residue in 
the lower hemisphere in the pole of the corresponding gauge-boson propagator. 
The rest of the calculation follows the steps of the one-loop calculation.
Making use of Eq.~(\ref{4ppP}) as well as the related identity
\bea
  i\,\Big[ k_2^2-M_2^2+i\eps\, \Big]\,4\,p_{\mu{\prime}}\,p_{\nu^{\prime}}\,
  P^{\mu^{\prime}\nu^{\prime}}(k_2,M_2) 
  &\approx & \frac{4\,k_{2\,0}\,p_0}{\vec{k_2}^2}\,
             \Big( (p-k_2)^2-m_f^2 \Big) \no \\
  &&\hspace*{-4cm}
  \approx \  \frac{4\,k_{2\,0}\,p_0}{\vec{k_2}^2}\,
             \bigg( \Big[ (p-(k_1+k_2))^2-m_f^2 \Big] 
                  - \Big[ (p-k_1)^2-m_f^2 \Big] \bigg) \,,
\ea
one obtains in leading logarithmic approximation 
\begin{align}
\label{angular}
 \delta Z_f^{\,(2)}(a) &\ \approx\ \left( -\,\frac{\alpha}{\pi} \right)\,
   \Gamma_{\!\!ff_1\!V_1}^{\,^{\scriptstyle 2}}\,\int_0^1 \frac{\dd y_1}{y_1} 
   \int_{y_1}^1\frac{\dd z_1}{z_1}\,\,\Theta\Bigl( y_1 z_1-\frac{M_1^2}{s} 
   \Bigr)\,\Theta\Bigl (y_1-\frac{m_f^2}{s}\,z_1\Bigr) \no \\
 & \hspace*{5mm} {\scriptstyle \times} \hspace*{2mm}  \left( -\,
   \frac{\alpha}{\pi} \right) \, \Gamma_{\!\!f_1f_2V_2}^{\,^{\scriptstyle 2}}\,
   \int_0^1\,\frac{\dd y_2}{(y_2+y_1)}\int_{y_2}^1 \frac{\dd z_2}{z_2}\,\, 
   \Theta\Bigl( y_2 z_2-\frac{M_2 ^2}{s} \Bigr)\,
   \Theta\Bigl (y_2-\frac{m_f^2}{s}\,z_2\Bigr) \no \\[2mm]
                       &\ \approx\ 
  \left( -\,\frac{\alpha}{\pi} \right)^2 \,\int_0^1\frac{\dd y_1}{y_1} 
  \int_{y_1}^1\frac{\dd z_1}{z_1}\,\int_0^1\frac{\dd y_2}{y_2} 
  \int_{y_2}^1 \frac{\dd z_2}{z_2}\,
  \Gamma_{\!\!ff_1\!V_1}^{\,^{\scriptstyle 2}}\,
  \cK^{\,(1)}(s,m_f^2,M_1,y_1,z_1) \no \\[2mm]
                       & \hspace*{15mm} {\scriptstyle \times} \hspace*{2mm}
  \Gamma_{\!\!f_1f_2V_2}^{\,^{\scriptstyle 2}}\, 
  \cK^{\,(1)}(s,m_f^2,M_2,y_2,z_2)\,\Theta(y_2-y_1)
\end{align}
with $\cK^{\,(1)}$ being defined in Eq.~(\ref{kernel}). 
As already hinted at above, the collinear region of the inner integral is
restricted by the collinear region of the outer integral ($y_2 \gg y_1$). 
So, the `rainbow' diagrams exhibit an explicit angular ordering.
\\[-5mm]

\ni
Similarly the `crossed rainbow' diagrams of set (b) yield
\begin{align}
 \delta Z_f^{\,(2)}(b) &\,=\,-\,(ie)^4\int\frac{\dd ^4 k_1}{(2\,\pi)^4}\,
   \int\frac{\dd ^4 k_2}{(2\,\pi)^4}\,\,\Gamma_{\!\!ff_1\!V_1}\,
   \Gamma_{\!\!f_1f_2V_2}\,\Gamma_{\!\!f_2f_3V_1}\,\Gamma_{\!\!ff_3V_2}
   \no \\[1mm]
 & \hspace*{5mm} {\scriptstyle \times} \hspace*{1mm} 
   \frac{4\,p_{\mu}\,p_{\nu}\,P^{\mu\nu}(k_1,M_1)\,\,4\,p_{\mu^{\prime}}\,
         p_{\nu^{\prime}}\,P^{\mu^{\prime}\nu^{\prime}}(k_2,M_2)}
        {[(p-k_1)^2-m_f^2+i\eps\,]\,[(p-k_1-k_2)^2-m_f^2+i\eps\,]\, 
         [(p-k_2)^2-m_f^2+i\eps\,]}
   \no \\[1mm]
 & \hspace*{5mm} {\scriptstyle \times} \hspace*{1mm} 
   \left( \frac{1}{[(p-k_1)^2-m_f^2+i\eps\,]} 
        + \frac{1}{[(p-k_1-k_2)^2-m_f^2+i\eps\,]} 
        + \frac{1}{[(p-k_2)^2-m_f^2+i\eps\,]} \right)
   \no \\[1mm]
 &\,=\,-\,\Gamma_{\!\!ff_1\!V_1}\,\Gamma_{\!\!f_1f_2V_2}\,
   \Gamma_{\!\!f_2f_3V_1}\,\Gamma_{\!\!ff_3V_2}\, 
   \left(-\,\frac{\alpha}{\pi} \right)\,\int_0^1\frac{\dd y_1}{y_1} 
   \int_{y_1}^1\frac{\dd z_1}{z_1}\,\cK^{\,(1)}(s,m_f^2,M_1,y_1,z_1)
   \no \\[1mm]
 & \hspace*{15mm} {\scriptstyle \times} \hspace*{1mm} 
   \left(-\,\frac{\alpha}{\pi} \right)\,\int_0^1\frac{\dd y_2}{y_2} 
   \int_{y_2}^1\frac{\dd z_2}{z_2}\,\,\cK^{\,(1)}(s,m_f^2,M_2,y_2,z_2) \,.
\end{align}

\ni
Obviously the `reducible' contribution from set (c) can only be the product of
the two corresponding one-loop contributions
\begin{align}
 \delta Z_f^{\,(2)}(c) & = \left(-\,\frac{\alpha}{\pi} \right)\, 
     \Gamma_{\!\!ff_1\!V_1}^{\,^{\scriptstyle 2}}\,                           
     \int_0^1 \frac{\dd y_1}{y_1} 
     \int_{y_1}^1\frac{\dd z_1}{z_1}  \,  \cK^{\,(1)}(s,m_f^2,M_1,y_1,z_1)
     \no \\
  &  \hspace*{15mm} {\scriptstyle \times} \hspace*{1mm} \left(-\,
     \frac{\alpha}{\pi} \right)\, \Gamma_{\!\!ff_2V_2}^{\,^{\scriptstyle 2}}\,
     \int_0^1 \frac{\dd y_2}{y_2} 
     \int_{y_2}^1 \frac{\dd z_2}{z_2}\,\, \,  \cK^{\,(1)}(s,m_f^2,M_2,y_2,z_2).
\end{align}
These `reducible' contributions follow from the fact that the irreducible
fermionic self-energy enters $Z_f$ in the denominator.
\\[-5mm]

\ni
The two remaining sets of diagrams turn out to be more difficult to calculate.
The main complication being that more than two gauge-boson propagators are
involved and hence a variety of possible on-shell combinations enlarges the
actual number of integrals to be performed. Defining $k_3=k_1-k_2$, the 
triple-gauge-boson (`TGB') diagrams of set (d) can be written in the following 
way
\begin{align}
\label{setd}
 \delta Z_f^{\,(2)}(d) &\,=\,i\,(ie)^4
     \int\frac{\dd ^4 k_1}{(2\,\pi)^4}\,\int\frac{\dd ^4 k_2}{(2\,\pi)^4}\,
     \frac{2\,p_{\mu}\,2\,p_{\nu}\,2\,p_{\rho}\,P^{\mu\mu^{\prime}}(k_1,M_1)\,
           P^{\rho\rho^{\prime}}(k_3,M_3)\,P^{\nu\nu^{\prime}}(k_2,M_2)\, 
           V_{\mu^{\prime}\rho^{\prime}\nu^{\prime}}}
          {[(p-k_1)^2-m_f^2+i\eps\,][(p-k_2)^2-m_f^2+i\eps\,]} \no \\[1mm]
&  \hspace*{5mm} {\scriptstyle \times} \hspace*{1mm} 
     \Gamma_{\!\!ff_1\!V_1}\,\Gamma_{\!\!f_1f_2V_3}\,\Gamma_{\!\!ff_2V_2}\,
     G_{132}\,\,\left( \frac{1}{[(p-k_1)^2-m_f^2+i\eps\,]}\,
                     + \frac{1}{[(p-k_2)^2-m_f^2+i\eps\,]} \right) \no \\[1mm]
& \,\equiv\,\,(i\,e^2)^2
     \int \frac{\dd ^4 k_1}{(2\,\pi)^4}\,\int\frac{\dd ^4 k_2}{(2\,\pi)^4}\, 
     \Gamma_{\!\!ff_1\!V_1}\,\Gamma_{\!\!f_1f_2V_3}\,\Gamma_{\!\!ff_2V_2}\,
     G_{132}\,{\rm IK}(d) \,,
\end{align}
with the integration kernel
\begin{align}
\label{IK1}
  {\rm IK}(d) &= -\,i\,\frac{2\,p_{\mu }\,2\,p_{\nu}\,2\,p_{\rho}\,
                             P^{\mu\mu^{\prime}}(k_1,M_1)\, 
                             P^{\rho\rho^{\prime}}(k_3,M_3)\, 
                             P^{\nu\nu^{\prime}}(k_2,M_2)\, 
                             V_{\mu^{\prime}\rho^{\prime}\nu^{\prime}}}
                           {[(p-k_1)^2-m_f^2+i\eps\,][(p-k_2)^2-m_f^2+i\eps\,]}
   \no \\[1mm]
&  \hspace*{9mm} {\scriptstyle \times} \hspace*{1mm}  
   \left( \frac{1}{[(p-k_1)^2-m_f^2+i\eps\,]}\,
        + \frac{1}{[(p-k_2)^2-m_f^2+i\eps\,]} \right)
\end{align}
and
\bea
 \,V_{\mu^{\prime}\rho^{\prime}\nu^{\prime}} &= & 
                          (2k_2-k_1)_{\mup}\,\G_{\nup\rhop}
                        + (-k_2-k_1)_{\rhop}\,\G_{\mup\nup} 
                        + (2\,k_1-k_2)_{\nup}\,\G_{\mup\rhop} \,.
\ea

\ni
The totally antisymmetric coupling $\,e\,G_{ijl}\,$ is the triple gauge-boson 
coupling with all three gauge-boson lines ($i,j,l$) defined to be incoming at 
the interaction vertex. In our convention this coupling is fixed according to 
$\,G_{\ga W^+W^-}=1\,$ and 
$\,G_{Z W^+W^-}=-\cos\theta_{\rm w}/\sin\theta_{\rm w}$.
\\[-5mm]

\ni
The integration kernel can be simplified by making use
of the fact that the following generic contributions will {\it not\,} lead to 
$(\log)^4$ corrections: \\[2mm]
\h{5} $\bullet$ terms with only one gauge-boson propagator \\[2mm]
\h{5} $\bullet$ terms with no fermion propagator \\[2mm]
\h{5} $\bullet$ terms with one fermion propagator and only two gauge-boson 
                propagators \\[2mm]
\h{5} $\bullet$ terms with two fermion propagators and two gauge-boson 
                propagators but only one $1/\vec{k}^2$ \\[2mm]
\h{5} $\bullet$ terms $\,\propto (1/k_i)^l\,$ with $\,l<8\,$ in the soft 
                $k_i$ limit; four of those powers will be compensated \\
\h{8.3}         by the loop integrals, hence four more are required to obtain
                four logarithms. \\[2mm]
Moreover we can make use of effective identities like
\bea
  \frac{(p-k_1)^2}{[(p-k_2)^2-m_f^2+i\eps\,]\,^2 
                   \prod\limits_{j=1}^3\,[k_j^2-M_j^2+i\eps\,]} 
  \to \,\frac{k_{1\,0}k_{2\,0}}{\vec{k_2}^2}\,
      \frac{1}{[(p-k_2)^2-m_f^2+i\eps\,]\,
      \prod\limits_{j=1}^3\,[k_j^2-M_j^2+i\eps\,]}\,, \no\\ 
\ea
because the part of $(p-k_1)^2$ that is proportional to the $\vec{k_1}$ 
component perpendicular to $\vec{k_2}$ will not survive the $\vec{k_1}$ 
integration. 

\ni
The integration kernel ${\rm IK}(d)$ of Eq.~(\ref{setd}) can now be written as
{\small
\begin{subequations}
\begin{align}
 & IK(d) \approx \no \\
\label{1}
& \approx  \hspace*{5mm} \frac{2 \,k_{1\,0}\, p_0}{\vec{k_1}^2}\,
 \frac{1}{[\,(p-k_1)^2-m_f^2+i\eps\,] \, [k_1^2-M_1^2+i\eps\,]}\,\,\,
\frac{2 \,k_{2\,0}\, p_0}{\vec{k_2}^2} \,
 \frac{1}{[\,(p-k_2)^2-m_f^2+i\eps\,] \, [k_2^2-M_2^2+i\eps\,]}\,  \\
\label{2}
 &  \hspace*{5mm} + \hspace*{2mm}
 \frac{2 \,k_{2\,0}\, p_0}{\vec{k_2}^2} \,
  \frac{1}{[\,(p-k_2)^2-m_f^2+i\eps\,] \, [k_2^2-M_2^2+i\eps\,]}\,\,\, 
  \frac{2 \,k_{3\,0}\, p_0}{\vec{k_3}^2}\,
  \frac{1}{[\,(p-k_1)^2-m_f^2+i\eps\,] \, [k_3^2-M_3^2+i\eps\,]}  \\
 \label{3}
&  \hspace*{5mm} - \hspace*{2mm}  \frac{2 \,k_{1\,0}\, p_0}{\vec{k_1}^2}\,
 \frac{1}{[\,(p-k_1)^2-m_f^2+i\eps\,] \, [k_1^2-M_1^2+i\eps\,]}\,\,\,
\frac{2 \,k_{3\,0}\, p_0}{\vec{k_3}^2} \,
 \frac{1}{[\,(p-k_2)^2-m_f^2+i\eps\,] \, [k_3^2-M_3^2+i\eps\,]}\,  \\
\label{4}
&  \hspace*{5mm} + \hspace*{2mm}  \frac{4 \,k_{1\,0}\, p_0}{\vec{k_1}^2}\,
 \frac{1}{[\,(p-k_2)^2-m_f^2+i\eps\,] \, [k_1^2-M_1^2+i\eps\,]\,
 [k_2^2-M_2^2+i\eps\,]\,  [k_3^2-M_3^2+i\eps\,]}\,  \\
 \label{6}
&  \hspace*{5mm} + \hspace*{2mm}  \frac{4 \,k_{2\,0}\, p_0}{\vec{k_2}^2}\,
 \frac{1}{[\,(p-k_1)^2-m_f^2+i\eps\,] \, [k_1^2-M_1^2+i\eps\,]\,
 [k_2^2-M_2^2+i\eps\,]\,  [k_3^2-M_3^2+i\eps\,]}\,  \\
\label{5}
 &  \hspace*{5mm} - \hspace*{2mm}  \frac{8 \,k_{3\,0}\, p_0}{\vec{k_3}^2}\,
  \frac{1}{[\,(p-k_2)^2-m_f^2+i\eps\,] \, [k_1^2-M_1^2+i\eps\,]\,
  [k_2^2-M_2^2+i\eps\,]\,  [k_3^2-M_3^2+i\eps\,]}\,  \\
 \label{7}
 &  \hspace*{5mm} + \hspace*{2mm}  \frac{8 \,k_{3\,0}\, p_0}{\vec{k_3}^2}\,
  \frac{1}{[\,(p-k_1)^2-m_f^2+i\eps\,] \, [k_1^2-M_1^2+i\eps\,]\,
  [k_2^2-M_2^2+i\eps\,]\,  [k_3^2-M_3^2+i\eps\,]}\,  \\
\label{n1}
  &  \hspace*{5mm} + \hspace*{2mm} \frac{2 \,k_{1\,0}\, p_0}{\vec{k_1}^2} \,
   \frac{1}{[\,(p-k_2)^2-m_f^2+i\eps\,] \, [k_2^2-M_2^2+i\eps\,]}\,\no \\
  &  \hspace*{20mm} {\scriptstyle \times}  \hspace*{2mm}
  \frac{2 \,k_{3\,0}\, p_0}{\vec{k_3}^2}\,\,
   \frac{1}{[\,(p-(k_2+k_3))^2-m_f^2+i\eps\,] \, [k_3^2-M_3^2+i\eps\,]}  \\
  \label{n2}
  &  \hspace*{5mm} - \hspace*{2mm} \frac{2 \,k_{2\,0}\, p_0}{\vec{k_2}^2} \,
   \frac{1}{[\,(p-k_1)^2-m_f^2+i\eps\,] \, [k_1^2-M_1^2+i\eps\,]}\,\no \\
  &  \hspace*{20mm} {\scriptstyle \times}  \hspace*{2mm}
  \frac{2 \,k_{3\,0}\, p_0}{\vec{k_3}^2}\,\,
   \frac{1}{[\,(p-(k_1-k_3))^2-m_f^2+i\eps\,] \, [k_3^2-M_3^2+i\eps\,]}  \\
  \label{n3}
 &  \hspace*{5mm} - \hspace*{2mm} \frac{4 \,k_{1\,0}\, p_0}{\vec{k_1}^2} \,
  \frac{1}{[\,(p-k_1)^2-m_f^2+i\eps\,] \, [k_1^2-M_1^2+i\eps\,]}\,\no \\
 &  \hspace*{20mm} {\scriptstyle \times}  \hspace*{2mm}
 \frac{2 \,(k_{1\,0}-k_{2\,0})\, p_0}{(\vec{k_1}-\vec{k_2})^2}\,\,
  \frac{1}{[\,(p-k_2)^2-m_f^2+i\eps\,] \, [k_2^2-M_2^2+i\eps\,]}  \\
 \label{n4}
&  \hspace*{5mm} + \hspace*{2mm} \frac{4 \,k_{2\,0}\, p_0}{\vec{k_2}^2} \,
 \frac{1}{[\,(p-k_2)^2-m_f^2+i\eps\,] \, [k_2^2-M_2^2+i\eps\,]}\,\no \\
&  \hspace*{20mm} {\scriptstyle \times}  \hspace*{2mm}
\frac{2 \,(k_{1\,0}-k_{2\,0})\, p_0}{(\vec{k_1}-\vec{k_2})^2}\,\,
 \frac{1}{[\,(p-k_1)^2-m_f^2+i\eps\,] \, [k_1^2-M_1^2+i\eps\,]}\,. 
 \end{align}
\end{subequations}}

\ni
Note here that the same result is obtained for the full gauge-boson
propagator $P_{\mu\nu}$ as well as for the purely transverse part 
$\propto Q_{\mu\nu}$, as expected in the collinear regime.
Apart from the coupling factor 
$\Gamma_{\!\!ff_1\!V_1}\,\Gamma_{\!\!f_1f_2V_3}\,\Gamma_{\!\!ff_2V_2}\,G_{132}$
the integrals in Eq.~(\ref{setd}) have been normalized in the usual way. 
Therefore the first term (\ref{1}) is easily identified as the product of two 
one-loop contributions (\ref{f1loop}) with momenta $k_1$ and $k_2$, \ie
\bea
(\ref{1}) \to   \frac{1}{4}\, \cK^{\,(1)}(s,m_f^2,M_1,y_1,z_1) \,
\cK^{\,(1)}(s,m_f^2,M_2,y_2,z_2) \,.
\ea 
In the second term (\ref{2}) the
momentum $k_1$ has to be expressed in terms of the momenta $k_2$ and $k_3$, \ie
$k_1=k_2+k_3$, in the fermion propagator as well as in the integration
variable. This is convenient since
those are the momenta appearing in the boson propagators of
(\ref{2}). (Remember that we have chosen
to take the residue in the lower hemisphere in the pole of the gauge-boson
propagators.) In doing this the `rainbow'-like structure can be immediately
recognized and upon integrating first the $u_3$ variable belonging to the
Sudakov parametrisation of $k_3$ we obtain instantly
\bea
(\ref{2}) \to  \frac{1}{4}\, \cK^{\,(1)}(s,m_f^2,M_2,y_2,z_2) \,
\cK^{\,(1)}(s,m_f^2,M_3,y_3,z_3) \, \Theta(y_3-y_2) \,.
\ea 
Similarly, replacing $k_2=k_1-k_3$ and subsequently reversing the sign of the
$k_3$ integration variable in (\ref{3}) leads again to a
`rainbow'-like structure and 
\bea
(\ref{3}) \to \frac{1}{4}\, \cK^{\,(1)}(s,m_f^2,M_1,y_1,z_1) \,
\cK^{\,(1)}(s,m_f^2,M_3,y_3,z_3) \, \Theta(y_3-y_1) \,.
\ea 

\ni
The following four terms are unique in the sense that they only contain one
fermion propagator and three gauge-boson propagators. As we will see later,
those can be identified as so-called `frog' contributions. Now having three
propagators serving as potential poles we have to sum over all three
possibilities of taking either two of them on-shell. Let us do this step by
step for the example of (\ref{4}). Starting by taking $k_1$ and $k_2$ as the
integration variables, \ie taking the corresponding propagators on-shell, the
third gauge-boson propagator becomes
\begin{equation}
  \frac{1}{k_3^2-M_3^2+i\eps} 
  \ =\ \frac{1}{k_1^2-2\,k_1\cdot k_2+k_2^2-M_3^2+i\eps} 
  \ \approx\ \frac{1}{-\,2\, k_1 \cdot k_2} 
  \ \to\ \frac{1}{-\,s(z_1\,y_2 + z_2\, y_1)} \,.
\end{equation}
We need a $\,1/(y_1\,z_2)\,$ contribution for a $\,(\log)^4\,$ correction, 
since from $\,k_{1\, 0}/\vec{k_1}^2\,$ and from $\,1/(p-k_2)^2\,$ we obtained 
$\,1/(z_1\,y_2)\,$ already. This leads to the $\Theta$-function 
$\,\Theta(z_2\,y_1-z_1\,y_2)$. Furthermore, performing the $u_1$ integration 
first, the third gauge-boson propagator restricts the $v_1$ integration range 
to $0\le v_1 \le v_2$. Hence we find for the first summand of kernel (\ref{4})
\bea
\label{frog1}
\hspace*{-2mm} \frac{1}{4}\, \cK^{\,(1)}(s,m_f^2,M_1,y_1,z_1) \,
\cK^{\,(1)}(s,m_f^2,M_2,y_2,z_2) \, \Theta(z_2 - z_1) \,\Theta(z_2\,y_1 - z_1\,
y_2).
\ea

\ni
Taking $k_1$ and $-k_3$ as the next two integration variables and performing
the $u_3$ integration first we obtain 
\begin{align}
\label{frog2}
  \hspace*{-2mm} -\,\frac{1}{4}\,\cK^{\,(1)}(s,m_f^2,M_1,y_1,z_1)\, 
  \cK^{\,(1)}(s,m_f^2,M_3,y_3,z_3)\,\Theta(y_3-y_1)\,\Theta(z_3-z_1)\,
  \Theta(z_3\,y_1-z_1\,y_3) \,.
\end{align}

\ni
Finally for $k_2$ and $k_3$ being the on-shell gauge-boson momenta
\begin{align}
  \hspace*{-2mm}-\,\frac{1}{4}\,\cK^{\,(1)}(s,m_f^2,M_2,y_2,z_2)\, 
  \cK^{\,(1)}(s,m_f^2,M_3,y_3,z_3)\,\Theta(z_3-z_2)\,\Theta(z_2-z_3) 
  \equiv 0 \,, \no
\end{align}
since the two $\Theta$-functions cannot be fulfilled simultaneously. Note
that the first $\Theta$-function originates from the $k_{1\, 0} = k_{3\, 0}+
k_{2\, 0} \approx k_{3\,0}$ constraint and the second $\Theta$-function
arises due to the restricted $v_3$ integration range.

\ni
In order to combine Eqs.~(\ref{frog1}) and (\ref{frog2}), we first relabel 
the integration variables of Eq.~(\ref{frog2})
\begin{align}
  \hspace*{-2mm}-\,\frac{1}{4}\,\cK^{\,(1)}(s,m_f^2,M_1,y_1,z_1)\,
  \cK^{\,(1)}(s,m_f^2,M_3,y_2,z_2)\,\Theta(y_2-y_1)\,\Theta(z_2-z_1)\, 
  \Theta(z_2\,y_1-z_1\,y_2)\,.
\end{align}
\ni
Adding to this the `one-way' double ordered part of Eq.~(\ref{frog1}) leads to 
\begin{align}
& \h{-15}\frac{1}{4} \, \cK^{\,(1)}(s,m_f^2,M_1,y_1,z_1) \, \Theta(z_2\,y_1 -
z_1\,y_2) \, \Theta(z_2-z_1) \, \Theta(y_2-y_1) \, \no \\[2mm]
& \h{10} {\scr \times}  \left[\cK^{\,(1)}(s,m_f^2,M_2,y_2,z_2)-
  \cK^{\,(1)}(s,m_f^2,M_3,y_2,z_2) \,\right] \no \\[2mm]
& \h{-15} = \frac{1}{4} \, \cK^{\,(1)}(s,m_f^2,M_1,y_1,z_1) \,\Theta(z_2\,y_1 -
z_1\,y_2) \, \Theta(z_2-z_1) \, \Theta(y_2-y_1) \, \no \\[1mm]
& \h{10} {\scr \times}\,\Theta \left(y_2-\frac{m_f^2}{s}\, z_2\right) \,
\left[ \Theta \left( y_2\, z_2 - \frac{M_2^2}{s} \right) 
     - \Theta \left(  y_2\, z_2 - \frac{M_3^2}{s}\right) \right]\,, 
\end{align}
which vanishes for all possible combinations of $M_i$ being the photon mass
or the generic mass $M$. This is trivial for $M_2=M_3$. For $\,M_2=\la\,$ and
$\,M_1=M_3=M\,$ the $\Theta$-functions $\,\Theta(z_2-z_1)\,\Theta(y_2-y_1)\,$
can be combined with $\,\Theta(y_1\,z_1-\frac{M^2}{s})$, restricting the 
$y_2, z_2$ integrations such that at least $\,y_2 z_2 \ge M^2/s\,$ 
and hence $\,\Theta \left(y_2\,z_2-\frac{M_2^2}{s} \right) 
           - \Theta \left(y_2\,z_2-\frac{M_3^2}{s} \right)\,$ vanishes. 
The same holds for $\,M_3=\la\,$ and $\,M_1=M_2=M$.

\ni
 Hence we find for (\ref{4})
\begin{align}
 \frac{1}{4} \, \cK^{\,(1)}(s,m_f^2,M_1,y_1,z_1) \,
\cK^{\,(1)}(s,m_f^2,M_2,y_2,z_2) \, \Theta(z_2-z_1) \, \Theta(y_1-y_2) \,,
\end{align}
with $\,\Theta(z_2\,y_1 - z_1\,y_2)\,$ being obsolete for this combination of
$\Theta$-functions. Analogously we find for (\ref{6})
\begin{align}
\label{efinal}
 \frac{1}{4} \, \cK^{\,(1)}(s,m_f^2,M_1,y_1,z_1) \,
\cK^{\,(1)}(s,m_f^2,M_2,y_2,z_2) \, \Theta(z_1-z_2) \, \Theta(y_2-y_1) \,,
\end{align}
 for (\ref{5}) 
\begin{align}
 \frac{1}{2} \, \cK^{\,(1)}(s,m_f^2,M_2,y_2,z_2) \,
\cK^{\,(1)}(s,m_f^2,M_3,y_3,z_3) \, \Theta(z_2-z_3) \, \Theta(y_3-y_2) \,,
\end{align}
and eventually for (\ref{7})
\begin{align}
 \frac{1}{2} \, \cK^{\,(1)}(s,m_f^2,M_1,y_1,z_1) \,
\cK^{\,(1)}(s,m_f^2,M_3,y_3,z_3) \, \Theta(z_1-z_3) \, \Theta(y_3-y_1) \,.
\end{align}
Next (\ref{n1}) can be identified as the
following double ordered contribution
\bea
(\ref{n1}) \to   \frac{1}{4}\, \cK^{\,(1)}(s,m_f^2,M_2,y_2,z_2) \,
\cK^{\,(1)}(s,m_f^2,M_3,y_3,z_3) \, \Theta(y_3-y_2)\,  \Theta(z_2-z_3) \,.
\ea
Similarly (\ref{n2}) becomes
\bea
(\ref{n2}) \to \frac{1}{4}\,  \cK^{\,(1)}(s,m_f^2,M_1,y_1,z_1) \,
\cK^{\,(1)}(s,m_f^2,M_3,y_3,z_3) \, \Theta(y_3-y_1)\,  \Theta(z_1-z_3) \,.
\ea
The remaining two contributions are 
\bea
(\ref{n3}) \to  \frac{1}{2}\, \cK^{\,(1)}(s,m_f^2,M_1,y_1,z_1) \,
\cK^{\,(1)}(s,m_f^2,M_2,y_2,z_2) \,  \Theta(z_2-z_1) \,,
\ea
and 
\bea
(\ref{n4}) \to  \frac{1}{2}\, \cK^{\,(1)}(s,m_f^2,M_1,y_1,z_1) \,
\cK^{\,(1)}(s,m_f^2,M_2,y_2,z_2) \,  \Theta(z_1-z_2) \,,
\ea
and can be combined to
\bea
(\ref{n3}) + (\ref{n4}) \to \frac{1}{2}\,  \cK^{\,(1)}(s,m_f^2,M_1,y_1,z_1) \,
\cK^{\,(1)}(s,m_f^2,M_2,y_2,z_2) \,.
\ea
After some relabeling, the two-loop correction factors originating from set~(d)
can be summarized as follows
\begin{align}
\label{Zd2}
\delta Z_f^{\,(2)}(d) & =  \frac{1}{4}\,\Gamma_{\!\!ff_1\!V_1}\,
                     \Gamma_{\!\!f_1f_2V_3}\,\Gamma_{\!\!ff_2V_2}\,G_{132}\, 
     \left( -\,\frac{\alpha}{\pi} \right)^2 \, \int_0^1 \frac{\dd y_1}{y_1} 
     \int_{y_1}^1\frac{\dd z_1}{z_1}  \,\int_0^1 \frac{\dd y_2}{y_2} 
     \int_{y_2}^1 \frac{\dd z_2}{z_2}\, \no \\[2mm]
                   &  \biggl\{ \Bigl[ \cK^{\,(1)}(s,m_f^2,M_1,y_1,z_1) 
                             + \cK^{\,(1)}(s,m_f^2,M_2,y_1,z_1) \Bigr]
                     \times 
                     \no\\[1mm]
                  & \!\!\!\!\!\hphantom{\frac{1}{4}\,\Gamma_{\!\!ff_1\!V_1}\,
                               \Gamma_{\!\!f_1f_2V_3}}
                     \times\,\,\cK^{\,(1)}(s,m_f^2,M_3,y_2,z_2)\,
                     \Theta(y_2-y_1)\,\Bigl[ 1 + 3\,\Theta(z_1-z_2) \Bigr]
                     \no\\[3mm]
                  & \!\!\!\!\!\hphantom{\frac{1}{4}\,\Gamma_{\!\!ff_1\!V_1}`\,
                               \Gamma_{\!\!f_1f_2V_3}\,\Gamma_{\!\!ff_2V_2}\,
                               G_{132}\,A}
                     {} \hspace*{-35mm} +\cK^{\,(1)}(s,m_f^2,M_1,y_1,z_1)\,
                        \cK^{\,(1)}(s,m_f^2,M_2,y_2,z_2)\,
                     \times 
                     \no\\[2mm]
                  & \!\!\!\!\!\hphantom{\frac{1}{4}\,\Gamma_{\!\!ff_1\!V_1}\,
                               \Gamma_{\!\!f_1f_2V_3}}
                     \times\,\,\Bigl[ 3 + \Theta(y_1-y_2)\,\Theta(z_2-z_1)
                                    + \Theta(y_2-y_1)\,\Theta(z_1-z_2) \Bigr]
                     \biggr\}~.                     \no\\[2mm]
\end{align}
  
\ni
Finally we calculate the `frog' diagrams of set (e): 
\begin{align}
\delta Z_f^{\,(2)}(e) &= -\,e^2\,\,\Gamma_{\!\!ff_1\!V_1}\,
     \Gamma_{\!\!ff_1\!V_2}\,\int\frac{\dd^4 k_1}{(2\,\pi)^4}\, 
     \frac{4\,p_{\mup}\,p_{\nup}\,P^{\mup \mu}(k_1,M_1)\,
           \Big( i\,\Sigma_{V_1V_2,\,\G}^{\,(1)}\,\G_{\mu\nu}\, \Big)\, 
           P^{\nu\nup}(k_1,M_2)\,}
          {[(p-k_1)^2-m_f^2+i\eps\,]^{\,2}} \no 
\end{align}
where the expressions for the various $\,\Sigma_{V_1V_2,\,\G}^{\,(1)}\,$ can be
found in Appendix~\ref{app:selfenergy}. Whenever the soft particle $V_3$ is a 
$W$ boson, the sum of the contributions from the gauge-boson propagator and 
the two mixed propagators is implicitly understood. With the purpose of making 
the bookkeeping as simple as possible for later summation over all possible 
combinations of particles in the various diagrams, we remove the explicit 
orientation in the inner loop and add the cases of both $V_3$ and $V_4$ being
the soft gauge boson. After the usual simplifications we obtain with the help 
of the result from (\ref{6})
\begin{align}
  \delta Z_f^{\,(2)}(e) & = \,+\,\frac{1}{2}\,\Gamma_{\!\!ff_1\!V_1}\,
    \Gamma_{\!\!ff_1\!V_2}\,G_{134}\,G_{243}\,
    \left( -\,\frac{\alpha}{\pi} \right)^2 \,\int_0^1\frac{\dd y_1}{y_1} 
    \int_{y_1}^1\frac{\dd z_1}{z_1}\,\int_0^1 \frac{\dd y_2}{y_2} 
    \int_{y_2}^1\frac{\dd z_2}{z_2}\, \no \\[2mm]
  & \hspace*{5mm} {\scriptstyle \times}  \hspace*{5mm}  
    \bigg[ \cK^{\,(1)}(s,m_f^2,M_1,y_1,z_1) 
         + \cK^{\,(1)}(s,m_f^2,M_2,y_1,z_1) \bigg] \no \\[2mm]
  & \hspace*{5mm} {\scriptstyle \times}  \hspace*{5mm}  
    \bigg[ \cK^{\,(1)}(s,m_f^2,M_3,y_2,z_2) 
         + \cK^{\,(1)}(s,m_f^2,M_4,y_2,z_2) \bigg] \no \\[3mm] 
  & \hspace*{5mm} {\scriptstyle \times}  \hspace*{5mm}  
    \Theta(y_2-y_1)\,\Theta(z_1-z_2) \,.
\end{align} 

\ni
To summarize everything we write the generic two-loop contribution of Sudakov 
logarithms to $\delta Z_f^{\,(2)}$ as: 
\bea
\label{twoloop}
  \delta Z^{\,(2)}_f &\approx& \Bigl( -\,\frac{\alpha}{\pi} \Bigr)^2\,
    \Gamma_f^{\,(2)}\, \int_0^1 \frac{\dd y_1}{y_1} 
    \int_{y_1}^1\frac{\dd z_1}{z_1} \int_0^1 \frac{\dd y_2}{y_2} 
    \int_{y_2}^1 \frac{\dd z_2}{z_2}\ \cK^{\,(2)}(y_1,z_1,y_2,z_2)~. 
\ea 
For the five different topologies the various products 
$\Gamma_f^{\,(2)}\,{\scriptstyle \times}\,\cK^{\,(2)}$ of coupling factors 
and integration kernels are given by
\bea
\label{f2loop}
  \mbox{set (a):} &&\!\!\!\!\!
                     \Bigl[ \Gamma_{\!\!ff_1\!V_1}^{\,^{\scriptstyle 2}}\,
                            \cK^{\,(1)}(s,m_f^2,M_1,y_1,z_1) \Bigr]\,
                     \Bigl[ \Gamma_{\!\!f_1f_2V_2}^{\,^{\scriptstyle 2}}\,
                            \cK^{\,(1)}(s,m_f^2,M_2,y_2,z_2) \Bigr]\,
                     \Theta(y_2-y_1)~, \no\\[2mm]
  \mbox{set (b):} && \!\!\!\!\!
                     -\,\Gamma_{\!\!ff_1\!V_1}\,\Gamma_{\!\!f_1f_2V_2}\,
                     \Gamma_{\!\!f_2f_3V_1}\,\Gamma_{\!\!ff_3V_2}\,
                     \cK^{\,(1)}(s,m_f^2,M_1,y_1,z_1)\,
                     \cK^{\,(1)}(s,m_f^2,M_2,y_2,z_2)~, \no\\[3mm]
  \mbox{set (c):} && \!\!\!\!\!
                     \Bigl[ \Gamma_{\!\!ff_1\!V_1}^{\,^{\scriptstyle 2}}\,
                            \cK^{\,(1)}(s,m_f^2,M_1,y_1,z_1) \Bigr]\,
                     \Bigl[ \Gamma_{\!\!ff_2V_2}^{\,^{\scriptstyle 2}}\,
                            \cK^{\,(1)}(s,m_f^2,M_2,y_2,z_2) \Bigr]~, 
                     \no\\[2mm]
 \mbox{set (d):} && \!\!\!\!\!
                     \frac{1}{4}\,\Gamma_{\!\!ff_1\!V_1}\,
                     \Gamma_{\!\!f_1f_2V_3}\,\Gamma_{\!\!ff_2V_2}\,G_{132}\,
                     \biggl\{ \Bigl[ \cK^{\,(1)}(s,m_f^2,M_1,y_1,z_1) 
                             + \cK^{\,(1)}(s,m_f^2,M_2,y_1,z_1) \Bigr]
                     \times 
                     \no\\[1mm]
                  && \!\!\!\!\!\hphantom{\frac{1}{4}\,\Gamma_{\!\!ff_1\!V_1}\,
                               \Gamma_{\!\!f_1f_2V_3}}
                     \times\,\,\cK^{\,(1)}(s,m_f^2,M_3,y_2,z_2)\,
                     \Theta(y_2-y_1)\,\Bigl[ 1 + 3\,\Theta(z_1-z_2) \Bigr]
                     \no\\[3mm]
                  && \!\!\!\!\!\hphantom{\frac{1}{4}\,\Gamma_{\!\!ff_1\!V_1}`\,
                               \Gamma_{\!\!f_1f_2V_3}\,\Gamma_{\!\!ff_2V_2}\,
                               G_{132}\,A}
                     {}+\cK^{\,(1)}(s,m_f^2,M_1,y_1,z_1)\,
                        \cK^{\,(1)}(s,m_f^2,M_2,y_2,z_2)\,
                     \times 
                     \no\\[2mm]
                  && \!\!\!\!\!\hphantom{\frac{1}{4}\,\Gamma_{\!\!ff_1\!V_1}\,
                               \Gamma_{\!\!f_1f_2V_3}}
                     \times\,\,\Bigl[ 3 + \Theta(y_1-y_2)\,\Theta(z_2-z_1)
                                    + \Theta(y_2-y_1)\,\Theta(z_1-z_2) \Bigr]
                     \biggr\}~,
                     \no\\[2mm]
  \mbox{set (e):} && \!\!\!\!\!
                     \,\frac{1}{2}\, \Gamma_{\!\!ff_1\!V_1}\,
                     \Gamma_{\!\!ff_1\!V_2}\,
                     G_{134}\,G_{243}\,\Bigl[ 
                     \cK^{\,(1)}(s,m_f^2,M_3,y_2,z_2) 
                     + \cK^{\,(1)}(s,m_f^2,M_4,y_2,z_2) \Bigr] 
                     \times 
                     \no\\[2mm]
                  && \!\!\!\!\! \times\,\, 
                     \Bigl[ \cK^{\,(1)}(s,m_f^2,M_1,y_1,z_1) +  
                            \cK^{\,(1)}(s,m_f^2,M_2,y_1,z_1) \Bigr]\,
                     \Theta(y_2-y_1)\,\Theta(z_1-z_2)~. 
\ea

\ni As a cross-check we also calculated the full two-loop result by means of 
the dispersion method. In that case the various $\Theta$-functions originate 
from the different two- and three-particle cuts that enter the calculation of
the absorptive parts. The final result agrees with Eq.~(\ref{f2loop}).
\\[-5mm]

\ni
In Appendix \ref{app:integrals} we have derived all relevant one- and two-loop
integrals. Here we give the results, using the generic notation
\begin{align}
  I^{(i)} = \Bigl( -\,\frac{\alpha}{\pi} \Bigr)^i
            \int_0^1 \frac{\dd y_1}{y_1} \,\int_{y_1}^1 \frac{\dd z_1}{z_1}\,
            \ldots            
            \int_0^1 \frac{\dd y_i}{y_i} \,\int_{y_i}^1 \frac{\dd z_i}{z_i}\,
            \cK^{(i)}(y_1,z_1,\ldots,y_i,z_i)~.
\end{align}
At one-loop level we found
\begin{align}
\label{App1LM}
  \cK^{\,(1)}(s,m_f^2,M,y_1,z_1)   &: I^{(1)} = {\rm L}(M,M)~, \\[1mm]
\label{App1Lla}
  \cK^{\,(1)}(s,m_f^2,\la,y_1,z_1) &: I^{(1)} = {\rm L_{\ga}}(\la,m_f)~.
\end{align}
The functions $\,{\rm L}(M_1,M_2)\,$ and $\,{\rm L_{\ga}}(\la,M_1)\,$ are the 
ones defined in Eqs.~(\ref{L}) and (\ref{Lgamma}). At two-loop level we found
for the angular ordered integrals:
\begin{align}
\label{two-loop/ang}
  \cK^{\,(1)}(s,m_f^2,M,y_1,z_1)\,\cK^{\,(1)}(s,m_f^2,M,y_2,z_2)\,
  \Theta(y_2-y_1) 
       &: I^{(2)} = \frac{1}{2}\,{\rm L}^2(M,M)~, \no\\[1mm]
  \cK^{\,(1)}(s,m_f^2,\la,y_1,z_1)\,\cK^{\,(1)}(s,m_f^2,\la,y_2,z_2)\,
  \Theta(y_2-y_1) 
       &: I^{(2)} = \frac{1}{2}\,{\rm L_{\ga}}^{\!\!2}(\la,m_f)~, \no\\[1mm]
  \cK^{\,(1)}(s,m_f^2,M,y_1,z_1)\,\cK^{\,(1)}(s,m_f^2,\la,y_2,z_2)\,
  \Theta(y_2-y_1) 
       &: I^{(2)} = \frac{7}{12}\,{\rm L}^2(M,M)~, \no\\[2mm]
  \cK^{\,(1)}(s,m_f^2,\la,y_1,z_1)\,\cK^{\,(1)}(s,m_f^2,M,y_2,z_2)\,
  \Theta(y_2-y_1) 
       &: I^{(2)} = {\rm L}(M,M)\,{\rm L_{\ga}}(\la,m_f) \no\\[1mm]
       &  \hphantom{I^{(2)} =} - \frac{7}{12}\,{\rm L}^2(M,M)~, 
\end{align}
and for the double ordered integrals:
\begin{align}
\label{two-loop/en}
  \cK^{\,(1)}(s,m_f^2,M,y_1,z_1)\,\cK^{\,(1)}(s,m_f^2,M,y_2,z_2)\,
  \Theta(y_2-y_1)\,\Theta(z_1-z_2)
       &: I^{(2)} = \frac{1}{4}\,{\rm L}^2(M,M)~, \no\\[1mm]
  \cK^{\,(1)}(s,m_f^2,M,y_1,z_1)\,\cK^{\,(1)}(s,m_f^2,\la,y_2,z_2)\,
  \Theta(y_2-y_1)\,\Theta(z_1-z_2)
       &: I^{(2)} = \frac{1}{3}\,{\rm L}^2(M,M)~, \no\\[1mm]
  \cK^{\,(1)}(s,m_f^2,\la,y_1,z_1)\,\cK^{\,(1)}(s,m_f^2,M,y_2,z_2)\,
  \Theta(y_2-y_1)\,\Theta(z_1-z_2)
       &: I^{(2)} = \frac{2}{3}\,{\rm L}(M,M)\,{\rm L}(M,m_f) \no\\[1mm]
       &  \hphantom{I^{(2)} =} {}-\frac{1}{4}\,{\rm L}^2(M,M)~.
\end{align}
Note that in the case of double ordering the collinear cut-off $m_f^2$ of the
$y_2$ integral is in fact redundant.
\\[-5mm]
 
\ni
Now the task at hand is to sum all possible contributions to obtain the full
two-loop correction to the external wave-function factor.
Using the abbreviations $\L \equiv \L(M,M)$, $\L_{m_f} \equiv \L(M,m_f)$
and $\L_{\ga} \equiv \L_{\ga}(\la,m_f)$ we find: 
\begin{align}
{\rm set(a)} &= \left\{ \begin{array}[H]{cc} 
    \dis{ \frac{1}{2}\left( \Cw{L} + \Cz{L} + \Cga{L} \right)^2 
        - \frac{1}{6}\,\frac{I_f^{\,3}\,Y_f^L}{2\,\swto}\,\L^2 }
    & \quad\quad\mbox{for $f_L,\bar{f}_R$} \\[5mm]
    \dis{ \frac{1}{2}\left( \Cz{R} + \Cga{R} \right)^2 }
    & \quad\quad\mbox{for $f_R,\bar{f}_L$}
                        \end{array} \right. \no  \\[3mm]
{\rm set(b)} &= \left\{ \begin{array}[H]{cc}
    \dis{ -\,\left( \Cw{L} + \Cz{L} + \Cga{L} \right)^2 } & \\[2mm]
    \dis{ {}+ 3\,\Cw{L}\Cw{L} + \Cw{L}\,\left( 1+2\,I_f^{\,3}Y_f^L \right)\, 
          \left[ \L_{\ga} - \L\right] }   
    & \quad\quad\mbox{for $f_L,\bar{f}_R$} \\[5mm]
    \dis{ -\,\left( \Cz{R} + \Cga{R} \right)^2 }
    & \quad\quad\mbox{for $f_R,\bar{f}_L$}
                        \end{array} \right. \no  \\[3mm]   
{\rm set(c)} &= \left\{ \begin{array}[H]{cc}
    \left(  \Cw{L} + \Cz{L} + \Cga{L} \right)^2 
    & \quad\quad\mbox{for $f_L,\bar{f}_R$} \\[5mm]   
    \left( \Cz{R} +\Cga{R} \right)^2 
    & \quad\quad\mbox{for $f_R,\bar{f}_L$}
                        \end{array} \right. \no  
\end{align}
and exclusively for left-handed fermions ($f_L,\bar{f}_R$)
\begin{align}
{\rm set(d)}_L & = -\,2\,\frac{I_f^{\, 3}\,Q_f}{\swto} 
  \left[ \, \L \,\L_{\ga} + \frac{2}{3}\,\L\,\L_{m_f} \right]  
  - \frac{9}{8\,\swfor}\,\L^2 \no \\[2mm]
  & \hspace*{2cm} {}+ \left[ 3 -\frac{1}{6} \right]\,\frac{1}{4\,\swto}\,\L^2 
  + \left[ 3 + \frac{1}{6} \right]\,\frac{I_f^{\,3}\,Y_f^L}{2\,\swto}\,\L^2  
  \no \\[3mm]
{\rm set(e)}_L & = \frac{7}{24\,\swto}\,\L^2 + \frac{1}{8\,\swfor}\,\L^2 
  + \frac{\cwto}{4\,\swfor}\,\L^2 
  + \frac{4}{3}\,\frac{I_f^{\,3}\,Q_f}{\swto}\,\L\,\L_{m_f} 
  - \frac{I_f^{\,3}\,Q_f}{\swto}\,\L^2 \,.\no 
\end{align}
\vspace*{5mm}

\ni 
Here $\,\delta Z_{f_{L/R}}^{\,(1)}(V)\,$ are the one-loop corrections to the
external wave-function factors given in Eq.~(\ref{delta1}).
Hence the full two-loop fermionic Sudakov correction factor reads 
\bea
\label{delta2ZF}
  \hspace*{1cm} \delta Z^{\,(2)}_{f} = \left\{ \begin{array}[H]{cc} 
  \dis{ \frac{1}{2}\left( \Cw{L} + \Cz{L} + \Cga{L} \right)^2 
        \ =\ \frac{1}{2}\left( \delta Z^{\,(1)}_{f_L} \right)^2 }
  & \quad\quad\mbox{for $f_L,\bar{f}_R$} \\[5mm] 
  \dis{ \frac{1}{2}\left( \Cz{R} + \Cga{R} \right)^2 
        \ =\ \frac{1}{2}\left( \delta Z^{\,(1)}_{f_R} \right)^2 }
  & \quad\quad\mbox{for $f_R,\bar{f}_L$}
                                               \end{array}\right. \,. 
\hspace*{1cm}
\ea

\ni
{}From Eq.~(\ref{delta2ZF}) we deduce our main statement,
namely that the virtual electroweak two-loop Sudakov correction factor is
obtained by a mere exponentiation of the one-loop Sudakov correction
factor. We also note that, in adding up all the contributions, we find that 
the `rainbow' diagrams of set (a) yield the usual exponentiating terms plus an 
extra term for left-handed fermions. This extra term originates from 
the charged-current interactions and is only non-vanishing as a result of the
mass gap between the massless photon and the massive $Z$ boson. It cancels 
against a specific term originating from the triple gauge-boson diagrams of 
set (d). Similar (gauge) cancellations take place between the 
`crossed rainbow' diagrams of set (b), the reducible diagrams of set (c), and 
another part of the triple gauge-boson diagrams of set (d). Finally,  
the left-over terms of set (d) get cancelled by the contributions from the
gauge-boson self-energy (`frog') diagrams of set (e).

\subsection{The bosonic self-energies at two-loop level}

\ni In a similar way, simply by adjusting the relevant couplings, we find that 
the two-loop Sudakov correction factors in the charged-boson sector can be 
obtained from the one-loop results by means of `exponentiation'. The same holds
in an equally straightforward way for the non-transverse neutral-boson sector.
Also the gauge cancellations conspire in a way very similar to what we already
saw in the fermionic case. For transverse $W$ bosons, for instance, we observe
the following. Due to the mass gap between the massive $Z$ boson and the 
massless photon, the `rainbow' contributions from set (a) exhibit an extra 
term, which in the transverse $W$-boson case is canceled in part by the
contributions from the triple gauge-boson diagrams of set (d) and in part by
contributions from the `frog' diagrams of set (e). The extra terms in the
`crossed rainbow' contribution of set (b), arising due to forbidden
combinations of one charged and one neutral particle, are in the case of the
photon compensated by contributions from the triple gauge-boson diagrams and 
in the case of the $Z$ boson by contributions from both the triple gauge-boson 
diagrams and the `frog' diagrams. Eventually we are left with the very simple 
result $\delta Z^{\,(2)}_{\sscr{W}_T} = 
\frac{1}{2} \left( \delta Z^{\,(1)}_{\sscr{W}_T}\right)^2$ for the two-loop 
contribution to the external wave-function factor.
\\[-5mm]

\ni
In the transverse neutral gauge-boson sector there is again the extra 
complication of having to determine $\,\tan\theta(k^2)\,$ at $\,k^2=0\,$ and
$\,k^2=\mz^2$. In analogy to the one-loop case discussed in 
Sect.~\ref{sec:boson1loop}, we expand Eqs.~(\ref{Cgamma}) and (\ref{CZ}) to 
two-loop order, which yields with the help of Eqs.~(\ref{taylor}) and 
(\ref{nonrene})
\begin{align}
  C_{\ga\ga}^{-1} & \myto{2-loop}{} 1 - \Sigma_{\ga\ga,\,g}^{\prime\,(1)} 
     - \Sigma_{\ga\ga,\,g}^{\prime\,(2)} 
     - \frac{3}{4}\,\left( \Sigma_{\ga\sscr{Z},\,g}^{\prime\,(1)} \right)^2
     \\[1mm]
  C_{\sscr{ZZ}}^{-1} & \myto{2-loop}{} 1 - \Sigma_{\sscr{ZZ},\,g}^{\prime\,(1)}
     - \Sigma_{\sscr{ZZ},\,g}^{\prime\,(2)} 
     - \frac{3}{4}\,\left( \Sigma_{\ga\sscr{Z},\,g}^{\prime\,(1)} \right)^2 \,.
\end{align}
The various self-energies can be calculated by means of the derivative
method explained in Sect.~\ref{sec:two-loop_fermions}, resulting in
\begin{equation}
  \Sigma_{\sscr{ZZ},\,g}^{\prime\,(2)} 
  \ =\ -\,\frac{\cw}{\sw}\,\Sigma_{\ga\sscr{Z},\,g}^{\prime\,(2)}
  \ =\ \frac{\cwto}{\swto}\,\Sigma_{\ga\ga,\,g}^{\prime\,(2)}
  \ =\ \cwto\,\biggl[ -\,\frac{2}{\sin^4\theta_{\rm w}}\,\L^2(M,M) \biggr] 
  \ \equiv\ \cwto\,\Sigma_{33,\,g}^{\prime\,(2)} \,.
\end{equation}
Next we derive from the ``non-renormalization'' condition (\ref{nonrene}) for 
the electromagnetic charge and Eq.~(\ref{Cgamma})
\bea 
  \tan\theta(0)_{\,\, \rm 2-loop} &=& C_{\ga\sscr{Z},\,g}^{\,(2)} 
     - C_{\ga\sscr{Z},\,g}^{\,(1)}\,C_{\sscr{ZZ},\,g}^{\,(1)}  
     \ =\ \frac{\cw}{2\sw}\,\biggl[ \Sigma_{\ga\ga,\,g}^{\prime\,(2)} 
          + \frac{1}{4}\,\Bigl( \Sigma_{\ga\ga,\,g}^{\prime\,(1)} \Bigr)^2
          + \frac{3}{4}\,\Bigl( \Sigma_{\ga\sscr{Z},\,g}^{\prime\,(1)} \Bigr)^2
                            \biggr] \no \\[1mm]
      &=& \frac{\cw}{2\sin^3\theta_{\rm w}}\,\left( 2\,\cwto - 1 \right)\, 
          \L^2(M,M) \,.\quad  
\ea
Since 
\begin{equation}
  C_{\sscr{ZZ},\,g}^{\,(1)} = -\,\Sigma_{\sscr{ZZ},\,g}^{\prime\,(1)}
  \qquad , \qquad
  C_{\ga\ga,\,g}^{\,(1)} = 0
\end{equation}
as a result of the ``non-renormalization'' condition (\ref{nonrenN}) for the 
masses, we find
\begin{equation}
  C_{\ga\sscr{Z},\,g}^{\,(2)} = {}-\frac{\cw}{2\sin^3\theta_{\rm w}}\, 
                                \left( 1+2\,\cwto \right)\,\L^2(M,M) \,. 
\end{equation}  
Now all ingredients are known and $\,\tan\theta(\mz^2)_{\,\, \rm 2-loop}\,$
can be determined trivially
\bea
  \tan\theta(\mz^2)_{\,\, \rm 2-loop} &=& C_{\ga\sscr{Z},\,g}^{\,(2)} 
       + \Sigma_{\ga\sscr{Z},\,g}^{\prime\,(2)} 
       + \Bigl( C_{\ga\sscr{Z},\,g}^{\,(1)}
                +\Sigma_{\ga\sscr{Z},\,g}^{\prime\,(1)} \Bigr)
         \Bigl( C_{\ga\ga,\,g}^{\,(1)}+\Sigma_{\ga\ga,\,g}^{\prime\,(1)} \Bigr)
       \no \\[2mm]
       &=&\ \frac{\cw}{2\sin^3\theta_{\rm w}}\,\left( 2\,\cwto - 1 \right)\, 
            \L^2(M,M) \,.
\ea
Replacing again the $Z$-boson and photon fields 
in the amputated Green's functions by the unbroken gauge 
fields $\,B\,$ and $\,W^3$, the final result in the transverse neutral 
gauge-boson sector reads:
\begin{align}
  \delta Z_{\sscr{N}_T,\,B}^{\,(2)} &\ =\ 0 \no \\[1mm]
  \delta Z_{\sscr{N}_T,\,W^3}^{\,(2)} &\ =\ 
       \frac{1}{2}\,\Bigl( \delta Z_{\sscr{N}_T,\,W^3}^{\,(1)} \Bigr)^2 \,.
\end{align}

\ni In fact, this step-wise procedure can be performed to all orders in 
perturbation theory, yielding
\bea
 \tan\theta(0) 
 \ =\ \frac{\sw\cw\,\Bigl[ 1-\sqrt{1-\Sigma_{33,\,g}^{\prime}}\, \Bigr]}
           {\swto+\cwto\,\sqrt{1-\Sigma_{33,\,g}^{\prime}}} 
 \quad &,& \quad
 C_{\ga\ga}^{\frac{1}{2}}
 \ =\ \cwto + \frac{\swto}{\sqrt{1-\Sigma_{33,\,g}^{\prime}}}
 \no \\[2mm]
 \tan\theta(\mz^2)
 \ =\ -\,\frac{\sw\cw\,\Bigl[ 1-\sqrt{1-\Sigma_{33,\,g}^{\prime}}\, \Bigr]}
              {\cwto+\swto\,\sqrt{1-\Sigma_{33,\,g}^{\prime}}}
 \quad &,& \quad 
 C_{\sscr{ZZ}}^{\frac{1}{2}} 
 \ =\ \swto + \frac{\cwto}{\sqrt{1-\Sigma_{33,\,g}^{\prime}}} \,. \quad\quad
\ea
This automatically leads to
\begin{equation}
  Z_{\sscr{N}_T,\,B}^{-1} \ =\ 1
  \qquad , \qquad
  Z_{\sscr{N}_T,\,W^3}^{-1} \ =\ 1-\Sigma_{33,\,g}^{\prime} \,.
\end{equation}

\subsection{General two-loop Sudakov logarithms}

We conclude this section with our general result. The full two-loop Sudakov 
correction to the external wave-function factor for an arbitrary 
on-shell/on-resonance particle is given by
\begin{equation}
\label{deltaZ2G}
  \delta Z^{\,(2)} = \frac{1}{2}\,\Big( \delta Z^{\,(1)} \Big)^2 \,,
\end{equation}
in terms of the one-loop correction given in Eq.~(\ref{deltaZ1}).
In summary we would like to point out that to calculate the two-loop Sudakov 
correction factor for an arbitrary species of initial- or final-state 
particles, \ie fermions, gauge bosons or would-be Goldstone bosons, the 
knowledge of the corresponding one-loop correction factor is sufficient. 
This is a well-known fact in massless or one-mass-scale theories, such as QED,
QCD or generally $SU(N)$, where in fact in covariant gauges the two-loop 
results are {\it effectively} obtained from so-called ladder diagrams, 
corresponding to our `rainbow' diagrams. We would like to stress again that 
for the SM, as a broken theory with two mass scales, the
result $\delta Z^{\,(2)} = \frac{1}{2}\left( \delta Z^{\,(1)} \right)^2$
is identical, but at all intermediate stages extra terms arise due to the
mass gap. Therefore the calculation of only one topology, \ie the
`rainbow' diagrams, does not lead (not even effectively) to the correct 
two-loop Sudakov correction factor.

\section{Conclusions} 

We have calculated the virtual electroweak Sudakov (double) logarithms at 
one- and two-loop level for arbitrary on-shell/on-resonance particles in the 
Standard Model. The associated Sudakov form factors apply, in principle in a 
universal way, to arbitrary non-mass-suppressed electroweak processes at high 
energies. We would like to stress that the universality of the Sudakov form 
factors has to be interpreted with care. Depending on the final state and the 
kinematical configuration, the process may possess various near-resonance 
subprocesses, which all have their own Sudakov correction factor. These 
correction factors are given (in leading-pole approximation) by the Sudakov 
form factors of the external particles involved in the subprocess. In this way
the Sudakov form factors for unstable particles, like the massive gauge bosons
and the Higgs boson, participate in the high-energy behaviour of reactions 
with exclusively stable particles in the final state. 
\\[-5mm]

\ni For the explicit calculation we adopted the temporal Coulomb gauge for both
massless and massive particles. About the latter case basically nothing is 
known in the literature, so in this paper we have tried to give as much detail 
as possible about its salient details. In view of the special status of the 
time-like components in this gauge, a careful analysis was required for the
derivation of the asymptotic fields of the theory, which are needed for a
proper description of the on-shell/on-resonance states. In particular the 
presence of lowest-order mixing between the massive gauge bosons and the 
corresponding would-be Goldstone bosons required special attention.
\\[-5mm]

\ni In the Sudakov limit, our calculation was significantly aided by a few 
special properties of the temporal Coulomb gauge. First of all, in this 
special gauge all the relevant contributions involve the exchange of 
effectively on-shell, transverse, collinear-soft gauge bosons. Moreover, these
relevant contributions are contained exclusively in the self-energies of the 
external on-shell/on-resonance particles (wave-function factors). This has
to be contrasted with covariant gauges, where 
all contributions are residing in vertex and higher-point diagrams. Second, 
the various self-energies are subject to explicit ``non-renormalization'' 
conditions to all orders in perturbation theory. This allows us to obtain the 
Sudakov wave-function factors through a combination of energy derivatives and 
projections by means of sources. As a result, we observe that 
the Standard Model behaves dynamically like an unbroken theory in the Sudakov
limit, in spite of the fact that the explicit particle masses are needed at 
the kinematical (phase-space) level while calculating the Sudakov correction 
factors. For instance, we obtain automatically a special version of the 
Equivalence Theorem, which states that the longitudinal degrees of freedom of 
the massive gauge bosons can be substituted by the corresponding 
Goldstone-boson degrees of freedom. As a result, the Sudakov form factors for 
longitudinal gauge bosons exhibit features that are typical for 
particles in the fundamental representation of $SU(2)$, whereas for the
transverse gauge bosons the usual adjoint features are obtained. Moreover, in
the transverse neutral gauge-boson sector the mass eigenstates decompose into 
the unbroken fields $W^3$ and $B$, each multiplied by the corresponding Sudakov
form factor. At the kinematical level, though, the large mass gap between the 
photon and the weak gauge bosons remains.
\\[-5mm]

\ni Our explicit one- and two-loop calculations of the Sudakov form factors in
the Standard Model reveal the following. The one-loop results are in agreement 
with the available calculations in the literature, including the distinctive 
terms originating from the mass gap between the photon and the weak gauge 
bosons. At two-loop level our findings are in agreement with an exponentiation 
of the one-loop results. We therefore conclude that also as far as the balance 
between the one- and two-loop virtual Sudakov logarithms are concerned, the  
Standard Model behaves like an unbroken theory at high energies. We would like 
to stress, though, that for the Standard Model, as a broken theory with two 
mass scales, all two-loop diagram topologies are needed to arrive at the 
correct result. In general it is not possible to get the correct result by 
singling out one particular topology, such as the `ladder'-like diagrams in 
unbroken theories.
\\[-5mm]
 
\ni All these conclusions can be extended to real-emission processes in a 
relatively straightforward way. After all, since the Sudakov logarithms 
originate from the exchange of soft, effectively on-shell transverse gauge 
bosons, many of the features derived for the virtual corrections will be 
intimately related to properties of the corresponding real-emission processes.
In this context we note that the Bloch--Nordsieck cancellation between 
virtual and real collinear-soft gauge-boson radiation~\cite{Bloch:1937pw} is 
violated in the Standard Model~\cite{Ciafaloni:2000rp} as soon as initial- or 
final-state particles carry 
an explicit weak charge (isospin) and summation over the partners within an
$SU(2)$ multiplet is not performed. In the case of final-state particles the
event-selection procedure might (kinematically) favour one of the partners 
within the $SU(2)$ multiplet, leading to a degree of `isospin-exclusiveness'.
In the initial state the situation is more radical. In that case the weak 
isospin is fixed by the accelerator, in contrast to QCD where confinement 
forces average over initial colour at hadron colliders. At an 
electron--positron collider, for instance, the Bloch--Nordsieck theorem is in 
general violated for left-handed initial states, even for fully inclusive 
cross-sections. The resulting electroweak effects can be very large, exceeding
the QCD corrections for energies in the TeV range. They are such that
at infinite energy the weak charges will become unobservable as asymptotic
states~\cite{Ciafaloni:2000rp}, which implies for instance an $SU(2)$ charge 
averaging of the initial-state beams.
\\[-5mm]

\ni
As a matter of fact, for a complete understanding of the perturbative 
structure of large logarithmic correction factors, subleading logarithms 
originating from soft, collinear, or ultraviolet singularities cannot be 
ignored~\cite{Denner:2000jv,Kuhn:2001hz}. For pure fermionic final states 
(numerical) cancellations can take place between leading and subleading 
logarithms~\cite{Kuhn:2001hz}. For on-shell bosons in the final state, however,
the Sudakov logarithms in general tend to be 
dominant~\cite{Beenakker:1993tt,Denner:2000jv}, being anyhow intrinsically
larger than the Sudakov logarithms for fermions owing to the larger adjoint 
$SU(2)$ factors. 

\begin{appendix}
\section*{Appendix}
\label{chap:app}
\addcontentsline{toc}{section}{Appendix}
\def\thesubsection{\Alph{subsection}}
\def\theequation{\thesubsection.\arabic{equation}}
\setcounter{section}{0}

\subsection{One- and two-loop integrals}
\label{app:integrals}
\setcounter{equation}{0}

\ni In this appendix we give the relevant one- and two-loop integrals that
occur in the Sudakov correction factors.
At one-loop accuracy we have to distinguish between two different cases, \ie
the exchanged soft gauge boson being a photon with the fictitious mass 
$\lambda$ or a massive gauge boson ($W$ or $Z$) with the generic mass $M$.
The exchanged gauge boson being massive, we extract from Eq.~(\ref{kernel}) 
with $M_1=M$
\begin{align}
J^{(1)}(M) = &\,\int_0^1\frac{\dd y_1}{y_1}\int_{y_1}^1\frac{\dd z_1}{z_1}\  
                \Theta\Bigl( y_1 z_1-\frac{M^2}{s} \Bigr)\,
                \Theta\Bigl( y_1-\frac{m_f^2}{s}\,z_1 \Bigr)\,.
\end{align}
{}From the first $\Theta$-function we obtain the integration boundaries
\begin{align}
J^{(1)}(M) = \int_{\,\frac{M}{\sqrt{s}}}^{\,1}\frac{\dd z_1}{z_1}
             \int_{\,\frac{M^2}{s\,z_1}}^{\,z_1}\frac{\dd y_1}{y_1}\
             \Theta\Bigl( y_1-\frac{m_f^2}{s}\,z_1 \Bigr)\,, \no 
\end{align}
which makes the second $\Theta$-function redundant, since $m_f\le {\cal O}(M)$.
Therefore 
\begin{align}
\label{App1LMA}
J^{(1)}(M)= & \int_{\,\frac{M}{\sqrt{s}}}^{\,1}\frac{\dd z_1}{z_1} 
              \int_{\,\frac{M^2}{s\,z_1}}^{\,z_1}\frac{\dd y_1}{y_1} 
        \ =\  \frac{1}{4}\,\log^2 \left( \frac{M^2}{s} \right)\,.
\end{align}
\ni
Similarly we obtain for $M_1=\lambda$
\begin{align}
\label{App1LlaA}
J^{(1)}(\la) &= \,\int_0^1\frac{\dd y_1}{y_1}\int_{y_1}^1\frac{\dd z_1}{z_1}\
                \Theta\Bigl( y_1 z_1-\frac{\lambda^2}{s} \Bigr)\, 
                \Theta\Bigl( y_1-\frac{m_f^2}{s}\,z_1 \Bigr) \no \\[1mm]
             &= \int_{\,\frac{\la}{m_f}}^{\,1}\z{1} 
                \int_{\,\frac{m_f^2\,z_1}{s}}^{\,z_1}\y{1} 
              + \int_{\,\frac{\la}{\sqrt{s}}}^{\,\frac{\la}{m_f}}\z{1} 
                \int_{\,\frac{\la^2}{s\,z_1}}^{\,z_1}\y{1} 
            \ =\ \frac{1}{4}\,\log^2 \left( \frac{\la^2}{s} \right)\, 
              -\,\frac{1}{4}\,\log^2 \left( \frac{\la^2}{m_f^2} \right)\,.
\end{align}
The two-loop integrals fall into two categories, namely the angular ordered 
integrals and the integrals that are (double) ordered in energy and angle
simultaneously. For the angular ordered two-loop integrals, see
Eq.~(\ref{angular}), we find for $M_1=M_2=M$
\begin{align}
J_{\rm angular}^{(2)}(M,M) &= \int_{0}^{1}\z{1}\int_{0}^{z_1}\y{1}
                              \int_{0}^{1}\z{2}\int_{0}^{z_2}\y{2}\   
                              \Theta\Bigl( y_1 z_1-\frac{M^2}{s} \Bigr)\,   
                              \Theta\Bigl( y_2 z_2-\frac{M^2}{s} \Bigr)\,
                              \Theta(y_2-y_1)\,. \no \\
\end{align}
By means of symmetry arguments, \ie $\Theta(y_2-y_1) \to \frac{1}{2}\left[
\Theta(y_2-y_1) +  \Theta(y_1-y_2) \right] = \frac{1}{2}$, we find
\begin{align}
J_{\rm angular}^{(2)}(M,M)&= \frac{1}{2}\,\left[ \frac{1}{4}\,
                             \log^2 \left( \frac{M^2}{s}\right) \right]^2\,,
\end{align} 
and similarly
\begin{align}
J_{\rm angular}^{(2)}(\la,\la) &= \frac{1}{2}\,\left[ 
            \frac{1}{4}\,\log^2 \left( \frac{\la^2}{s} \right)\, 
        - \,\frac{1}{4}\,\log^2 \left( \frac{\la^2}{m_f^2}\right) 
                                              \right]^2\,.
\end{align} 
Furthermore
\normalsize
\begin{align}
J_{\rm angular}^{(2)}(M,\la) &= \int_{0}^{1}\z{1}\int_{0}^{z_1}\y{1}
          \int_{0}^{1}\z{2}\int_{0}^{z_2}\y{2}\ 
          \Theta\Bigl( y_1 z_1-\frac{M^2}{s} \Bigr)\,   
          \Theta\Bigl( y_2 z_2-\frac{\la^2}{s} \Bigr) \no \\[3mm]
                             & \hphantom{= \int_{0}^{1}\z{1}\int_{0}^{z_1}\y{1}
                                        \int_{0}^{1}\z{2}\int_{0}^{z_2}\y{2}\!}
          {\scriptstyle\times}\,\Theta\Bigl( y_2-\frac{m_f^2\,z_2}{s} \Bigr)\, 
          \Theta(y_2-y_1) \no \\[3mm]
                             &=  \int_{\,\frac{M}{\sqrt{s}}}^{\,1}\z{1} 
          \int_{\,\frac{M^2}{s\,z_1}}^{\,z_1}\y{1}
          \int_{y_1}^{1}\z{2}\int_{y_1}^{z_2}\y{2} 
          \ =\ \frac{7}{12}\,\log^4 \left( \frac{M}{\sqrt{s}} \right)\,,
\end{align}
and hence  by means of $\,\Theta (y_2-y_1) = 1- \Theta(y_1-y_2)\,$ we find for 
$\,M_1=\la\,$ and $\,M_2=M\,$
\begin{align}
J_{\rm angular}^{(2)}(\la,M) &= 
          \left[ \frac{1}{4}\,\log^2 \left( \frac{M^2}{s}\right) \right]\,  
          {\scriptstyle \times}\, 
          \left[ \frac{1}{4}\,\log^2 \left( \frac{\la^2}{s} \right)\,
          - \,\frac{1}{4}\,\log^2 \left( \frac{\la^2}{m_f^2} \right) \right]  
          - \frac{7}{12}\,\log^4 \left( \frac{M}{\sqrt{s}} \right)\,.
\end{align}

\ni
For the double (energy and angular) ordered integrals, see Eq.~(\ref{efinal}), 
we find for the $\,M_1=M_2=M\,$ case 
\begin{align}
& J_{\rm double\,\,ordered}^{(2)}(M,M) = \no \\[2mm]
& \h{10}= \int_{0}^{1}\z{1}\int_{0}^{z_1}\y{1}\int_{0}^{1}\z{2}
          \int_{0}^{z_2}\y{2}\ \Theta\Bigl( y_1 z_1-\frac{M^2}{s} \Bigr)\,  
          \Theta\Bigl( y_2 z_2-\frac{M^2}{s} \Bigr)\,
          \Theta(y_2-y_1)\,\Theta(z_1-z_2) \no \\[3mm]
& \h{10}= \int_{\,\frac{M}{\sqrt{s}}}^{\,1}\z{2}
          \int_{\,\frac{M^2}{s\,z_2}}^{\,z_2}\y{2}\int_{z_2}^{1}\z{1} 
          \int_{\,\frac{M^2}{s\,z_1}}^{\,y_2}\y{1} 
     \ =\ \frac{1}{4}\left[ \frac{1}{4}\,\log^2 \left( \frac{M^2}{s}\right) 
                     \right]^2 \,.
\end{align}
For $\,M_1=M\,$ and $\,M_2=\la\,$ we obtain
\begin{align}
& J_{\rm double\,\,ordered}^{(2)}(M,\la) = \no \\[2mm]
& \h{10}= \int_{0}^{1}\z{1}\int_{0}^{z_1}\y{1}\int_{0}^{1}\z{2} 
          \int_{0}^{z_2}\y{2}\ \Theta\Bigl( y_1 z_1-\frac{M^2}{s} \Bigr)\,   
          \Theta\Bigl( y_2 z_2-\frac{\la^2}{s} \Bigr)\,
          \Theta(y_2-y_1)\,\Theta(z_1-z_2) \no \\[3mm]
& \h{10}= \int_{\,\frac{M}{\sqrt{s}}}^{\,1}\z{1}
          \int_{\,\frac{M^2}{s\,z_1}}^{\,z_1}\y{1}\int_{y_1}^{z_1}\z{2} 
          \int_{y_1}^{z_2}\y{2} 
     \ =\ \frac{1}{3}\,\log^4 \left( \frac{M}{\sqrt{s}} \right)\,. 
\end{align}
Finally for $\,M_1=\la\,$ and $\,M_2=M\,$ (note here that in the `frog'
configurations no two photons can appear in the integration kernel) the
double-ordered integral reads
{\small
\begin{align}
& J_{\rm double\,\,ordered}^{(2)}(\la,M) = \no \\[2mm] 
& = \int_{0}^{1}\z{1}\int_{0}^{z_1}\y{1}\int_{0}^{1}\z{2}\int_{0}^{z_2}\y{2}\
    \Theta\Bigl( y_1 z_1-\frac{\la^2}{s} \Bigr)\, 
    \Theta\Bigl( y_1-\frac{m_f^2}{s}z_1 \Bigr)\, 
    \Theta\Bigl( y_2 z_2-\frac{M^2}{s} \Bigr)\, 
    \Theta(y_2-y_1)\,\Theta(z_1-z_2) \no \\[3mm]
& = \int_{0}^{1}\z{1}\int_{0}^{z_1}\y{1}\int_{0}^{1}\z{2}\int_{0}^{z_2}\y{2}\
    \Theta\Bigl( y_1 z_1-\frac{M^2}{s} \Bigr)\,   
    \Theta\Bigl( y_2 z_2-\frac{\la^2}{s} \Bigr)\,
    \Theta\Bigl( y_2-\frac{m_f^2}{s}z_2 \Bigr)\,  
    \Theta(y_1-y_2)\,\Theta(z_2-z_1).
\end{align}}
Now we can write
\begin{align}
\Theta(y_1-y_2)\,\Theta(z_2-z_1) &= \left[ 1- \Theta(y_2-y_1) \right]\,
                                    \Theta(z_2- z_1) \no \\[2mm]
                                 &= \Theta(z_2- z_1) - \Theta(y_2-y_1) 
                                    + \Theta(y_2-y_1)\,\Theta(z_1- z_2)
\end{align}
and hence
\begin{align}
& J_{\rm double\,\,ordered}^{(2)}(\la,M) = 
  -\,\frac{7}{12}\,\log^4 \left( \frac{M}{\sqrt{s}} \right) 
  + \frac{1}{3}\,\log^4 \left( \frac{M}{\sqrt{s}} \right) \no \\[3mm]
& \h{5}+ \int_{0}^{1}\z{1}\int_{0}^{z_1}\y{1}\int_{0}^{1}\z{2}
  \int_{0}^{z_2}\y{2}\ \Theta\Bigl( y_1 z_1-\frac{M^2}{s} \Bigr)\,   
  \Theta\Bigl( y_2 z_2-\frac{\la^2}{s} \Bigr)\, 
  \Theta\Bigl( y_2 -\frac{m_f^2\, z_2}{s} \Bigr)\,\Theta( z_2-z_1) \no \\[2mm]
&= -\,\frac{1}{4}\,\log^4 \left( \frac{M}{\sqrt{s}} \right) 
   + \int_{\,\frac{M}{\sqrt{s}}}^{\,1}\frac{\dd z_1}{z_1}  
     \int_{\,\frac{M^2}{s\,z_1}}^{\,z_1}\frac{\dd y_1}{y_1}
     \int_{z_1}^{1}\z{2}\int_{\,\frac{m_f^2\,z_2}{s}}^{\,z_2}\y{2} 
     \no \\[2mm]
&= -\,\frac{1}{4}\,\log^4 \left( \frac{M}{\sqrt{s}} \right) 
   + \frac{2}{3}\,\log^3 \left( \frac{M}{\sqrt{s}} \right)\,
     \log \left( \frac{m_f}{\sqrt{s}} \right)\,.
\end{align}

\vspace*{5mm}

\subsection{Bosonic one-loop self-energies}
\label{app:selfenergy}
\setcounter{equation}{0}

For completeness we give in this appendix the full (\ie `non-derivative') 
bosonic self-energies at one-loop level. As an example we present the
explicit derivation of the $\G_{\mu\nu}$ part of the $W$-boson self-energy, 
which is a basic building block in the calculation of all two-loop Sudakov 
correction factors (see Sect.~\ref{sec:two-loop}). We first have to select the 
contributing diagrams. In principle there are $4 \scr{\times} 4$ possible 
generic combinations of scalar, mixed and gauge-boson particle states in the 
upper and lower part of the loop, which are displayed in Fig.~(\ref{16dia}). 
As explained in Sect.~\ref{sec:one-loop_fermions}, the fermion- and ghost-loop 
contributions do not have the right pole structure for producing Sudakov 
logarithms.

\begin{minipage}[c]{15.5cm}
\vspace{-3mm}
\begin{center}
\begin{figure}[H]
  \begin{fmffile}{1loopnew}
  \begin{fmfgraph*}(80,80) \fmfpen{thin} \fmfleft{i1} \fmfright{o1}
  \fmf{boson,tension=0.9,width=0.6}{i1,v1} 
  \fmf{boson,tension=0.5,width=0.6}{v1,v3} 
  \fmf{boson,tension=0.9,width=0.6}{v3,o1} 
  \fmfdot{v1} \fmfdot{v3} \fmfv{l=$1$,l.d=13,l.a=65}{i1}
  \fmf{boson,left,tension=0.1,width=0.6,label=$k_2$,l.s=left}{v1,v3} 
  \fmfv{l=$k_1$,l.d=8,l.a=-135}{v1}
  \fmffreeze
  \end{fmfgraph*} \hspace*{4mm}
  \begin{fmfgraph*}(80,80) \fmfpen{thin} \fmfleft{i1} \fmfright{o1}
  \fmf{boson,tension=0.9,width=0.6}{i1,v1} 
  \fmf{boson,tension=0.5,width=0.6}{v1,v3} 
  \fmf{boson,tension=0.9,width=0.6}{v3,o1} 
  \fmfdot{v1} \fmfdot{v3} \fmfv{l=$2$,l.d=13,l.a=65}{i1}
  \fmf{dashes,left,tension=0.1,width=0.6}{v1,v3}
  \fmffreeze
  \end{fmfgraph*} \hspace*{4mm}
  \begin{fmfgraph*}(80,80) \fmfpen{thin} \fmfleft{i1} \fmfright{o1}
  \fmf{boson,tension=0.9,width=0.6}{i1,v1} 
  \fmf{boson,tension=0.5,width=0.6}{v1,v3} 
  \fmf{boson,tension=0.9,width=0.6}{v3,o1} 
  \fmfdot{v1} \fmfdot{v3} \fmfv{l=$3$,l.d=13,l.a=65}{i1}
 \fmf{phantom,left,tension=0.1,width=0.6,tag=1}{v1,v3}
 \fmfposition
 \fmfipath{p[]}
 \fmfiset{p1}{vpath1(__v1,__v3)}
 \fmfi{dashes,width=0.6}{subpath (0,length(p1)/2) of p1}
 \fmfi{boson,width=0.6}{subpath (length(p1)/2,length(p1)) of p1}
  \fmf{phantom,left}{v1,v3}
  \fmffreeze
  \end{fmfgraph*} \hspace*{4mm}
 \begin{fmfgraph*}(80,80) \fmfpen{thin} \fmfleft{i1} \fmfright{o1}
  \fmf{boson,tension=0.9,width=0.6}{i1,v1} 
  \fmf{boson,tension=0.5,width=0.6}{v1,v3} 
  \fmf{boson,tension=0.9,width=0.6}{v3,o1} 
  \fmfdot{v1} \fmfdot{v3} \fmfv{l=$4$,l.d=13,l.a=65}{i1}
 \fmf{phantom,left,tension=0.1,width=0.6,tag=1}{v1,v3}
 \fmfposition
 \fmfipath{p[]}
 \fmfiset{p1}{vpath1(__v1,__v3)}
 \fmfi{boson,width=0.6}{subpath (0,length(p1)/2) of p1}
 \fmfi{dashes,width=0.6}{subpath (length(p1)/2,length(p1)) of p1}
  \fmf{phantom,left}{v1,v3}
  \fmffreeze
  \end{fmfgraph*}\\[-12mm]
 \begin{fmfgraph*}(80,80) \fmfpen{thin} \fmfleft{i1} \fmfright{o1}
  \fmf{boson,tension=0.9,width=0.6}{i1,v1} 
  \fmf{dashes,tension=0.5,width=0.6}{v1,v3} 
  \fmf{boson,tension=0.9,width=0.6}{v3,o1} 
  \fmfdot{v1} \fmfdot{v3} \fmfv{l=$5$,l.d=13,l.a=65}{i1}
  \fmf{boson,left,tension=0.1,width=0.6}{v1,v3}
  \fmffreeze
  \end{fmfgraph*} \hspace*{4mm}
  \begin{fmfgraph*}(80,80) \fmfpen{thin} \fmfleft{i1} \fmfright{o1}
  \fmf{boson,tension=0.9,width=0.6}{i1,v1} 
  \fmf{dashes,tension=0.5,width=0.6}{v1,v3} 
  \fmf{boson,tension=0.9,width=0.6}{v3,o1} 
  \fmfdot{v1} \fmfdot{v3} \fmfv{l=$6$,l.d=13,l.a=65}{i1}
  \fmf{dashes,left,tension=0.1,width=0.6}{v1,v3}
  \fmffreeze
  \end{fmfgraph*} \hspace*{4mm}
  \begin{fmfgraph*}(80,80) \fmfpen{thin} \fmfleft{i1} \fmfright{o1}
  \fmf{boson,tension=0.9,width=0.6}{i1,v1} 
  \fmf{dashes,tension=0.5,width=0.6}{v1,v3} 
  \fmf{boson,tension=0.9,width=0.6}{v3,o1} 
  \fmfdot{v1} \fmfdot{v3} \fmfv{l=$7$,l.d=13,l.a=65}{i1}
 \fmf{phantom,left,tension=0.1,width=0.6,tag=1}{v1,v3}
 \fmfposition
 \fmfipath{p[]}
 \fmfiset{p1}{vpath1(__v1,__v3)}
 \fmfi{dashes,width=0.6}{subpath (0,length(p1)/2) of p1}
 \fmfi{boson,width=0.6}{subpath (length(p1)/2,length(p1)) of p1}
  \fmf{phantom,left}{v1,v3}
  \fmffreeze
  \end{fmfgraph*} \hspace*{4mm}
 \begin{fmfgraph*}(80,80) \fmfpen{thin} \fmfleft{i1} \fmfright{o1}
  \fmf{boson,tension=0.9,width=0.6}{i1,v1} 
  \fmf{dashes,tension=0.5,width=0.6}{v1,v3} 
  \fmf{boson,tension=0.9,width=0.6}{v3,o1} 
  \fmfdot{v1} \fmfdot{v3} \fmfv{l=$8$,l.d=13,l.a=65}{i1}
 \fmf{phantom,left,tension=0.1,width=0.6,tag=1}{v1,v3}
 \fmfposition
 \fmfipath{p[]}
 \fmfiset{p1}{vpath1(__v1,__v3)}
 \fmfi{boson,width=0.6}{subpath (0,length(p1)/2) of p1}
 \fmfi{dashes,width=0.6}{subpath (length(p1)/2,length(p1)) of p1}
  \fmf{phantom,left}{v1,v3}
  \fmffreeze
  \end{fmfgraph*} \\[-12mm]
  \begin{fmfgraph*}(80,80) \fmfpen{thin} \fmfleft{i1} \fmfright{o1}
  \fmf{boson,tension=0.9,width=0.6}{i1,v1} 
  \fmf{boson,left,tension=0.1,width=0.6}{v1,v3}
  \fmf{boson,tension=0.9,width=0.6}{v3,o1} 
  \fmfdot{v1} \fmfdot{v3} \fmfv{l=$9$,l.d=13,l.a=65}{i1}
  \fmf{phantom,tension=0.5,width=0.6,tag=1}{v1,v3} 
  \fmfposition
  \fmfipath{p[]}
  \fmfiset{p1}{vpath1(__v1,__v3)}
  \fmfi{dashes,width=0.6}{subpath (0,length(p1)/2) of p1}
  \fmfi{boson,width=0.6}{subpath (length(p1)/2,length(p1)) of p1}
   \fmf{phantom}{v1,v3}
  \fmffreeze
  \end{fmfgraph*} \hspace*{4mm}
  \begin{fmfgraph*}(80,80) \fmfpen{thin} \fmfleft{i1} \fmfright{o1}
  \fmf{boson,tension=0.9,width=0.6}{i1,v1} 
  \fmf{dashes,left,tension=0.1,width=0.6}{v1,v3}
  \fmf{boson,tension=0.9,width=0.6}{v3,o1} 
  \fmfdot{v1} \fmfdot{v3} \fmfv{l=$10$,l.d=13,l.a=65}{i1}
  \fmf{phantom,tension=0.5,width=0.6,tag=1}{v1,v3} 
  \fmfposition
  \fmfipath{p[]}
  \fmfiset{p1}{vpath1(__v1,__v3)}
  \fmfi{dashes,width=0.6}{subpath (0,length(p1)/2) of p1}
  \fmfi{boson,width=0.6}{subpath (length(p1)/2,length(p1)) of p1}
   \fmf{phantom}{v1,v3}
  \fmffreeze
  \end{fmfgraph*} \hspace*{4mm}
  \begin{fmfgraph*}(80,80) \fmfpen{thin} \fmfleft{i1} \fmfright{o1}
  \fmf{boson,tension=0.9,width=0.6}{i1,v1} 
  \fmf{boson,tension=0.9,width=0.6}{v3,o1} 
  \fmfdot{v1} \fmfdot{v3} \fmfv{l=$11$,l.d=13,l.a=65}{i1}
  \fmf{phantom,left,tension=0.5,width=0.6,tag=1}{v1,v3}
 \fmfposition
 \fmfipath{p[]}
 \fmfiset{p1}{vpath1(__v1,__v3)}
 \fmfi{dashes,width=0.6}{subpath (0,length(p1)/2) of p1}
 \fmfi{boson,width=0.6}{subpath (length(p1)/2,length(p1)) of p1}
  \fmf{phantom,left}{v1,v3}\fmfv{l=$11$,l.d=13,l.a=65}{i1}
  \fmf{phantom,tension=0.5,width=0.6,tag=2}{v1,v3} 
  \fmfposition
  \fmfipath{p[]}
  \fmfiset{p2}{vpath2(__v1,__v3)}
  \fmfi{dashes,width=0.6}{subpath (0,length(p2)/2) of p2}
  \fmfi{boson,width=0.6}{subpath (length(p2)/2,length(p2)) of p2}
   \fmf{phantom}{v1,v3}
  \fmffreeze
  \end{fmfgraph*} \hspace*{4mm}
  \begin{fmfgraph*}(80,80) \fmfpen{thin} \fmfleft{i1} \fmfright{o1}
  \fmf{boson,tension=0.9,width=0.6}{i1,v1} 
  \fmf{boson,tension=0.9,width=0.6}{v3,o1} 
  \fmfdot{v1} \fmfdot{v3} \fmfv{l=$12$,l.d=13,l.a=65}{i1}
  \fmf{phantom,left,tension=0.5,width=0.6,tag=1}{v1,v3}
 \fmfposition
 \fmfipath{p[]}
 \fmfiset{p1}{vpath1(__v1,__v3)}
 \fmfi{boson,width=0.6}{subpath (0,length(p1)/2) of p1}
 \fmfi{dashes,width=0.6}{subpath (length(p1)/2,length(p1)) of p1}
  \fmf{phantom,left}{v1,v3}
  \fmf{phantom,tension=0.5,width=0.6,tag=1}{v1,v3} 
  \fmfposition
  \fmfipath{p[]}
  \fmfiset{p1}{vpath1(__v1,__v3)}
  \fmfi{dashes,width=0.6}{subpath (0,length(p1)/2) of p1}
  \fmfi{boson,width=0.6}{subpath (length(p1)/2,length(p1)) of p1}
   \fmf{phantom}{v1,v3}
  \fmffreeze
  \end{fmfgraph*} \\[-12mm]
  \begin{fmfgraph*}(80,80) \fmfpen{thin} \fmfleft{i1} \fmfright{o1}
  \fmf{boson,tension=0.9,width=0.6}{i1,v1} 
  \fmf{boson,left,tension=0.1,width=0.6}{v1,v3}
  \fmf{boson,tension=0.9,width=0.6}{v3,o1} 
  \fmfdot{v1} \fmfdot{v3} 
  \fmf{phantom,tension=0.5,width=0.6,tag=1}{v1,v3} 
  \fmfposition
  \fmfipath{p[]}
  \fmfiset{p1}{vpath1(__v1,__v3)}
  \fmfi{boson,width=0.6}{subpath (0,length(p1)/2) of p1}
  \fmfi{dashes,width=0.6}{subpath (length(p1)/2,length(p1)) of p1}
   \fmf{phantom}{v1,v3} \fmfv{l=$13$,l.d=13,l.a=65}{i1}
  \fmffreeze
  \end{fmfgraph*} \hspace*{4mm}
  \begin{fmfgraph*}(80,80) \fmfpen{thin} \fmfleft{i1} \fmfright{o1}
  \fmf{boson,tension=0.9,width=0.6}{i1,v1} 
  \fmf{dashes,left,tension=0.1,width=0.6}{v1,v3}
  \fmf{boson,tension=0.9,width=0.6}{v3,o1} 
  \fmfdot{v1} \fmfdot{v3} 
  \fmf{phantom,tension=0.5,width=0.6,tag=1}{v1,v3} 
  \fmfposition
  \fmfipath{p[]}
  \fmfiset{p1}{vpath1(__v1,__v3)}
  \fmfi{boson,width=0.6}{subpath (0,length(p1)/2) of p1}
  \fmfi{dashes,width=0.6}{subpath (length(p1)/2,length(p1)) of p1}
   \fmf{phantom}{v1,v3} \fmfv{l=$14$,l.d=13,l.a=65}{i1}
  \fmffreeze
  \end{fmfgraph*} \hspace*{4mm}
  \begin{fmfgraph*}(80,80) \fmfpen{thin} \fmfleft{i1} \fmfright{o1}
  \fmf{boson,tension=0.9,width=0.6}{i1,v1} 
  \fmf{boson,tension=0.9,width=0.6}{v3,o1} 
  \fmfdot{v1} \fmfdot{v3} 
  \fmf{phantom,left,tension=0.5,width=0.6,tag=1}{v1,v3}
 \fmfposition
 \fmfipath{p[]}
 \fmfiset{p1}{vpath1(__v1,__v3)}
 \fmfi{dashes,width=0.6}{subpath (0,length(p1)/2) of p1}
 \fmfi{boson,width=0.6}{subpath (length(p1)/2,length(p1)) of p1}
  \fmf{phantom,left}{v1,v3}
  \fmf{phantom,tension=0.5,width=0.6,tag=2}{v1,v3} 
  \fmfposition
  \fmfipath{p[]}
  \fmfiset{p2}{vpath2(__v1,__v3)}
  \fmfi{boson,width=0.6}{subpath (0,length(p2)/2) of p2}
  \fmfi{dashes,width=0.6}{subpath (length(p2)/2,length(p2)) of p2}
   \fmf{phantom}{v1,v3}\fmfv{l=$15$,l.d=13,l.a=65}{i1}
  \fmffreeze
  \end{fmfgraph*} \hspace*{4mm}
  \begin{fmfgraph*}(80,80) \fmfpen{thin} \fmfleft{i1} \fmfright{o1}
  \fmf{boson,tension=0.9,width=0.6}{i1,v1} 
  \fmf{boson,tension=0.9,width=0.6}{v3,o1} 
  \fmfdot{v1} \fmfdot{v3} 
  \fmf{phantom,left,tension=0.5,width=0.6,tag=1}{v1,v3}
 \fmfposition
 \fmfipath{p[]}
 \fmfiset{p1}{vpath1(__v1,__v3)}
 \fmfi{boson,width=0.6}{subpath (0,length(p1)/2) of p1}
 \fmfi{dashes,width=0.6}{subpath (length(p1)/2,length(p1)) of p1}
  \fmf{phantom,left}{v1,v3}
  \fmf{phantom,tension=0.5,width=0.6,tag=1}{v1,v3} 
  \fmfposition
  \fmfipath{p[]}
  \fmfiset{p1}{vpath1(__v1,__v3)}
  \fmfi{boson,width=0.6}{subpath (0,length(p1)/2) of p1}
  \fmfi{dashes,width=0.6}{subpath (length(p1)/2,length(p1)) of p1}
   \fmf{phantom}{v1,v3} \fmfv{l=$16$,l.d=13,l.a=65}{i1}
  \fmffreeze
  \end{fmfgraph*} 
  \end{fmffile}
\vspace*{-1.2cm}
\caption{\label{16dia}{Diagrams that can give rise to Sudakov logarithms in 
                       the one-loop gauge-boson self-energy}}
\end{figure}
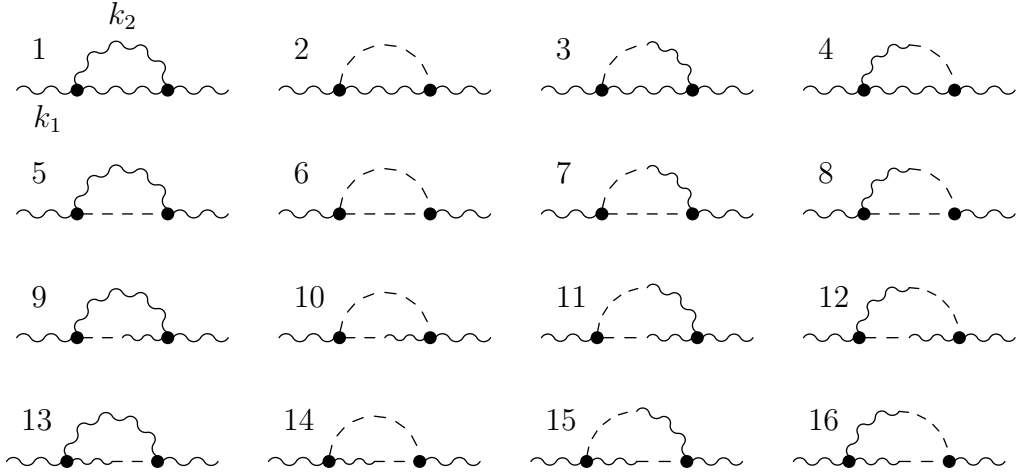
\end{center}
\end{minipage} 

\vspace*{8mm}
\ni
The only way to obtain $\G_{\mu\nu}$ contributions from the $W$-boson 
self-energy is by having a gauge boson as lower (energetic) particle in the 
loop. The mixed and scalar propagators do not contain a $\G_{\alpha\beta}$ 
term and hence there is no way to contract the Lorenz index $\nu$ through to 
the other side of the diagram along the lower line. In principle this can be
circumvented by having a gauge boson as upper (soft) particle in the loop. 
However, as we have seen in Sect.~\ref{sec:one-loop_fermions}, the presence of
$1/|\vec{k}_2|$ is crucial for obtaining Sudakov logarithms. This eliminates
the $\G_{\alpha\beta}$ term in the propagator of the soft gauge boson as well 
as the $\G_{\mu\nu}$ term from the tensor reduction of 
$k_{2\,\mu}k_{2\,\nu}/\vec{k}_2^2 \Myto{\G_{\mu\nu}}{} k_{2\,\bot\,\mu} 
k_{2\,\bot\,\nu}/\vec{k}_2^2 \le 1$.
Hence we are left with the first four diagrams, out of which diagram (2) does 
not contribute. In diagram (2) the scalar propagator exhibits the required pole
structure only at subleading order $\,\propto \mw^2\,$ and on top of that an 
additional factor of $\,\mw^2\,$ originates from the two vertices, leading to 
an overall $\,\mw^4\,$ suppression.  
\\[-5mm]

\ni
We start by calculating the $\G_{\mu\nu}$ part of diagram (1):

\vspace*{6mm}
 \begin{minipage}[c]{15.5cm}
\vspace{-1.3cm}
\begin{figure}[H]
\hspace{7cm}
  \begin{fmffile}{selfbnew}
  \begin{fmfgraph*}(140,140) \fmfpen{thin} \fmfleft{i1} \fmfright{o1}
  \fmf{boson,tension=0.7,label=$\scriptstyle{W^{\pm}(k_1,,\mw)}$,l.side=right,
       width=0.6}{i1,v1} 
  \fmf{boson,tension=0.5,label=$\scriptstyle{V_4(k_1\!-\!k_2,,M_4)}$,
       l.side=right,width=0.6,l.d=15}{v1,v3} 
  \fmf{boson,tension=0.7,label=$\scriptstyle{W^{\pm}(k_1,,\mw)}$,l.side=right,
       width=0.6}{v3,o1} 
  \fmfdot{v1} \fmfdot{v3} 
  \fmf{boson,left,tension=-1,label=$\scriptstyle{V_3(k_2,,M_3)}$,l.side=left,
       width=0.6}{v1,v3}
 \fmf{phantom,left}{v1,v3}
  \fmffreeze
\fmfv{label=$\mu$,l.a=45,}{i1} \fmfv{label=$\nu$,l.a=135}{o1}
  \end{fmfgraph*}
  \end{fmffile}
\end{figure}
{}\vspace*{-4.3cm}
\bea
\hspace{-5cm}  \,i\, \Sigma_{\sscr{WW},\,\mu\nu}^{\,(1)\,, V_4}\,(V_3)
               \,\,\,\,\, =\,   \no
\ea
{}\vspace*{0.2cm}
\end{minipage}\\

\ni
yielding 
\begin{align}
 \,i\,\Sigma_{\sscr{WW},\,\mu\nu}^{\,(1)\,,V_4}(V_3)\,& = 
     \int\frac{\dd^4 k_2}{(2\,\pi)^4}\,P^{\sigma \sigma^{\prime}}(k_2,M_3)\,
     P^{\rho\rhop}(k_1\!-\! k_2,M_4)\,(ie\,G_{134})\,V_{\mu\sigma\rho}\, 
     (ie\,G_{243})\,V_{\nu\rho^{\,\prime}\sigma^{\prime}}\,.
\end{align}
The totally antisymmetric coupling
$\,e\,G_{ijl}\,$ is the triple gauge-boson coupling with all three gauge-boson
lines ($i,j,l$) defined to be incoming at the interaction vertex, \ie we are
dealing with 
$G_{134}\,G_{243} = G_{W^{\pm} V_3 V_4}\,G_{W^{\mp} V_4^{\dagger}V_3^{\dagger}}
= G_{W^{\pm} V_3 V_4}^{\,2}$ 
in the above expression. In our convention this coupling is fixed according to 
$\,G_{\ga W^+ W^-}=1\,$ and 
$\,G_{Z W^+ W^-}=-\cos\theta_{\rm w}/ \sin\theta_{\rm w}$. 
The tensor structures of the two triple gauge-boson interactions read
\bea
\label{vertex1full}
  V_{\mu \sigma \rho} &=& (k_1+k_2)_\rho \,\G_{\mu\sigma}
       + (-k_2+k_1-k_2)_{\mu} \,\G_{\sigma\rho} 
       + (-k_1+k_2-k_1)_{\sigma} \,\G_{\mu\rho} \no \\[4mm]
& \to & -\,(2\,k_1-k_2)_{\sigma} \,\G_{\mu\rho} \\[2mm]
\label{vertex2full}
  V_{\nu\rho^{\,\prime}\sigma^{\prime}} &=& 
       -\,\biggl[ (k_1-k_2+k_1)_{\sigma^{\prime}} \,\G_{\rho^{\,\prime}\nu}
                + (k_2-k_1+k_2)_\nu \,\G_{\sigma^{\prime}\rho^{\,\prime}}
                + (-k_1-k_2)_{\rhop} \,\G_{\sigma^{\prime}\nu} \biggr]
  \no \\[3mm]
& \to & -\,(2\,k_1-k_2)_{\sigmap} \,\G_{\rhop\nu}\,, 
\ea
where we have selected the part that will eventually lead to $\G_{\mu\nu}$ 
once we consider the effective replacement $P^{\rho\rhop}(k_1\!-\! k_2,M_4) \to
-\,i\,\G^{\rho\rhop}/[(k_1-k_2)^2-M_4^2+i\eps\,]$ (see discussion above).
Let us leave the soft particle $V_3$ unspecified for the time being, \ie $V_3
= \ga, Z, W$ are all possible.  With  
\begin{align}
\label{4ppPmass}
 &\h{-10} i\,\Big[ k_2^2-M_3^2+i\eps\,\Big]\,(2\,k_1-k_2)_{\sigma}\,
          (2\,k_1-k_2)_{\sigmap}\,\,P^{\sigma\sigmap}(k_2,M_3) \no \\[1mm]
 &\approx \,\,\, -\,(2\,k_1-k_2)_{\sigma}\,(2\,k_1-k_2)_{\sigmap}\, 
          \frac{k_{2\,0}}{\vec{k_2}^2}\,\Big( k_{2\,\sigma}\,n_{\sigmap} 
          + n_{\sigma}\,k_{2\,\sigmap} \Big) \no \\
 &\approx \,\,\, -\,\frac{2\,k_{2\,0}}{\vec{k_2}^2}\,
          \Big( 2\,k_1\c k_2-k_2^2 \Big)\,(2\,k_{1\,0}-k_{2\,0}) 
 \ \approx\ \frac{4\,k_{1\,0}\,k_{2\,0}}{\vec{k_2}^2}\,\,
            \Big[ (k_1-k_2)^2-k_1^2 \Big]\,,
\end{align}
we obtain for the $\G_{\mu\nu}$ part of the $W$-boson self-energy
\begin{align}
\label{sigmaW}
 i\,\Sigma_{\sscr{WW},\,\G}^{\,(1)\,,V_4}\,(V_3) &\approx \,{\,e^2}\, 
       G_{134}\,G_{243}\,\int\frac{\dd^4 k_2}{(2\,\pi)^4}\, 
       \frac{(2\,k_1-k_2)_{\sigma}\,(2\,k_1-k_2)_{\sigmap}\,
             i\,P^{\sigma\sigmap}(k_2,M_3)}
            {[(k_1-k_2)^2-M_4^2+i\eps\,]} \no \\[2mm]
                                                 &\approx \,{\,e^2}\, 
       G_{134}\,G_{243}\,\int\frac{\dd^4 k_2}{(2\,\pi)^4}\,
       \frac{4\,k_{1\,0}\,k_{2\,0}}{\vec{k_2}^2}\, 
       \frac{(k_1-k_2)^2-M_4^2\, +M_4^2-k_1^2}
            {[(k_1-k_2)^2-M_4^2+i\eps\,]\,[ k_2^2-M_3^2+i\eps\,]} \no \\[1mm]
                                                 &\approx \,-{\,e^2}\, 
       G_{134}\,G_{243}\,\int\frac{\dd^4 k_2}{(2\,\pi)^4}\,
       \frac{4\,k_{1\,0}\,k_{2\,0}}{\vec{k_2}^2}\, 
       \frac{k_1^2-M_4^2}{[(k_1-k_2)^2-M_4^2+i\eps\,]\,[k_2^2-M_3^2+i\eps\,]}
       \,,
\end{align}
in the Sudakov limit.
\\[-5mm]

\ni
In the case of $V_3$ being a neutral gauge boson $(N)$ and hence $V_4$ being
the $W$ boson ($M_4=\mw$) we are left with
\begin{subequations} 
\begin{align}
\label{sigmaWga}
  i\,\Sigma_{\sscr{WW},\,\G}^{\,(1)\,}\,(\ga) 
  & = [k_1^2 -\mw^2]\,\,{\cal F}(\la,\mw) \\[1mm]
\label{sigmaWZ}
  i\,\Sigma_{\sscr{WW},\,\G}^{\,(1)\,}\,(Z) 
  & = \frac{\cwto}{\swto}\,[k_1^2 -\mw^2]\,\,{\cal F}(\mz,\mw)\,,
\end{align}
\end{subequations}
with 
\bea
\label{calF}
  {\cal F}(M_3,M_4) = -\,e^2\,\int\frac{\dd^4 k_2}{(2\,\pi)^4}\,
          \frac{4\,k_{1\,0}\,k_{2\,0}}{\vec{k_2}^2}\, 
          \frac{1}{[(k_1-k_2)^2-M_4^2+i\eps\,]\,[k_2^2-M_3^2+i\eps\,]}\,.
\ea

\ni
Now from Eq.~(\ref{ZW}) we recall that 
\bea
  \delta Z^{\,(1)}_{\sscr{W}_T} 
  = 
  \left. \frac{\Sigma_{\sscr{WW},\,\G}^{\,(1)}}{k_1^2-\mw^2}
  \right|_{_{\scr{\,k_1^2=\mw^2}}} \,,
\ea

\ni
hence the Sudakov correction factor reads
\begin{align}
\label{b1loopa}
\delta Z_{\sscr{W}_T}^{\,(1)}(N) & = (-i)\,\,G_{134}\,G_{243}\,\, 
                                     {\cal F}(\mn,\mw) \,.
\end{align}

\vspace*{3mm}
\ni
Note here that diagrams (3) and (4) do {\it not}  contribute in the above
case of $V_3$ being a neutral particle. First of all, the photon does not have
a would-be Goldstone boson partner. Secondly, the $Z\chi$ and $\chi Z$ mixing
propagator is at both ends attached to two $W$ bosons, leading to a vanishing
contribution since in the SM the $\chi$ does not couple to two $W$ bosons. 
Apart from the couplings, Eq.~(\ref{b1loopa}) is identical to 
Eq.~(\ref{f1loop}). Hence the required steps to eventually obtain the Sudakov
logarithms are identical to the ones given explicitly in the fermion sector. 
The one-loop Sudakov correction factors for transverse $W$ bosons and a soft
neutral gauge boson in the loop are then given by
\begin{subequations} 
\label{bdelta1}
\bea
\label{bphoton1}
  \delta Z^{\,(1)}_{\sscr{W}_T}\,(\gamma) &=& \,Q_{\sscr{W}}^2\, 
                                              {\rm L_{\ga}}(\la,M) \\[2mm]
\label{bZ1}
  \delta Z^{\,(1)}_{\sscr{W}_T}\,(Z) &=& \,\frac{\cwto}{\swto}\,{\rm L}(M,M)
  \ =\ \left[ \frac{1}{\swto} -  Q_{\sscr{W}}^2 \right]\,{\rm L}(M,M) \,,
\ea
\end{subequations} 
with ${\rm L_{\ga}}(\la,M)$ and $ {\rm L}(M,M)$ being defined in
Eqs.~(\ref{Lgamma}) and (\ref{L}), respectively.
\\[-5mm]

\ni
In the case of $V_3$ being the $W$ boson and hence $V_4$ being either a photon
or a $Z$ boson we also have to calculate diagrams (3) and (4). Leaving the 
charge of the mixed propagator general, we obtain

\begin{minipage}[c]{15.5cm}
\vspace*{-8mm} 
\begin{figure}[H]
\hspace*{6.8cm}
  \begin{fmffile}{selfmixnew}
  \begin{fmfgraph*}(140,140) \fmfpen{thin} \fmfleft{i1} \fmfright{o1}
  \fmf{boson,tension=0.9,label={$\scr{W^{\pm}(k_1,,\mw)}$},l.side=right,
       width=0.6}{i1,v1} 
  \fmf{boson,tension=0.5,label={$\scr{N(k_1-k_2,,\mn)}$},l.side=right,
       width=0.6,l.d=15}{v1,v3} 
  \fmf{boson,tension=0.9,label={$\scr{W^{\pm}(k_1,,\mw)}$},l.side=right,
       width=0.6}{v3,o1} 
  \fmfdot{v1} \fmfdot{v3} 
\fmf{phantom,left,tension=0.15,width=0.6,tag=1}{v1,v3}
\fmfposition
\fmfipath{p[]}
\fmfiset{p1}{vpath1(__v1,__v3)}
\fmfi{dashes,label={\small $\phi(k_2)$},l.side=left,
      width=0.6}{subpath (0,length(p1)/2) of p1}
\fmfi{boson,label={\small $W(k_2)$},l.side=left,
      width=0.6}{subpath (length(p1)/2,length(p1)) of p1}
  \fmf{phantom,left}{v1,v3}
  \fmffreeze
\fmfv{label=$\mu$,l.a=45,}{i1} \fmfv{label=$\nu$,l.a=135}{o1}
  \end{fmfgraph*}
  \end{fmffile}
\end{figure}
{}\vspace*{-4.3cm}
\begin{align}
\hspace{2.8cm}  \,i\,\Sigma_{\sscr{WW},\,{\mu\nu}}^{\,(1)\,,\,N}\,(\phi W)
                \,\,\,\,\, & = \, \no \\[10mm]
&\hspace*{-65mm} = \int\frac{\dd^4 k_2}{(2\,\pi)^4}\,(-\,ie\,\widetilde{G}_N 
                   \,\mw)\,\G_{\mu\rho}\,P^{\rho\rhop}(k_1\!-\! k_2,\mn)
                   \,\frac{\mp\,i\mw}{[k_2^2-\mw^2+i\eps\,]}\, 
                   \frac{k_{2\,0}}{\vec{k_2}^2}\,n^{\sigmap}
                   (ie\,G_{W^{\mp}NW^{\pm}})\,V_{\nu\rhop\sigmap}\,,
\end{align}
\vspace*{0.2cm}
\end{minipage}\\
where we have introduced the abbreviation $\,\widetilde{G}_{\gamma}= 1\,$ and
$\,\widetilde{G}_{Z}= \sw/\cw$. Again the triple gauge-boson vertex can be 
simplified according to Eq.~(\ref{vertex2full}):
$V_{\nu\rhop\sigmap} \to -\,(2\,k_1-k_2)_{\,\sigmap} \,\G_{\rhop\nu}$.
Bearing in mind that $k_{2\,\sigmap}\,n^{\sigmap}$ together with the already 
present factor $k_{2\,0}$ will kill the crucial factor $1/\vec{k_2}^2$, we can
safely ignore the $k_{2\,\sigmap}$ term. Selecting the $\G_{\mu\nu}$ part we 
obtain
\begin{align}
\label{sigmaWmix3}
  \,i\,\Sigma_{\sscr{WW},\,\G}^{\,(1)\,,\,N}\,(\phi W)\, & =
  -\,e^2\,\widetilde{G}_N\,G_{W^{\mp}NW^{\pm}}\,
  \int\frac{\dd^4 k_2}{(2\,\pi)^4}\,\frac{2\,k_{2\,0}\,k_{1\,0}}{\vec{k_2}^2}\,
  \frac{\mp\,\mw^2}{[(k_1-k_2)^2-\mn^2+i\eps\,]\,[k_2^2-\mw^2+i\eps\,]} \,,
\end{align}
and for the contribution from diagram (4) we can immediately write
\begin{align}
\label{sigmaWmix4}
  \,i\,\Sigma_{\sscr{WW},\,\G}^{\,(1)\,,\,N}\,(W \phi)\, & =
  -\,e^2\,\widetilde{G}_N\,G_{W^{\pm}W^{\mp}N}\, 
  \int\frac{\dd^4 k_2}{(2\,\pi)^4}\,\frac{2\,k_{2\,0}\,k_{1\,0}}{\vec{k_2}^2}\,
  \frac{\mp\,\mw^2}{[(k_1-k_2)^2-\mn^2+i\eps\,]\,[k_2^2-\mw^2+i\eps\,]} \,.
\end{align}
Note that Eqs.~(\ref{sigmaWmix3}) and (\ref{sigmaWmix4}) are identical as a 
result of $G_{W^{\pm}W^{\mp}N}=G_{W^{\mp}NW^{\pm}}$.  
Hence upon adding the contributions corresponding to diagrams (1), (3) and
(4), we find for the neutral particle being a photon 
[\ie $G_{W^{\pm}W^{\mp}\ga} = G_{W^{\mp}\ga W^{\pm}} = \pm 1$, 
$\widetilde{G}_{\ga} = 1$]
\begin{align}
\label{sigmaga}
   \,i\,\Sigma_{\sscr{WW},\,\G}^{\,(1)\,,\,\ga}\,( {\scriptstyle W + \phi W 
   + W\phi} )\, & = \Big( [k_1^2-\la^2] - \mw^2 \Big)\,{\cal F}(\mw,\la) \,,
\end{align}
where we can neglect the photon mass $\la$ in the prefactor.
For the neutral particle being the $Z$ boson 
[\ie $G_{W^{\pm}W^{\mp}Z} = G_{W^{\mp} ZW^{\pm}} = \mp\cw/\sw$,
$\widetilde{G}_Z= \sw/\cw$] we find
\begin{align}
\label{sigmaZ}
  \,i\,\Sigma_{\sscr{WW},\,\G}^{\,(1)\,,\,Z}\,( {\scriptstyle W + \phi W 
  + W\phi} ) & = \!\left( \frac{\cwto}{\swto}\,[k_1^2\!-\!\mz^2]+\mw^2 \right)
                 {\cal F}(\mw,\mz) 
             \,=\,\frac{\cwto}{\swto}\,[k_1^2\!-\!\mw^2]\,{\cal F}(\mw,\mz)\,,
\end{align}
making use of $\,\mz^2\,\cwto= \mw^2$. In all situations we find 
$\,\Sigma_{\sscr{WW},\,\G}^{\,(1)} \propto (k_1^2-\mw^2)$, in agreement with 
the ``non-renormalization'' condition (\ref{nonrenMW}) in Sect.~\ref{sec:wave}.
This proves that for the $W$-boson self-energy the derivative method gives the 
correct results, which in fact is true to all orders in perturbation theory. 
\\[-5mm]

\ni
The $\G_{\mu\nu}$ part of the full self-energy for neutral particles can be
obtained from Eqs.~(\ref{sigmaW}), (\ref{sigmaWmix3}) and (\ref{sigmaWmix4}),
bearing in mind that we have two identical contributions from the two soft
limits:
\begin{subequations} 
\begin{align}
  i\,\Sigma_{\gamma\gamma,\,\G}^{\,(1)}( {\scriptstyle W + \phi W 
  + W\phi} )\, & =\,2\,\Big( [k_1^2-\mw^2]+\mw^2 \Big)\,{\cal F}(\mw,\mw) 
  \\[3mm]
  i\,\Sigma_{\sscr{ZZ},\,\G}^{\,(1)}( {\scriptstyle W + \phi W 
  + W\phi} )\, & =\,2\left( \frac{\cwto}{\swto}\,[k_1^2-\mw^2] - \mw^2 \right)
                  {\cal F}(\mw,\mw) 
  \\[2mm]
  i\,\Sigma_{\gamma\sscr{Z},\,\G}^{\,(1)}( {\scriptstyle W + \phi W 
  + W\phi} )\, & =\,2\left( -\,\frac{\cw}{\sw}\,[k_1^2-\mw^2] 
                  - \frac{1}{2}\left( \frac{\cw}{\sw}-\frac{\sw}{\cw} \right) 
                  \mw^2 \right) {\cal F}(\mw,\mw)
  \\[2mm]
  i\,\Sigma_{\sscr{Z}\gamma,\,\G}^{\,(1)}( {\scriptstyle W + \phi W 
  + W\phi} )\, & =\,2\left( -\,\frac{\cw}{\sw}\,[k_1^2-\mw^2] 
                  + \frac{1}{2}\left( \frac{\sw}{\cw}-\frac{\cw}{\sw} \right)
                  \mw^2 \right) {\cal F}(\mw,\mw) \,.
\end{align}
\end{subequations} 
Hence we find 
\begin{subequations}  
\begin{align}
  i\,\Sigma_{\gamma\gamma,\,\G}^{\,(1)}( {\scriptstyle W + \phi W 
  + W\phi} )\, & =\,2\,[\, k_1^2 \,]\,{\cal F} (\mw,\mw)  
  \\[3mm]
  i\,\Sigma_{\sscr{ZZ},\,\G}^{\,(1)}( {\scriptstyle W + \phi W 
  + W \phi} )\, & =\,2\,\frac{\cwto}{\swto}\,[\, k_1^2-\mz^2 \,]\,
                   {\cal F}(\mw,\mw) 
  \\[2mm]
  i\,\Sigma_{\gamma\sscr{Z},\,\G}^{\,(1)}( {\scriptstyle W + \phi W 
  + W\phi} )\,=\,i\,\Sigma_{\sscr{Z}\gamma,\,\G}^{\,(1)}( {\scriptstyle W 
  + \phi W + W\phi} )\  & =\, -\,\frac{\cw}{\sw}\,\Big( [\, k_1^2 \,] 
                            + [\, k_1^2 -\mz^2\,] \Big)\,{\cal F}(\mw,\mw)\,,
\end{align}
\end{subequations} 
or generically
\begin{align}
\label{sigmaN}
  \,i\,\Sigma_{\sscr{N_1 N_2},\,\G}^{\,(1)}( {\scriptstyle W + \phi W 
  + W\phi} )\, & = G_{134} G_{243}\,\bigg( [\, k_1^2-M_{\sscr{N_1}}^2 \,] 
                 + [\, k_1^2-M_{\sscr{N_2}}^2 \,] \bigg)\,{\cal F}(\mw,\mw) \,,
\end{align}
with $G_{134} G_{243} = G_{N_1W^{\pm}W^{\mp}}\, G_{N_2 W^{\pm}W^{\mp}}$. From 
this generic expression we see that the diagonal self-energies 
(with $N_1=N_2=N$) are proportional to the inverse pole $(k_1^2-\mn^2)$. This
confirms the ``non-renormalization'' condition (\ref{nonrenN}) at the one-loop
level. The explicit calculation of the full mixed $\ga\,$--$\,Z$ self-energy
confirms the prediction that we made in Sect.~\ref{sec:boson1loop}, which was 
based on ``non-renormalization'' of the masses as well as the electromagnetic 
charge. Since those ``non-renormalization'' conditions hold to all orders,
we can make use of them to calculate the full two-loop mixed $\ga\,$--$\,Z$ 
self-energy based on the knowledge of the derivatives of 
$\,\Sigma_{\sscr{N_1 N_2},\,\G}$.
\\[-5mm] 

\ni We finish this appendix by giving the relevant set of bosonic 
one-loop (non-derivative) self-energies in the SM, using 
$\,{\cal F}(M,M) = {\cal F}(M,\la) = i\,\L(M,M)\,$
and $\,{\cal F}(\la,M) = i\,\L_{\gamma}(\la,M)$. Denoting the momentum of the 
external particles by $k$, the list reads\\

{\h{-8} $\bullet$ \h{1} Charged-boson sector}
\bea
  \Sigma_{\sscr{WW},\,\G}^{\,(1)\,} &=& [k^2 -\mw^2]\left[ \,
    \L_{\gamma}(\la,M) + \left( \frac{2}{\swto}-1 \right)\L(M,M) \right] 
    \\[2mm]
  \Sigma_{\phi\phi}^{\,(1)\,}       &=& -\,[k^2 -\mw^2]\left[ \,
    \L_{\gamma}(\la,M)+ \frac{\cwto}{\swto}\,\L(M,M) \right] \no \\[2mm]
                                    & & -\,k^2\left[  
    \frac{1}{4\,\cwto} - \frac{1}{4\,\swto} \right]\L(M,M) \\[2mm]
  \Sigma_{\sscr{W}^+\phi^+,\,n}^{\,(1)\,} &=& -\,k_0\,\mw\left[ \,
    \frac{1}{8\,\cwto} - \frac{1}{8\,\swto} \right]\L(M,M) 
\ea

{\h{-8} $\bullet$ \h{1} Neutral-boson sector}
\bea
  \Sigma_{\gamma\gamma,\,\G}^{\,(1)}   &=& 2\,\big[\, k^2 \,\big]\,\L(M,M) 
    \\[3mm]
  \Sigma_{\sscr{ZZ},\,\G}^{\,(1)}      &=& 2\,\frac{\cwto}{\swto}\, 
    \big[\, k^2-\mz^2 \,\big]\,\L(M,M) \\[3mm]
  \Sigma_{\gamma\sscr{Z},\,\G}^{\,(1)} &=& 
  \Sigma_{\sscr{Z}\gamma,\,\G}^{\,(1)}\ =\ -\,\frac{\cw}{\sw}\, 
    \Big( k^2 + [\, k^2 -\mz^2\,] \Big)\,\L(M,M) \\[3mm]
  \Sigma_{\gamma\chi,\,n}^{\,(1)}      &=&
  -\,\Sigma_{\chi\gamma,\,n}^{\,(1)}\   =\ -\,i\,k_0\,\mz\,\frac{\cw}{2\,\sw}\,
    \L(M,M) \\[3mm]
  \Sigma_{\sscr{Z}\chi,\,n}^{\,(1)}    &=&
  -\,\Sigma_{\chi\sscr{Z},\,n}^{\,(1)}\ =\ \frac{i}{4}\,k_0\,\mz 
    \left( \frac{\cwto}{\swto }-\frac{1}{2\cwto\,\swto}-1 \right) \L(M,M)
    \\[3mm]
  \Sigma_{\chi\chi}^{\,(1)}            &=& -\,\frac{1}{4\,\swto}
    \left( 2\,k^2+\frac{k^2}{\cwto}-4\,\mz^2\,\cwto \right) \L(M,M) \\[3mm]
  \Sigma_{\sscr{HH}}^{\,(1)}           &=& -\,[k^2 -\mh^2]\left( 
    \frac{3}{4\,\swto} + \frac{1}{4\,\cwto} \right) \L(M,M)\,.
\ea
These expressions are in agreement with the various ``non-renormalization'' 
conditions listed in Sect.~\ref{sec:wave}.

\end{appendix}


\def\np#1#2#3{{\sl  Nucl.~Phys.\/}~{\bf B#1} {(#2) #3}}
\def\spj#1#2#3{{\sl Sov.~Phys.~JETP\/}~{\bf #1} {(#2) #3}}
\def\plb#1#2#3{{\sl Phys.~Lett.\/}~{\bf B#1} {(#2) #3}}
\def\pl#1#2#3{{\sl Phys.~Lett.\/}~{\bf #1} {(#2) #3}}
\def\prd#1#2#3{{\sl Phys.~Rev.\/}~{\bf D#1} {(#2) #3}}
\def\pr#1#2#3{{\sl Phys.~Rep.\/}~{\bf #1} {(#2) #3}}
\def\epjc#1#2#3{{\sl Eur.~Phys.~J.\/}~{\bf C#1} {(#2) #3}}
\def\ijmp#1#2#3{{\sl Int.~J.~Mod.~Phys.\/}~{\bf A#1} {(#2) #3}}
\def\ptps#1#2#3{{\sl Prog.~Theor.~Phys.~Suppl.\/}~{\bf #1} {(#2) #3}}
\def\npps#1#2#3{{\sl Nucl.~Phys.~Proc.~Suppl.\/}~{\bf #1} {(#2) #3}}
\def\sjnp#1#2#3{{\sl Sov.~J.~Nucl.~Phys.\/}~{\bf #1} {(#2) #3}}
\def\hepph#1{{\sl hep--ph}/{#1}}

\end{document}